\documentclass[%
 reprint,
nofootinbib,
 amsmath,amssymb,
 aps,
]{revtex4-2}
\usepackage{graphicx}% Include figure files
\usepackage{dcolumn}% Align table columns on decimal point
\usepackage{bm}% bold math
\usepackage{amsmath}
\usepackage{amsfonts}
\usepackage{amssymb}
 \usepackage{mathrsfs}
\usepackage{ tipa }
\usepackage{tikz}
\usepackage{tensor}
\usepackage{upgreek}
\usepackage{subfig}

\DeclareSymbolFont{toneletters}{T1}{\familydefault}{m}{it}
\DeclareMathSymbol\ethm{\mathord}{toneletters}{"F0}

\usepackage{hyperref}
\usepackage{xcolor}
\usepackage{bbold}
\definecolor{darkblue}{rgb}{0, 0.1, 0.6}
\hypersetup{
    colorlinks=true,
    linkcolor=darkblue,
    filecolor=darkblue,      
    urlcolor=darkblue,
    citecolor = darkblue,
    pdftitle={Polcar \& Witzany: Towards relativistic inspirals into black holes surrounded by matter},
    pdfpagemode=FullScreen,
    }
\usepackage{orcidlink}
\usepackage{fancyhdr}
\usepackage{extramarks}
\makeatletter

\begin{document}

\title{Towards relativistic inspirals into black holes surrounded by matter}% Force line breaks with \\

\author{Luk\'{a}\v{s} Polcar$^{1}$ \orcidlink{0000-0002-9464-2834}}
 \email{polcar.vm@seznam.cz}
\author{Vojt\v{e}ch Witzany$^{1}$ \orcidlink{0000-0002-9209-5355}}%
\affiliation{ Institute of Theoretical Physics, Faculty of Mathematics and Physics, Charles University in Prague, 18000 Prague, Czech Republic}

\begin{abstract}
Extreme mass ratio inspirals, compact objects spiraling into massive black holes, represent key sources for future space-based gravitational-wave detectors such as LISA. The inspirals will occur within rich astrophysical environments containing gravitating matter. Motivated by this, we develop a fully relativistic framework for inspirals under the gravitational influence of matter environments. Our approach employs a two-parameter perturbation expansion in the mass ratio and an environmental parameter. This yields a modified Teukolsky equation capturing the leading cross-order.
We then implement a simple pole-dipole approximation of an axisymmetric environment through a thin matter shell and restrict to non-rotating black holes. As a result, we obtain a piecewise type D spacetime. This enables the use of Teukolsky-based methods while accounting for junction physics. The presence of the matter shell leads to effectively non-separable boundary conditions for the Teukolsky scalar and introduces mode mixing between adjacent multipoles. Additionally, the shell oscillates under the wave perturbation of the inspiral, contributing to the overall flux. The framework provides novel insights into the global dynamics of gravitational radiation in tidal environments. Furthermore, it represents a complete theoretical foundation for a future computation of inspirals and waveforms in our environmental model.
\end{abstract}

\maketitle

\tableofcontents
%%%%%%%%%%%%%%%%%%%%%%%%%%%%%%%%%%%%%%%%%%%%%%%%%%%%%%%%%%%%%%%%%%%%%%%%%%%%%%%%%%%%%%%%%%%%%%%%%%%%%%%%%%%%%%%%%
\section{Introduction}\label{sec:intro}
%%%%%%%%%%%%%%%%%%%%%%%%%%%%%%%%%%%%%%%%%%%%%%%%%%%%%%%%%%%%%%%%%%%%%%%%%%%%%%%%%%%%%%%%%%%%%%%%%%%%%%%%%%%%%%%%%

%%%%%%%%%%%%%%%%%%%%%%%%%%%%%%%%%%%%%%%%%%%%%%%%%%%%%%%%%%%%%%%%%%%%%%%%%%%%%%%%%%%%%%%%%%%%%%%%%%%%%%%%%%%%%%%%%

The detections of gravitational waves (GWs) by the LIGO-Virgo-KAGRA collaboration \cite{LIGOScientific:2018mvr,LIGOScientific:2020ibl,KAGRA:2021vkt} have established GW astronomy as a powerful tool for probing the strong-field regime of general relativity. Building on these ground-based successes, future space-based detectors such as the Laser Interferometer Space Antenna (LISA) \cite{LISA:2017pwj,LISA:2024hlh} will extend observations to the millihertz frequency band. In this band, one of the most prominent and scientifically valuable source classes are the so-called extreme mass ratio inspirals (EMRIs). These involve stellar-mass compact objects ($m \sim 1$--$100 M_\odot$) spiraling into massive black holes ($M \sim 10^4$--$10^7 M_\odot$) \cite{Babak:2017tow}. The long inspiral phase, lasting $\sim 10^4$--$10^6$ orbital cycles within LISA's sensitivity band, enables unprecedented precision in black hole parameter measurement and tests of general relativity in the extreme gravity regime \cite{Gair:2012nm,Berry:2019wgg}. The stellar-mass compact objects in EMRIs will come from dense nuclear clusters in the environments of the massive black holes. Understanding this environment and its interactions with massive black holes in the centers of galaxies is one of the 8 science objectives identified in the LISA definition study report \cite{LISA:2024hlh}.

%%%%%%%%%%%%%%%%%%%%%%%%%%%%%%%%%%%%%%%%%%%%%%%%%%%%%%%%%%%%%%%%%%%%%%%%%%%%%%%%%%%%%%%%%%%%%%%%%%%%%%%%%%%%%%%%%
\subsection{The Teukolsky equation and environmental effects} 

The theoretical modeling of EMRIs has traditionally relied upon the Teukolsky formalism \cite{Teukolsky1,Teukolsky:1974yv}, which provides a separable master equation for curvature perturbations of Kerr black holes. This formalism exploits the remarkable algebraic properties of the Kerr spacetime, specifically its Petrov type D character \cite{Stephani:2003tm,Griffiths:2009dfa}. Using these properties, it separates the four-dimensional perturbation problem into decoupled ordinary differential equations. Superficially, the Teukolsky equation only addresses perturbations to the curvature tensor. However, one can use its solutions to reconstruct the full spacetime metric using the so-called Chrzanowski-Cohen-Kegeles (CCK) formalism \cite{Chrzanowski:1975wv,Kegeles:1979an}. The separability of the Teukolsky equation has enabled extensive calculations of GW emission \cite{Drasco_2006}, orbital evolution under radiation reaction \cite{Sago_2006,Hughes:2021exa}, and black hole quasinormal modes \cite{Berti:2009kk}, forming the theoretical foundation for current GW astronomy. 

However, the standard Teukolsky formalism applies only to a special class of space-times: the vacuum, Petrov type D spacetimes. For asymptotically flat spacetimes without electromagnetic fields or external singularities, this class essentially contains only isolated Kerr black holes \cite{Stephani:2003tm,Griffiths:2009dfa}. 

Such a restriction becomes problematic when modeling realistic astrophysical environments, since massive black holes in galactic nuclei are invariably embedded within complex matter distributions. These include the already mentioned nuclear star clusters \cite{merritt2013dynamics}, various gas and dust structures \cite{neumayer2020nuclear,genzel2010galactic}, dark matter halos \cite{Bertone:2016nfn}, accretion flows, and debris from tidal disruption events \cite{Gezari:2021bmb}. More exotic possibilities for black hole environments include, e.g., clouds from currently undiscovered scalar fields \cite{Ferreira17,Dyson:2025dlj}. These environments influence the inspiral through gravitational perturbations and direct matter interactions. The resulting GW phase shifts accumulate over the long inspiral, potentially reaching observable levels \cite{Narayan:1999ay,Yunes:2011ws,Barausse:2014tra,Polcar_2022}. 

 In our case, we focus exclusively on gravity-mediated effects. Even more specifically, as one draws closer to the black hole, various physical effects tend to disrupt, dilute, and thin out the objects in its vicinity. These effects include tidal forces, increasing rotational shearing and associated instabilities, and gravitational radiation. As a result, only a few massive objects remain in the inner few hundred Schwarzschild radii \citep{merritt2013dynamics}. In most physical cases of interest, the external matter distribution can then be considered as far from the black hole and its field expanded only to a handful of leading tidal multipoles, as we have already done in our preliminary analysis in Ref. \cite{Polcar_2022}. 

%%%%%%%%%%%%%%%%%%%%%%%%%%%%%%%%%%%%%%%%%%%%%%%%%%%%%%%%%%%%%%%%%%%%%%%%%%%%%%%%%%%%%%%%%%%%%%%%%%%%%%%%%%%%%%%%%
\subsection{Phase contribution analysis} 

How, then, can we quantify that we have correctly included all environmental effects in a waveform model? A useful estimate of the waveform accuracy is the accuracy of the phase elapsed in the frequency band of the GW detector \cite{Flanagan:1997sx}. Specifically, one needs to identify all the sources of the phase shift that are induced by the environment over the course of the long inspiral.

To analyze the elapsed phase, one typically resorts to a two-timescale expansion \cite{kevorkian2012multiple}. This approach exploits the fact that radiation reaction acts significantly only over many orbital periods in EMRIs \cite{Hinderer:2008dm,Pound_2021}. At leading order, the system is then viewed as a conservative system that is led to adiabatic decay through a ``leakage'' of energy and angular momentum through GW fluxes. This picture is sufficient for us to characterize the leading effects of external environmental perturbations in EMRIs during the inspiral stage but would have to be refined for plunge and merger. 

Where do environmental effects appear in adiabatic inspirals? The key insight is that environmental corrections appear in \emph{both} the conservative dynamics (orbital energy as a function of frequency) and the dissipative dynamics (GW energy flux). Consider the phase $\phi_{\rm el}$ elapsed when passing through the frequency band $\Omega \in [\Omega_1,\Omega_2]$ of a detector during an adiabatic quasi-circular inspiral. We choose to parametrize the dynamics by the relations of the energy of the system as a function of angular orbital frequency $E(\Omega)$ and the GW energy flux as a function of frequency $\dot{E}_{\rm GW} (\Omega)$. Then we have
\begin{align}
    \phi_{\rm el} = \int_{\Omega_1}^{\Omega_2} {\rm d} \phi(\Omega) = \int_{\Omega_1}^{\Omega_2} \frac{{\rm d}E}{{\rm d} \Omega} \frac{\Omega}{\dot{E}_{\rm GW} } {\rm d} \Omega\,.
\end{align}
The $E(\Omega)$ relation is related to the conservative dynamics, while $\dot{E}_{\rm GW}$ corresponds to the overall radiative, or ``dissipative'' dynamics. Even though $\dot{E}_{\rm GW}$ is a small quantity, its leading order term appears directly in the leading-order phase. Consequently, the sub-leading corrections to $\dot{E}_{\rm GW}$ due to an environmental perturbation contribute at the same order as corrections to the $E(\Omega)$ relation. In the context of EMRIs in non-Kerr space-times, one often sees works that treat only the conservative dynamics ($\sim$ the $E(\Omega)$ relation) while essentially ignoring the full set of corrections to the GW fluxes ($\sim$ the $\dot{E}_{\rm GW}(\Omega)$ relation). 

What, then, is needed to obtain all the leading environmental contributions to the GW phase in EMRIs? The conservative dynamics are simply obtained by corrections to the motion of test particles in the field perturbed by the environment. For example, one can analyze the motion in fields where the far-away gravitating matter is parametrized by tidal multipoles superposed with the black hole \cite{Vieira:1996zf,Vieira:1998hc,deMoura:1999wf,Bonga:2019ycj,Polcar_2022}, or specific sources can be considered in the close environment \cite{Semerak:2010lzj,Semerak:2012dx,Sukova:2013jxa,Witzany:2015yqa,Polcar:2019wfi,Polcar:2019kwu}. The dissipative terms are more complicated. Given that we focus only on gravity-mediated interactions and ignore direct matter interactions between the inspiraling objects and the environment, one must compute two different corrections to the GW flux:
\begin{enumerate}
    \item The correction caused by the fact that the test-particle motion is different due to the environment's gravity.
    \item The correction caused by the fact that the dynamics of the waves emerging from the inspiral \textit{themselves} are modified in the background deformed by the environment.
\end{enumerate}
Only when \textit{all} of the contributions above have been computed can we say that all leading-order gravitational effects of the environment have been included. The second correction to the GW flux is the most challenging one, and it is also the main focus of our paper. In particular, finding this correction requires solving a new wave equation that goes beyond the usual Teukolsky formalism in pure Kerr space-time.

Perhaps the conceptually closest calculations to ours were performed by \citet{Cardoso:2022whc}, \citet{Duque:2023seg}, \citet{Datta} and \citet{Rahman}, who studied EMRIs into black holes surrounded by matter using fully relativistic and self-consistent black hole perturbation theory. However, these works assume a spherically symmetric background, whereas this paper develops a framework that should apply in environments with no particular symmetries, and we build a specific example for an axisymmetric matter distribution.

Other related relativistic studies include those of \citet{Dyson:2025dlj}, \citet{liscalar} and \citet{Brito:2023pyl}, which consider EMRIs into black holes surrounded by scalar clouds. These clouds emit scalar radiation during the inspiral, but the corresponding gravitational wave fluxes include only partial contributions from the scalar field. As such, these treatments are less complete than those in Refs \cite{Cardoso:2022whc,Duque:2023seg,Datta,Rahman} and the present work. Another relativistic model that considers the interaction between an EMRI and a matter halo is that of \citet{Vicente:2025gsg}. In this study, the authors examine the effects of the matter through a semi-local model for the dynamical friction exerted on the secondary by the halo. However, they do not include other effects of the matter halo on the space-time background and wave generation as considered in this work.

%%%%%%%%%%%%%%%%%%%%%%%%%%%%%%%%%%%%%%%%%%%%%%%%%%%%%%%%%%%%%%%%%%%%%%%%%%%%%%%%%%%%%%%%%%%%%%%%%%%%%%%%%%%%%%%%%
\subsection{Previous work on extensions of the Teukolsky formalism} 

Let us note that a number of works have already moved towards an extension of the linear Teukolsky formalism, even though often in a different context or with a different application in mind. The second-order perturbation theory framework developed by \citet{Campanelli_1999} and refined by \citet{spiersmoxon} provides a systematic approach to nonlinear perturbations, though its application remains restricted to Kerr backgrounds. 

The modified Teukolsky equations derived by \citet{Li_2023} for perturbations of spinning black holes in modified gravity theories demonstrate how the formalism can be extended to non-Einstein equations, while exploiting the separability of the perturbations within Einstein gravity. \citet{Hussain:2022ins} and \citet{Cano_2023} have also derived Teukolsky-like equations in modified-gravity settings and used them to compute the shifted quasi-normal mode spectrum of black holes (see also Ref. \cite{Ghosh:2023etd}). 

%%%%%%%%%%%%%%%%%%%%%%%%%%%%%%%%%%%%%%%%%%%%%%%%%%%%%%%%%%%%%%%%%%%%%%%%%%%%%%%%%%%%%%%%%%%%%%%%%%%%%%%%%%%%%%%%%
\subsection{Research goal and organization of this paper}

In conclusion, to obtain complete first-principle models of the phase contribution of environments to EMRIs, we must develop a perturbation theory of black holes surrounded by gravitating matter. This is also necessary to provide meaningful constraints on the environment through observatories such as LISA and to disentangle these effects from the deviations possibly caused by modified gravity and other exotic scenarios. 

In our previous Ref.~\cite{Polcar_2022}, we analyzed the orbital motion in a tidal environment and used a simplified (or ``kludge'') Newtonian quadrupole formula to drive inspirals. In this work, we carry out a fully relativistic, strong-field analysis of the generation of GW fluxes, which should replace the aforementioned kludge with an accurate first-principle model. In particular, we develop a modified Teukolsky formalism and demonstrate in detail how to deal with each aspect of the global problem of the GW flux computation. This paper is the first in a two-part series. The second paper will include the results of a complete implementation of the formalism, along with concrete examples of how tidal environments affect EMRIs.

The remainder of this paper is organized as follows. In Section \ref{sec:Teukolsky formalism} we introduce the Teukolsky equation, and its modifications in a double expansion in both the environmental perturbation and the mass of the secondary in the EMRI. Section~\ref{sec:holeringpert} then restricts the focus to a concrete simple model of the environment, a ``pole-dipole ring'', and specializes to slowly rotating black holes. The Section also establishes the Newman-Penrose quantities for the background and decomposes the modified Teukolsky equation into pieces that can be computed separately. In Section~\ref{sec:solteuk} we deal with the fact that the outgoing GWs from the EMRI also act locally on the environment. This requires local metric reconstruction and solving the equations of the perturbed matter in the environment. Finally, we present a summary of our results in Section \ref{sec:summary} and discuss future directions in Section~\ref{sec:discussion}. The paper also contains a number of appendices that provide more technical details. We also provide Mathematica notebooks with our computations  on github \cite{MmaNB}. Throughout, we employ geometrized units with $G = c = 1$ and the metric signature $(-,+,+,+)$.

%%%%%%%%%%%%%%%%%%%%%%%%%%%%%%%%%%%%%%%%%%%%%%%%%%%%%%%%%%%%%%%%%%%%%%%%%%%%%%%%%%%%%%%%%%%%%%%%%%%%%%%%%%%%%%%%%
\section{Teukolsky formalism and its modification}\label{sec:Teukolsky formalism}
%%%%%%%%%%%%%%%%%%%%%%%%%%%%%%%%%%%%%%%%%%%%%%%%%%%%%%%%%%%%%%%%%%%%%%%%%%%%%%%%%%%%%%%%%%%%%%%%%%%%%%%%%%%%%%%%%

Is there a way to take advantage of the symmetries of isolated black holes when formulating the perturbation theory of black holes in environments? The answer is affirmative, as we will see in this section with the introduction of our modified Teukolsky formalism. We will start by introducing the basics of the perturbations of Kerr space-time and then derive the equation governing perturbations of black holes in slightly non-Kerr space-times. This section develops our modified Teukolsky formalism in three steps: first reviewing the standard approach in Sections \ref{subsec:higlysym} to \ref{Teukolsky Kerr}, then deriving our generalization in Section \ref{modifTueeqderivation}, and finally analyzing its key components in Section \ref{modifTeukterms}. This will serve as a basis for the concrete physical setup we shall deal with in Section \ref{sec:holeringpert}.

The perturbation theory we consider here shares many similarities with the second-order black-hole perturbation theory as discussed by Refs \cite{Campanelli_1999,spiersmoxon}. In particular, we used many of the technical results of the second-order Teukolsky equation treatment by \citet{spiersmoxon}. From the context of modified-gravity extensions of the Teukolsky formalisms, our notation and discussion follow to some degree that of \citet{Li_2023}. 

%%%%%%%%%%%%%%%%%%%%%%%%%%%%%%%%%%%%%%%%%%%%%%%%%%%%%%%%%%%%%%%%%%%%%%%%%%%%%%%%%%%%%%%%%%%%%%%%%%%%%%%%%%%%%%%%%
\subsection{Perturbation formalism in   type D vacuum spacetimes }
\label{subsec:higlysym}

To find the metric perturbation $h^{(1)}_{\mu\nu}$ corresponding to the gravitational field sourced by an inspiraling particle, one needs to solve the linearized Einstein equations
\begin{align}
\label{linEineq}
\delta G_{\mu\nu}[ h^{(1)}]=8 \uppi \hspace{1pt} T^{(1)}_{\mu\nu}\,,
\end{align}
where $\delta G_{\mu\nu}$ is the linearized Einstein operator while the particle is moving in the background geometry described by the metric $g^{(0)}_{\mu\nu}$  along the geodesic $x^\mu(\tau)$ with four-velocity $u^{\mu}(\tau)$, is represented by the  stress-energy tensor
\begin{align}
\label{particleTmunu}
T^{(1)}_{\mu\nu}=m\frac{u_{\mu}(\tau)u_{\nu}(\tau)}{u^{0}(\tau)\sqrt{-g}}\delta^{3}(x^i-x^i(\tau)).
\end{align}
Naturally, the system of  equations  \eqref{linEineq} is  complicated to solve in the generic case; however, if the background metric $g^{(0)}_{\mu\nu}$ is highly symmetric, the problem simplifies significantly. In particular, for spacetimes in the Petrov type D vacuum class, Teukolsky \cite{Teukolsky1} employed Riemann tensor identities within the Newman-Penrose formalism to derive a single dynamical equation equivalent to the system \eqref{linEineq}. This equation can then be written in the form \cite{Pound_2021}
\begin{align}
\label{Teukabstract}
\mathcal{O}^{(0)}\Psi^{(1)}=\mathcal{T}^{(1)} \,,
\end{align}
where $\mathcal{O}^{(0)}$ is a second-order differential operator, and $\mathcal{T}^{(1)}$ is obtained from a second-order differential operator acting on $T^{(1)}_{\mu\nu}$. The unknown variable $\Psi^{(1)}$  is then a perturbation of a Weyl scalar (either $\Psi_0$ or $\Psi_4$ ) from which the complete metric   $h^{(1)}$ can be reconstructed. The Teukolsky equation \eqref{Teukabstract} is typically solved on the Kerr background as it is the most physically relevant type-D black hole spacetime. The Teukolsky equation can be written in a fully separable form, thus decomposing the variable $\Psi^{(1)}$ into modes, where each satisfies an ordinary differential equation. 

For type D spacetimes like Kerr, the Teukolsky framework provides an elegant solution method through mode decomposition. However, beyond type D backgrounds, analytical perturbative techniques become significantly more challenging. A natural approach is to extend the Teukolsky formalism by treating non-type D backgrounds as perturbations of type D vacuum spacetimes, thus expanding relevant quantities in two perturbation parameters. In the remaining parts of this section, we review the Newman-Penrose and Geroch-Held-Penrose formalisms as well as the Teukolsky formalism in the Kerr spacetime, after which we derive the modified Teukolsky equation for more general backgrounds.

%%%%%%%%%%%%%%%%%%%%%%%%%%%%%%%%%%%%%%%%%%%%%%%%%%%%%%%%%%%%%%%%%%%%%%%%%%%%%%%%%%%%%%%%%%%%%%%%%%%%%%%%%%%%%%%%%
\subsection{The NP and the GHP formalism}
\label{subsec:NPGHP}

The Newman-Penrose (NP) formalism provides a powerful framework for analyzing spacetime curvature by using projections into null tetrads (for more details, see Appendix \ref{app:NPGHP}). In return, one obtains simplifications of the Einstein equations in highly symmetric spacetimes. The formalism starts by defining the complex null tetrad \cite{Newman:1961qr,Chandrasekhar:1984siy}
\begin{align}
\label{NPtetrad}
e^{\mu}_{a}=\lbrace l^\mu, n^\mu, m^\mu, \bar{m}^\mu \rbrace,
\end{align}
where the first two vectors $l$ and $n$ are real null vectors, while $m$ is a complex null vector which completes the basis along with its conjugate $\bar{m}$. The nonzero scalar products of these vectors are
\begin{align}
\label{NPtetradscalproduct}
l^\mu  n_\mu=-1, \hspace{15pt}   m^\mu \bar{m}_\mu=1.
\end{align}
We can now project any tensor onto the NP tetrad. For example, for $A_{\mu\nu}$ we have
\begin{align}
A_{\mu\nu}\rightarrow A_{ab}=A_{\mu\nu}e^{\mu}_{a}e^{\nu}_{b}.
\end{align}
From now on, we shall use Greek letters for abstract/coordinate indices while Latin indices are reserved for the NP-tetrad components. The Christoffel symbols $\Gamma^{\mu}{}_{\nu\rho}$ are represented by the Ricci rotation coefficients $\gamma_{cba}$. Using the components of $\gamma_{cba}$ and their linear combinations, one can define the Newman-Penrose spin coefficients (see eq. (4.1b) in Ref. \cite{Newman:1961qr})
\begin{align*}
\gamma_{abc}\Leftrightarrow\text{spin coefficients}=\lbrace\kappa,\tau,\sigma,\rho,\pi,\nu,\mu,\lambda,\epsilon,\gamma,\beta,\alpha\rbrace.
\end{align*}
The NP formalism also has special symbols for derivatives in the direction of the tetrad vectors
\begin{align}
\label{NPderivatives}
\begin{split}
D=l^\mu\nabla_\mu,\hspace{15pt}  \Delta=n^\mu\nabla_\mu, \\   \delta=m^\mu\nabla_\mu, \hspace{15pt} \bar{\delta}=\bar{m}^\mu\nabla_\mu .
\end{split}
\end{align}

Having established the representation of the connection in the NP formalism, one can turn to the projection of the Riemann curvature tensor onto the NP tetrad, which has a specific notation.
\begin{align}
\label{Riemannscalars}
R_{\mu\nu\rho\sigma}\rightarrow\lbrace\Psi_A,\Phi_{AB}, \Lambda\rbrace \,,
\end{align}
where the complex scalars $\Psi_A$ represent the Weyl tensor, $\Phi_{AB}$ represent the traceless Ricci tensor, and $\Lambda$ is proportional to the scalar curvature. Of these variables, $\Psi_4$  defined as
\begin{align*}
\Psi_4=C_{\mu\nu\lambda\rho}n^{\mu}\bar{m}^\nu n^{\lambda}\bar{m}^\rho,
\end{align*}
is of particular importance to us as it contains all the information needed to extract the GW fluxes through the horizon and infinity \cite{Teukolsky:1974yv}. 

The Geroch-Held-Penrose (GHP) formalism \cite{Geroch:1973am} expands upon the NP formalism while bringing about some further benefits for our calculations, which we shall discuss at the end of this section.
The NP tetrad is not unique; the scalar products \eqref{NPtetradscalproduct} remain unchanged under tetrad rotation. If we restrict ourselves only to the subgroup that preserves the direction of the null vectors $l$ and $n$ (the so-called class III transformation \cite{Chandrasekhar:1984siy}), the tetrad rotation reads 
\begin{align}
\label{tetradtransform}
\begin{split}
& l^\mu\rightarrow  \eta\bar{\eta} l^\mu \,,  
\\  
& n^\mu\rightarrow  \eta^{-1}\left( \bar{\eta}\right)^{-1} n^\mu \,,
\\
& m^\mu\rightarrow  \eta\left(\bar{\eta}\right)^{-1} m^\mu \,,
\end{split}
\end{align}
where the transformation is parametrized by a single complex function $\eta$. From \eqref{tetradtransform} one can deduce how the derived quantities, such as the curvature scalars \eqref{Riemannscalars}, transform. A general quantity $\Theta$ then transforms as $\Theta\rightarrow\eta^p  \bar{\eta}^q\Theta$ where the pair $\lbrace p,q\rbrace$ is called the GHP weight of $\Theta$.  The tetrad vectors $l$, $n$, $m$ and $\bar{m}$ have GHP weight $\lbrace 1,1\rbrace$,$\lbrace -1,-1\rbrace$,$\lbrace 1,-1\rbrace$ and $\lbrace -1,1\rbrace$, respectively, from which one can, for example, find that the GHP weight of $\Psi_4$ is $\lbrace -4,0\rbrace$. Instead of the pair $\lbrace p,q\rbrace$, one can also use the boost weight $b=\left( p+q\right)/2$  and the spin weight $s=\left( p-q\right)/2$. This will become useful in the mode decomposition of the GHP quantities.

The GHP formalism also introduces a transformation that exchanges the tetrad vectors as $l\leftrightarrow n$ and $m\leftrightarrow \bar{m}$  in a general quantity $\chi$, so we can write this transformation as $\chi\rightarrow \chi'$. A new notation for the spin coefficients can then be based on this transformation. From their definitions (see  Appendix \ref{app:NPGHP}) we have, for example, $\beta'=-\alpha$, $\epsilon'=-\gamma$; these four spin coefficients are also the only ones that do not have a well-defined GHP weight. This can be amended by combining them with the NP derivatives and introducing the GHP derivatives
\begin{align}
\begin{split}
\label{GHPderivatives}
\text{\textthorn}=D-p\epsilon-q\bar{\epsilon},\hspace{15pt}  \text{\textthorn}'=\Delta+p\epsilon'+q\bar{\epsilon}', \\  \ethm=\delta-p\beta+q\bar{\beta'}, \hspace{15pt} \ethm'=\bar{\delta}+p\beta'-q\bar{\beta}.
\end{split}
\end{align}
Here $p$ and $q$ are the GHP weights of the object on which the GHP derivatives act. The derivatives then have a definite spin weight (the same as the corresponding tetrad vectors).

 The advantages of using the GHP instead of the NP formalism in our context are threefold. First, expressions tend to be shorter in the GHP form than in the NP one. Second, it serves as a good consistency check throughout our calculations, as any expression constructed from GHP quantities has a definite GHP weight, which is preserved when we manipulate the expression. Last, the GHP operators $\ethm$ and $\ethm'$ act as raising and lowering operators on the spin-weighted spherical harmonics in the Schwarzschild background, which is discussed in Appendix \ref{app:SWSHs} while more details on GHP formalism are included in Appendix \ref{app:NPGHP}.

%%%%%%%%%%%%%%%%%%%%%%%%%%%%%%%%%%%%%%%%%%%%%%%%%%%%%%%%%%%%%%%%%%%%%%%%%%%%%%%%%%%%%%%%%%%%%%%%%%%%%%%%%%%%%%%%%
\subsection{Teukolsky formalism in Kerr spacetime}\label{Teukolsky Kerr}

We shall now briefly review the Teukolsky equation in the Kerr spacetime (see, e.g., Ref. \cite{Pound_2021} for more). When formulated as an equation for the perturbation of Weyl scalar $\Psi_4$ the Teukolsky equation reads
\begin{align}
\label{TeukKerr}
 \mathcal{O}^{(0)}\Psi_4^{(1)}=\mathcal{T}^{(1)}.
      \end{align} 

This equation is derived directly from the Bianchi and Ricci identities in the NP formalism (Section \ref{modifTueeqderivation}) but is not separable in the variable $\Psi_4^{(1)}$; in fact, we have to rescale the variable $\Psi_4^{(1)}$ by a factor to obtain a separable equation as follows. 

 In the Kerr spacetime we can define the following functions   
\begin{align}
\begin{split}
\xi&=r-\mathrm{i} a \cos\theta,\enspace  \Sigma=r^2+a^2\cos^2\theta,\\  \Delta&=r^2-2Mr+a^2.
\end{split}
      \end{align}   
We can put the Teukolsky equation into its separable form   by multiplying it by  $\mathfrak{G}(a)=-2\xi^{4}\Sigma$ and by introducing the variable ${}_{-2}\Psi^{(1)}$ 
\begin{align}
     & {}_{-2}\mathcal{O}^{(0)}=\mathfrak{G} \mathcal{O}^{(0)}, \; \,
     \\
     & {}_{-2}\mathcal{T}^{(1)}=\mathfrak{G} \mathcal{T}^{(1)}, \; \, 
     \\
     & {}_{-2}\Psi^{(1)}=\xi^{4} \Psi_4^{(1)}.
\end{align}   
The transformed Teukolsky equation then reads
\begin{align}
 {}_{-2}\mathcal{O}^{(0)} {}_{-2}\Psi^{(1)}= {}_{-2}\mathcal{T}^{(1)}.
      \end{align} 
 This equation is now fully separable and is in fact a special case of the so-called Teukolsky master equation \cite{Teukolsky1} for a scalar with spin weight $s=-2$. For $s=+2$, the Teukolsky master equation also describes a gravitational perturbation, specifically $ {}_{2}\Psi^{(1)}=\Psi_0^{(1)}$. Regardless of the spin weight $s$, the Teukolsky master equation can be solved by the following ansatz     
\begin{align}
  \label{Teukolskymaster}
 {}_{s}\Psi^{(1)}_{lm\omega}={}_{s}R_{lm\omega}(r){}_{s}S_{lm\omega}(\theta,\phi)e^{-\mathrm{i}\omega t},
      \end{align}  
where ${}_{s}S_{lm\omega}$ are the spin-weighted spheroidal harmonics  which themselves can be separated as ${}_{s}S_{lm\omega}(\theta,\phi)={}_{s}S_{lm\omega }(\theta)e^{\mathrm{i}m \phi}$, the indices $l$ and $m$ satisfy  $ l\geq \vert s\vert$ and $\vert m\vert\leq l$.
The functions $S_{lm\omega}(\theta)$ obey an angular equation, which we do not include here; we should however note that it depends on the black hole spin parameter $a$ and so do the functions $S_{lm\omega}(\theta)$.

The radial function, on the other hand, is a solution to the equation

  \begin{align}
  \label{TeukolskyKerrRadial}
{}_{s}\mathfrak{D}_{lm\omega}{}_{s}R_{lm\omega}(r)=\,_{s}{\mathcal{T}^{(1)}_{lm\omega}}(r).
      \end{align} 
where $\mathcal{T}^{(1)}_{lm\omega}(r)$ is the radial part of the source  (obtained by decomposition into ${}_{s}S_{lm}$ modes and a Fourier transform), while the radial operator reads \cite{Drasco_2006,Hughes}
\begin{align}
\begin{split}
{}_{s}\mathfrak{D}_{lm\omega}{}_{s}R_{lm\omega}(r) =& \Delta^{-s}\frac{\mathrm{d}}{\mathrm{d}r}\left(\Delta^{s+1}\frac{\mathrm{d}{}_{s}R_{lm\omega}(r)}{\mathrm{d}r}\right)\\& -_{s}V_{lm\omega}(r){}_{s}R_{lm\omega}(r) \,,
\end{split}
      \end{align}   
 where  $_{s}V_{lm\omega}(r)$  is the Teukolsky effective potential (see for example in Ref. \cite{Pound_2021}). The homogeneous radial equation has a basis of solutions $\lbrace R^{(H)},R^{(\infty)}\rbrace$ defined by their boundary conditions on the Kerr horizon $r=r_+$ and infinity. For $R^{(H)}$, we have an ingoing wave at the horizon
    \begin{align}
\label{TeukolskyRadialbasisin}
    \begin{split}
{}_sR^{(H)}_{lm\omega}(r\rightarrow r_+)&\sim {}_sB^{\rm hole}_{lm\omega} \Delta^{-s}e^{-\mathrm{i}p r_{*}},\\{}_sR^{(H)}_{lm\omega}(r\rightarrow \infty)&\sim {}_sB^{\rm  out}_{lm\omega}r^{-1-2s}e^{\mathrm{i}\omega r_{*}}\\&+{}_sB^{ \rm in}_{lm\omega}r^{-1}e^{-\mathrm{i}\omega r_{*}},
    \end{split}
\end{align} 
while $R^{(\infty)}$ describes an outgoing wave at infinity  
\begin{align}
\label{TeukolskyRadialbasisup}
    \begin{split}
{}_sR^{(\infty)}_{lm\omega}(r\rightarrow r_+)&\sim {}_sD^{\rm  out}_{lm\omega} e^{\mathrm{i}p r_{*}}\\
&+ {}_sD^{\rm  in}_{lm\omega} \Delta^{-s}e^{-\mathrm{i}p r_{*}},\\ 
 {}_sR^{(\infty)}_{lm\omega}(r\rightarrow \infty)&\sim {}_sD^{\infty}_{lm\omega}   r^{-1-2s}e^{\mathrm{i}\omega r_{*}} .
     \end{split}
\end{align} 
Here $r_{*}$ is the Kerr tortoise coordinate, $p=\omega-m\Omega_+$  with $\Omega_+$ being the angular velocity of the horizon $\Omega_+=\frac{a}{2Mr_+}$ with $B$ and $D$ being the reflection and transmission coefficients determined by our convention and equation \eqref{TeukolskyKerrRadial}. 
Knowing the fundamental system $\lbrace R^{(H)},R^{(\infty)}\rbrace$, the inhomogeneous equation \eqref{TeukolskyKerrRadial} can then be solved using the method of variation of constants \cite{Drasco_2006} 
  \begin{align}
\label{TeukolskyRadialsolution}
 \begin{split}
{}_sR_{lm\omega}(r)&={}_sC^{(H)}_{lm\omega}(r){}_s R^{(\infty)}_{lm\omega}(r)\\&+{}_s C^{(\infty)}_{lm\omega}(r){}_sR^{(H)}_{lm\omega}(r)\,,
 \end{split}
\end{align} 
  where $C^{(H)}$ and $C^{(\infty)}$ are given by  an integral over the source $\mathcal{T}_{lm}$ and the respective homogeneous  solution  
 \begin{align}  
\label{TeukolskyRadialvariationofconstants}
 \begin{split}
{}_sC^{(H)}_{lm\omega}(r)&=\displaystyle \int_{r_+}^{r}  \frac{{}_sR^{(H)}_{lm\omega}(r')}{W(r')\Delta(r')}{}_s\mathcal{T}^{(1)}_{lm}(r')  \mathrm{d}r' ,
\\
{}_sC^{(\infty)}_{lm\omega}(r)&=\displaystyle \int_{r}^{\infty}  \frac{{}_sR^{(\infty)}_{lm\omega}(r')}{W(r')\Delta(r')}{}_s\mathcal{T}^{(1)}_{lm}(r')  \mathrm{d}r'.
   \end{split}
\end{align} 
Here $W(r)$ is the Wronskian of the homogeneous equation. As a result, the complete solution \eqref{TeukolskyRadialsolution} thus has proper boundary conditions describing outgoing radiation at infinity and ingoing at the horizon of the black hole. 

The bound geodesic motion in the Kerr (Schwarzschild) spacetime is fully integrable \cite{carter1968global}. If we use the Boyer-Lindquist time as our evolution parameter, we can characterize the bound motion using three fundamental frequencies $(\Omega^\phi,\Omega^r,\Omega^\theta)$. It can be shown \cite{Drasco_2006}  that since the source $\mathcal{T}^{(1)}$ is a functional of the geodesic motion, it can be decomposed into modes with discrete frequencies which  are linear combinations of the fundamental frequencies 
\begin{align}
\omega_{mnk}=m\Omega^\phi+n\Omega^r+k\Omega^\theta.
   \end{align} 
This form is valid for a generic (inclined, eccentric) orbit, while for special orbits such as circular orbits, fewer fundamental frequencies may appear in the decomposition.
   
 Consequently, any quantity depending on $\omega$ can be specified by two, three, or four indices depending on the type of bound orbit.
   Then the complete solution  $\Psi_4^{(1)}$ of the equation \eqref{TeukKerr} can be written as a sum over a discrete set of frequencies%
   \footnote{Here we also use that the $m$ mode number in the $e^{i m \phi}$ decomposition of the Teukolsky equation coincides with the $\sim m \Omega^\phi$ mode number of the source.}
\begin{align}   
 \xi^4\Psi_4^{(1)}=\displaystyle\sum_{lm\omega} {}_{-2}R_{lm\omega}(r){}_{-2}S_{lm\omega}(\theta)e^{\mathrm{i}m\phi-\mathrm{i}\omega_{nkm} t}.
    \end{align} 
It is important to note that the particle is moving on the bound geodesics between a pericenter and an apocenter, $r_{\rm min}<r(t)<r_{\rm max}$; outside of this interval the  functions   $C^{(A)}_{lmnk}(r)$  become constants from which one can calculate the GW fluxes \cite{Drasco_2006, Pound_2021}.

On the Schwarzschild or more generally a spherically symmetric background, the angular part of the Teukolsky master equation   \eqref{Teukolskymaster} is solved by the spin-weighted  spherical harmonics ${}_{s}Y_{lm}$  \cite{Goldberg}. The spin-weighted spherical harmonics can be thought of as a special case of the \textit{spheroidal} harmonics with zero spin or more precisely 
\begin{align}   
 {}_{s}S_{lm\omega}(\theta,\phi)\vert_{a\omega=0}= {}_{s}Y_{lm}(\theta,\phi)={}_{s}Y_{lm}(\theta)e^{im \phi}.
    \end{align} 
Identities for the spin-weighted spherical harmonics useful for our computations are discussed in Appendix \ref{app:SWSHs}.

%%%%%%%%%%%%%%%%%%%%%%%%%%%%%%%%%%%%%%%%%%%%%%%%%%%%%%%%%%%%%%%%%%%%%%%%%%%%%%%%%%%%%%%%%%%%%%%%%%%%%%%%%%%%%%%%%
\subsection{Deriving a modified Teukolsky equation}
\label{modifTueeqderivation}
How does one deal with curvature perturbations outside of the special class of Petrov type D vacuum spacetimes? Answering that question is the purpose of the following section. 

We start with obtaining the Teukolsky equations for the curvature scalar $\Psi_4$ from the identities of the Riemann tensor in the relevant perturbation orders. Analogous equations can be written for $\Psi_0$ if we apply the ``prime'' transformation $'$ defined in the previous section. In the following, we shall generally follow the notation presented in the paper of \citet{Li_2023} with a few key differences.

We shall start non-perturbatively from three identities
 \begin{align}
 & E_3\nu-E_4\lambda-\Psi_4=0 \,, \label{eq:RiccId}\\
 & F_3\Psi_4-J_3\Psi_3+3\nu\Psi_2=S_3 \,, \label{eq:BianchId1} \\
 & F_4\Psi_4- J_4\Psi_3+3\lambda\Psi_2=S_4, \label{eq:BianchId2}
\end{align}
where 
\begin{align}
 & E_3\equiv \bar{\delta}+3\alpha+\bar{\beta}+\pi-\bar{\tau} \,,  
 \\
 & E_4 \equiv \Delta+\mu+\bar{\mu}+3\gamma-\bar{\gamma}  \,, 
 \\
 & F_3 \equiv \delta+4\beta-\tau \,, 
 \;\  F_4 \equiv D+4\varepsilon-\rho \,, 
 \\
 & J_3 = \Delta+ 2\gamma+4\mu \,,
 \;\ J_4 = \bar{\delta}+4\pi+2\alpha\,.
\end{align} 

The first equation \eqref{eq:RiccId} is a Ricci identity, while the other equations \eqref{eq:BianchId1}, \eqref{eq:BianchId2} are Bianchi identities, all of which are satisfied independently of the Einstein equations. The ``sources'' $S_A$ represent homogeneous linear differential operators acting on the Ricci scalars  $S_A=\mathcal{S}_A(\Phi_{AB}, \Lambda)$ (see Appendix \ref{app:modifTeuk}). In the context of GR, Einstein's  equations imply   $S_A=\mathcal{S}_A(T_{ab})$. As a result, in Einstein vacua we necessarily have $S_A=0$. 

We now multiply the Ricci identity \eqref{eq:RiccId} by $3\Psi_2$. Then, we substitute for $\lambda$ and $\nu$ from the Bianchi identities \eqref{eq:BianchId1} and \eqref{eq:BianchId2}. This yields a master equation for $\Psi_4$
\begin{align}
\label{mastereq}
\mathcal{O} \Psi_4+\mathcal{K}\Psi_3=\mathcal{T}\,.
\end{align} 
This equation is, unlike the Teukolsky equation \eqref{Teukabstract}, completely general, as it is just a consequence of the Bianchi and Ricci identities. The explicit form of the operators present in equation \eqref{mastereq} is
\begin{align}
\begin{split}
\label{mastereqoperators}
& \mathcal{O}=\left( E_4-\frac{\Delta\Psi_2}{\Psi_2}\right)F_4-\left( E_3-\frac{\bar{\delta}\Psi_2}{\Psi_2}\right)F_3-3\Psi_2 \,,
\\
& \mathcal{K}=-\left( E_4-\frac{\Delta\Psi_2}{\Psi_2}\right)J_4+\left( E_3-\frac{\bar{\delta}\Psi_2}{\Psi_2}\right)J_3 \,,
\\
& \mathcal{T}=\left( E_4-\frac{\Delta\Psi_2}{\Psi_2}\right)S_4-\left( E_3-\frac{\bar{\delta}\Psi_2}{\Psi_2}\right)S_3 \,.
\end{split}
\end{align}
Here we can see that $\mathcal{T}$ is expressed as an operator acting on $S_A$, thus playing the role of the matter source on the right-hand side of equation \eqref{mastereq} while the purely curvature terms are located on the left-hand side.

One can perturbatively solve equation \eqref{mastereq} by expanding around a suitable zeroth-order solution, specifically, a Petrov type D vacuum spacetime. For example, \citet{Campanelli_1999} introduced an arbitrary-order expansion in the perturbation parameter $\varepsilon$. However, we aim to compute GWs forming in a background which itself is \textit{not} a vacuum type D solution but is close to it. This closeness will be assumed to be controlled by a separate parameter $\zeta$. Specifically, in our case $\zeta$ will be proportional to the mass of the cloud or halo in the environment of the massive black hole; the parameter $\varepsilon$ will then be proportional to the mass of the light compact object in the EMRI.

Motivated by the set-up above, we resort to a double expansion in both $\varepsilon$ and $\zeta$.\footnote{In our calculations, we treat these parameters as \textit{bookkeeping} devices. They indicate the origin and magnitude of the respective terms in any of the equations.} 
To keep track of the origin of the terms in our expansion, we use a superscript $(i,j)$ indicating the order $\sim \zeta^i \varepsilon^j$ at which the respective term or operator arises. This approach allows us to decompose the problem into separate tractable computations.

The metric takes the following form in the double expansion
 \begin{align}
 \label{expansionscheme}
 g_{\mu\nu}=g^{(0,0)}_{\mu\nu}+\zeta h^{(1,0)}_{\mu\nu}+\varepsilon \left(h^{(0,1)}_{\mu\nu}+\zeta h^{(1,1)}_{\mu\nu}\right)+...\,,
 \end{align}
 where $g^{(0,0)}_{\mu\nu}$ is a vacuum type D metric. For the purpose of our calculations, we assume that $h^{(1,0)}_{\mu\nu}$  is a known perturbation (ideally expressed using elementary functions), while $h^{(0,1)}_{\mu\nu}$ is found from the $\mathcal{O}(\varepsilon)$ Teukolsky-like equation. We are thus interested in the cross term $h^{(1,1)}_{\mu\nu}$ which is nontrivial and for which an $\mathcal{O}(\varepsilon\zeta)$ or $(1,1)$ Teukolsky-like equation is to be derived. 
 
Similarly to the expansion of the metric \eqref{expansionscheme}, we now expand other relevant quantities, most notably the NP tetrad 
  \begin{align}
  \begin{split}
  \label{tetradexpansion}
e^{\mu}_{a}=& (e^{\mu}_{a})^{(0,0)}+\zeta (e^{\mu}_{a})^{(1,0)}+\varepsilon (e^{\mu}_{a})^{(0,1)}+\varepsilon \zeta(e^{\mu}_{a})^{(1,1)}
 \\& + \mathcal{O}(\varepsilon^2,\zeta^2)\,.
\end{split}
 \end{align} 
A Petrov type D spacetime described by the metric $g^{(0,0)}_{\mu\nu}$ is defined by the existence of two principal null directions (PNDs) of multiplicity two \cite{Stephani:2003tm}. It is thus natural for the tetrad  $(e^{\mu}_{a})^{(0,0)}$  to respect this symmetry by aligning the null vectors $(l^{\mu})^{(0,0)}$ and $(n^{\mu})^{(0,0)}$  along the two PNDs. In such a tetrad  we have
 \begin{align} 
\Psi_{A\neq 2}^{(0,0)}=0 \,.
 \end{align} 
Since $g^{(0,0)}_{\mu\nu}$ is a vacuum spacetime (${T}^{(0,0)}_{ab}=0$) of type D, the Goldberg-Sachs theorem applies \cite{Stephani:2003tm}. This means that both vectors $(l^{\mu})^{(0,0)}$ and $(n^{\mu})^{(0,0)}$ are tangent to a null shear-free geodesic congruence and that four of the spin coefficients vanish \cite{Stephani:2003tm}
 \begin{align}
 \label{zerospincoeffs}
 \kappa^{(0,0)}=\sigma^{(0,0)}=\lambda^{(0,0)}=\nu^{(0,0)}=0\,.
 \end{align} 
 
 Now that we have made our assumptions about the $(0,0)$ order, we can first go through the $(0,1)$ order, which results in the standard Teukolsky equation \eqref{TeukKerr} corresponding to the linearized Einstein equations \eqref{linEineq}.
 If we now linearize the master equation in $\varepsilon$ and set $\zeta=0$, we arrive at
 \begin{align}
 \label{Teuk01} 
 \mathcal{O}^{(0,0)} \Psi_4^{(0,1)}=\mathcal{T}^{(0,1)}.
\end{align}
The operator $ \mathcal{O}$ and the source term $ \mathcal{T}$ from \eqref{mastereq} reduce to
 \begin{align}
 \begin{split}
  \label{O00op} 
 & \mathcal{O}^{(0,0)}=\mathcal{E}_4^{(0,0)} F_4^{(0,0)}-\mathcal{E}_3^{(0,0)} F_3^{(0,0)} -3\Psi_2^{(0,0)}\,,
 \\
 & \mathcal{T}^{(0,1)} =\mathcal{E}_4^{(0,0)} S_4^{(0,1)}-\mathcal{E}_3^{(0,0)} S_3^{(0,1)}\,, 
 \end{split}
 \end{align}
 where we have defined operators $ \mathcal{E}_3$ and $ \mathcal{E}_4$ as\footnote{Note that \citet{Li_2023} use a definition of the $\mathcal{E}_A$ operators that agrees with ours only at the $(0,0)$ order.}
 \begin{align}
  \label{Epsop} 
 \mathcal{E}_3=E_3+3\pi,\hspace{15pt}    \mathcal{E}_4=E_4+3\mu.
\end{align}
Since $S_3$ and  $S_4$ have the form of first-order differential operators acting on the Ricci scalars or alternatively the components of the stress-energy tensor ($S_A=\mathcal{S}^{ab}_A T_{ab}$ see Appendix \ref{app:modifTeuk}), we can then define a second-order differential operator  relating the source term directly to $T^{(0,1)}_{\mu\nu}$.
 \begin{align}
 \begin{split}
  \label{T00op} 
 \mathcal{T}^{(0,1)} &=\mathfrak{T}^{(0,0)ab} T^{(0,1)}_{ab} 
\\
& \equiv \left(\mathcal{E}_4^{(0,0)} \mathcal{S}_4^{(0,0)ab}-\mathcal{E}_3^{(0,0)} \mathcal{S}_3^{(0,0)ab}\right) T^{(0,1)}_{ab} .
\end{split}
\end{align}

One may notice that $\mathcal{K}\Psi_3$ is entirely absent from the Teukolsky equation \eqref{Teuk01}. This is due to the operator identity 
 \begin{align}
  \label{operidentity00} 
\mathcal{E}_3^{(0,0)}J_3^{(0,0)}-\mathcal{E}_4^{(0,0)}J_4^{(0,0)}=0\,,
\end{align}
which acts on the scalar $\Psi_3^{(0,1)}$. This can be proven using a commutation relation for the NP derivatives  and several Ricci identities (details are in the Supplemental notebooks \cite{MmaNB}), as was shown already by \citet{Teukolsky1}. Alternatively, instead of using the identity    \eqref{operidentity00}  one can set $\Psi_3^{(0,1)}=0$ using a tetrad rotation (Class I transformation of the tetrad $(e^{\mu}_{a})^{(0,1)}$, see Appendix \ref{app:NPGHP}). 
In our derivation of the Teukolsky equations, however, we do not make any gauge choice regarding the tetrad apart from $(e^{\mu}_{a})^{(0,0)}$, as mentioned above. Although both of these approaches produce the same $(0,1)$ Teukolsky equation, at higher orders we get a different form of the equation depending on whether $\Psi_3^{(0,1)}$ is present or not.

Obviously, one can derive the $(1,0)$ equation, which has the same form as  \eqref{Teuk01} with $(0,1)\rightarrow (1,0)$. This equation is, however, of no use to us as we are  assuming to know the metric $h^{(1,0)}$ explicitly. We can thus move on to the order $(1,1)$ where the metric $h^{(1,1)}$ satisfies the  field equations 
\begin{align}
\label{Einsteineq11} 
\delta G_{\mu\nu}(h^{(1,1)})+\delta^2G_{\mu\nu}(h^{(1,0)},h^{(0,1)})=8 \pi T^{(1,1)}_{\mu\nu}.
\end{align}
These are, in a sense, second-order Einstein equations with $h^{(1,0)}\neq h^{(0,1)}$. From the master equation \eqref{mastereq} we can obtain the $(1,1)$ Teukolsky equation for $\Psi_4^{(1,1)}$
\begin{align}
\label{Teukeq11} 
 \mathcal{O}^{(0,0)} \Psi_4^{(1,1)}+\mathcal{G}^{(1,1)}=\mathcal{T}^{(1,1)}\,,
\end{align}
where $\mathcal{G}^{(1,1)}$ is a pure curvature term and $\mathcal{T}^{(1,1)}$ is derived from matter sources. This is what we call the \textit{modified Teukolsky equation}, which was put into a form closely resembling Einstein's equations \eqref{Einsteineq11}. We can notice that the same operator $ \mathcal{O}^{(0,0)}$ acts on the $ \Psi_4^{(1,1)}$ as in the ordinary Teukolsky equation, while $\mathcal{G}^{(1,1)}$ has no analogue in the original Teukolsky equation. Explicitly, the curvature term can be written as
\begin{align}
\begin{split}
\label{G11} 
\mathcal{G}^{(1,1)}= & \mathcal{O}^{(1,0)} \Psi_4^{(0,1)}+ \mathcal{O}^{(0,1)} \Psi_4^{(1,0)}
\\
& +\mathcal{K}^{(1,1)}(\Psi_3^{(0,1)},\Psi_3^{(1,0)})\,,
\end{split}
\end{align} 
where we have separated it into the part containing $\Psi_4$ and a $\mathcal{K}$-part given by symmetric operators acting on $\Psi_3^{(0,1)}$ and  $\Psi_3^{(1,0)}$. The equation \eqref{Teukeq11} is symmetric with respect to $(1,0)\leftrightarrow (0,1)$ since we treated both $h^{(1,0)}$ and $h^{(0,1)}$ on equal footing during our derivation.

%%%%%%%%%%%%%%%%%%%%%%%%%%%%%%%%%%%%%%%%%%%%%%%%%%%%%%%%%%%%%%%%%%%%%%%%%%%%%%%%%%%%%%%%%%%%%%%%%%%%%%%%%%%%%%%%%
\subsection{Terms in the modified Teukolsky equation}
\label{modifTeukterms}
\begin{figure*}
\begin{tikzpicture}[scale=1.45]

    \tikzstyle{box} = [rectangle, draw, fill=red!15, text centered, rounded corners]

    \node[box, text width=3.5 cm, minimum height=1.0cm] (box1)  at (0,0) {$\Psi_4^{(0,1)}$ from eq. \eqref{Teuk01}     };
    \node[box, text width=2.0cm, minimum height=1.0cm ] (box2)  at (2.5,0) { reconstruction of $h^{(0,1)}_{\mu\nu}$};
    \node[box, text width=3.5cm, minimum height=1.5cm] (box3)  at (5.5,0) {NP quantities $(e^{\mu}_{a})^{(0,1)},\Psi_A^{(0,1)}$ and $\gamma_{abc}^{(0,1)}$};
    \node[box, text width=1.5cm, minimum height=1.5cm]  (box4)  at (8.5,0) { $\mathcal{S}^{(1,1)}_{\rm eff}$};
    \node[box, text width=3.5cm, minimum height=1.5cm] (box5) at (2.5,-1.5) {$T^{(0,1)}_{\mu\nu},T^{(1,0)}_{\mu\nu}$ and $ h^{(1,0)}_{\mu\nu}$};
    \node[box, text width=3.5cm, minimum height=1.7cm] (box6) at (5.5,-1.5) {\small{$ T^{(1,1)}_{\mu\nu}$ from $\nabla_\mu T^{\mu\nu}=0$}};

    \draw[->, thick, >=stealth, line width=0.5mm] (box1) -- (box2);
    \draw[->, thick, >=stealth, line width=0.5mm] (box2) -- (box3);
    \draw[->, thick, >=stealth, line width=0.5mm] (box3) -- (box4);
    \draw[->, thick, >=stealth, line width=0.5mm] (box5) -- (box6);
    \draw[->, thick, >=stealth, line width=0.5mm] (box2) -- (box6);
    \draw[->, thick, >=stealth, line width=0.5mm] (box6) -- (box4);

\end{tikzpicture}
\caption{A flowchart describing necessary steps  to construct the source of the modified Teukolsky equation \eqref{Teukeq11source} .}
\label{fig:Teukeqscheme}
 \end{figure*}
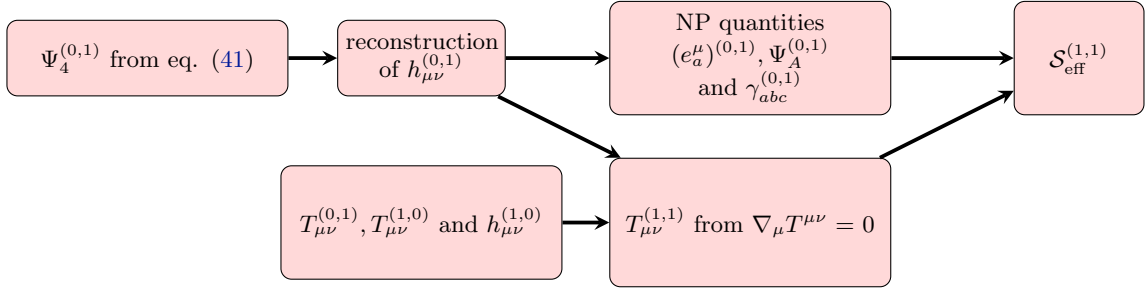

Having derived the modified Teukolsky equation \eqref{Teukeq11}, we now examine its individual terms to understand their physical origin and formulate a solution strategy. As was already noted, the operator $ \mathcal{O}^{(0,0)}$ acting on our unknown variable $\Psi_4^{(1,1)}$ remains unchanged from the $(0,1)$  order. If we then put the curvature term $\mathcal{G}^{(1,1)}$ on the right-hand side, we can rewrite the equation \eqref{Teukeq11} into the form of a standard Teukolsky equation with an effective source $\mathcal{S}^{(1,1)}_{\rm eff}=\mathcal{T}^{(1,1)}-\mathcal{G}^{(1,1)}$.
\begin{align}
\label{Teukeq11source} 
 \mathcal{O}^{(0,0)} \Psi_4^{(1,1)} =\mathcal{S}^{(1,1)}_{\rm eff}.
\end{align}
All we thus have to do is to construct the source $\mathcal{S}^{(1,1)}_{\rm eff}$, and then we can in principle use the well-known techniques from the standard Teukolsky equation, most notably the fact that $ \mathcal{O}^{(0,0)} $ is fully separable in the Kerr/Schwarzschild background. We summarize the steps to do so in a flowchart in Figure \ref{fig:Teukeqscheme} and the following paragraphs.

The entire source $\mathcal{S}^{(1,1)}_{\rm eff}$ is expressed in terms of NP quantities, which can be calculated from the spacetime metric. While the  $(1,0)$ quantities can be directly calculated given the knowledge of the  explicit form of $ h^{(1,0)}_{\mu\nu}$, obtaining the quantities of the  $(0,1)$ 
order is far less trivial. Naturally, one can start by solving the Teukolsky equation \eqref{Teuk01} to get $\Psi_4^{(0,1)}$. However, to obtain the other  $(0,1)$  Weyl scalars, spin coefficients, and NP derivatives, one first needs the metric $ h^{(0,1)}_{\mu\nu}$  and the associated tetrad $(e^{\mu}_{a})^{(0,1)}$. For this purpose, we will use the CCK metric reconstruction method \cite{Chrzanowski:1975wv,Kegeles:1979an} (see, e.g.,  \cite{Pound_2021} for other methods of obtaining the metric perturbation).

Apart from the NP spin coefficients and Weyl scalars present in $\mathcal{S}^{(1,1)}$, we also have to consider the Ricci scalars constructed from the components of the stress-energy tensor, in particular the NP components of $T^{(1,1)\mu\nu}$.  This part of $T^{\mu\nu}$ is to be determined from the lower-order equations of motion. To clarify this part more, let us separate the stress-energy tensor as follows
\begin{align} 
 T^{\mu\nu}= T^{\mu\nu}_{\text{particle}}+ T^{\mu\nu}_{\text{matter}} ,
\end{align}
where $T^{\mu\nu}_{\text{particle}}$ is the stress-energy tensor of the inspiraling  particle  \eqref{particleTmunu} while    $T^{\mu\nu}_{\text{matter}}$ represents the environmental matter surrounding the central black hole. At this point, we shall assume that there is no direct interaction (no direct exchange of momentum and energy) between the particle and the matter, which is equivalent to
\begin{align} 
\label{divTmunu} 
\nabla_\mu T^{\mu\nu}_{\text{particle}}=0,\hspace{15pt} \nabla_\mu T^{\mu\nu}_{\text{matter}}=0.
\end{align}
However, the particle and the matter still influence each other gravitationally, which manifests in the $T^{(1,1)\mu\nu}$ contributions. To clarify the perturbative orders, it is convenient to write down the expansion for each part of the stress-energy tensor
\begin{align} 
\label{Tmunusexpansions} 
 T^{\mu\nu}_{\text{particle}}& =\varepsilon  T^{(0,1)\mu\nu}+\varepsilon\zeta T^{(1,1)\mu\nu}_{\text{particle}} \,,\\
  T^{\mu\nu}_{\text{matter}}& =\zeta  T^{(1,0)\mu\nu}+\varepsilon\zeta T^{(1,1)\mu\nu}_{\text{matter}} \,,
\end{align}
where the subscript matter/particle is necessary only at $(1,1)$ order.
Thus, to obtain $T^{(1,1)\mu\nu}_{\text{particle}}$, one has to solve a perturbed geodesic equation in the background $g^{(0,0)}_{\mu\nu}+\zeta h^{(1,0)}_{\mu\nu}$ and then insert the result into the formula \eqref{particleTmunu} whereas for the $T^{(1,1)\mu\nu}_{\text{matter}}$, one needs to solve the equations of motion of the matter at the $(1,1)$ order and then use the  formula for the respective stress-energy tensor.

Let us now describe in more detail the individual terms present in the $\mathcal{G}^{(1,1)}$ curvature contribution to the modified Teukolsky equation given in eq. \eqref{G11}. Let us start with the operator $\mathcal{O}^{(1,0)}$ constructed from $h^{(1,0)}_{\mu\nu}$, which acts on $\Psi_4^{(0,1)}$. This operator can then be split into two parts
   \begin{align}
     \mathcal{O}^{(1,0)}=\mathcal{O}^{(1,0)}_\mathrm{D\mbox{-}vac}+\mathcal{O}^{(1,0)}_{\mathrm{corr}} \,,
   \end{align} 
where the first part is defined as
\begin{align}
    \begin{split}
     \mathcal{O}^{(1,0)}_\mathrm{D\mbox{-}vac} \equiv  \,
     & \mathcal{E}_4^{(1,0)} F_4^{(0,0)}+\mathcal{E}_4^{(0,0)} F_4^{(1,0)}-\mathcal{E}_3^{(1,0)} F_3^{(0,0)}
     \\ & -\mathcal{E}_3^{(0,0)} F_3^{(1,0)} -3\Psi_2^{(1,0)}.
    \end{split}
\end{align}
  By comparing with equation \eqref{O00op} one sees that we defined $ \mathcal{O}^{(1,0)}_\mathrm{D\mbox{-}vac}$ so that it has the same structure as $  \mathcal{O}^{(0,0)}$. This part can thus be interpreted as a perturbation of the operator $  \mathcal{O}^{(0,0)}$ in the case where $g^{(0,0)}_{\mu\nu}+\zeta h^{(1,0)}_{\mu\nu}$ remains in the family of type-D vacuum spacetimes. In most cases of interest, this would correspond to $g_{\mu\nu}^{(0,0)}$ being a Kerr metric and $\zeta h_{\mu\nu}^{(0,1)}$ being a linearized shift of the mass and angular-momentum parameters. The other ``correction'' part $ \mathcal{O}^{(1,0)}_{\mathrm{corr}}$ is then
\begin{align}
\begin{split}
   \mathcal{O}^{(1,0)}_{\mathrm{corr}}\equiv
    & \left(\Psi_2^{(0,0)}\right)^{-1}\Big[G_3^{(0,0)}(\Psi_3^{(1,0)})F_3^{(0,0)}
    \\ 
    & -G_4^{(0,0)}(\Psi_3^{(1,0)})F_4^{(0,0)}+S_5^{(1,0)}F_3^{(0,0)}
    \\
    & -S_6^{(1,0)}F_4^{(0,0)}\Big] \,,
\end{split}
\end{align}
where $G_3^{(0,0)}$ and $G_4^{(0,0)}$ are linear differential operators (see eq. \eqref{operatordefapp} in Appendix \ref{app:modifTeuk})  It can be easily seen that  $\mathcal{O}^{(1,0)}_{\mathrm{corr}}$  vanishes in the type D ($\Psi_3^{(1,0)}=0$) vacuum ($S_5^{(1,0)}=0=S_6^{(1,0)}$) background.\footnote{At this point, we can see one of the benefits of expanding  $\bar{\delta}( \Psi_2)/ \Psi_2$ and  $\Delta( \Psi_2)/ \Psi_2$ in  \eqref{mastereqoperators} as this enables us to distinguish a purely type D vacuum background from a general background, which is useful in our model introduced in the following section.}
 Another term in the modified Teukolsky equation \eqref{Teukeq11} is $\mathcal{K}^{(1,1)}(\Psi_3^{(0,1)},\Psi_3^{(1,0)})$, which is composed of operators acting on $\Psi_3^{(0,1)}$ and $\Psi_3^{(1,0)}$. One could also expect $\Psi_3^{(1,1)}$ to be present, but this part identically vanishes due to the identity \eqref{operidentity00}.  We can then separate this term into two parts
 \begin{align}
 \begin{split}
    \mathcal{K}^{(1,1)}=
    & \mathcal{K}^{(1,1)}_1(\Psi_3^{(0,1)},\Psi_3^{(1,0)})  
     +\mathcal{K}^{(1,0)}_2\Psi_3^{(0,1)} \\& +\mathcal{K}^{(0,1)}_2\Psi_3^{(1,0)} \,,
\end{split}
\end{align} 
 where  $\mathcal{K}^{(1,1)}_1(\Psi_3^{(0,1)},\Psi_3^{(1,0)})$ contains only first derivatives and also terms quadratic in $\Psi_3$. On the other hand,   $\mathcal{K}^{(1,0)}_2$ contains second-order differential operators acting on the  $\Psi_3^{(0,1)}$ scalar while having the same structure as the $\Psi_3^{(0,1)}$  part (eq. \eqref{operidentity00})    in the $(0,1)$ Teukolsky equation
 \begin{align}
 \begin{split}
\mathcal{K}^{(1,0)}_2=
& \mathcal{E}_3^{(1,0)}J_3^{(0,0)}+\mathcal{E}_3^{(0,0)}J_3^{(1,0)}  
\\
& -\mathcal{E}_4^{(1,0)}J_4^{(0,0)}-\mathcal{E}_4^{(0,0)}J_4^{(1,0)}.
\end{split}
\end{align} 
This operator is not identically zero in general, but we can use the same commutation relations and Ricci identities as for the proof of eq. \eqref{operidentity00} to eliminate the second derivatives to arrive at an expression of the form   
\begin{align}
\label{K2tilde}
 \begin{split}
   & \mathcal{K}^{(1,0)}_2\Psi_3^{(0,1)}  + \mathcal{K}^{(0,1)}_2\Psi_3^{(1,0)}=
  \\
   & \quad \quad \mathcal{\tilde{K}}^{(1,0)}_2\Psi_3^{(0,1)} +\mathcal{\tilde{K}}^{(0,1)}_2\Psi_3^{(1,0)} +20 \Psi_3^{(1,0)}\Psi_3^{(0,1)} ,
  \end{split}
\end{align} 
where the new  $\tilde{\mathcal{K}}$  operators are now proportional to $\lambda^{(1,0)}$,    $\nu^{(1,0)}$ ,   $\lambda^{(0,1)}$  and   $\nu^{(0,1)}$  (explicitly in eq.\eqref{K2tildeapp}), which will be convenient in our particular model. This expression is then quadratic in $\Psi_3$ and we can define 
 \begin{align}
 \begin{split}
& \mathcal{K}^{(1,1)}_2(\Psi_3^{(0,1)},\Psi_3^{(1,0)}) := \\& \quad \quad\mathcal{\tilde{K}}^{(1,0)}_2\Psi_3^{(0,1)}+\mathcal{\tilde{K}}^{(0,1)}_2\Psi_3^{(1,0)} +20 \Psi_3^{(1,0)}\Psi_3^{(0,1)}\,,
\end{split}
\end{align} 
as a counterpart of $\mathcal{K}^{(1,1)}_1(\Psi_3^{(0,1)},\Psi_3^{(1,0)})$. In total, we can write 
 \begin{align}\begin{split}
& \mathcal{K}^{(1,1)}(\Psi_3^{(0,1)},\Psi_3^{(1,0)})=
\\& \quad \quad \mathcal{K}^{(1,1)}_1(\Psi_3^{(0,1)},\Psi_3^{(1,0)})+\mathcal{K}^{(1,1)}_2(\Psi_3^{(0,1)},\Psi_3^{(1,0)})\,.
   \end{split}\end{align} 
   The explicit forms of  $\mathcal{K}^{(1,1)}_A$ can be found in  Appendix \ref{app:modifTeuk}. Let  us also point out that by a suitable tetrad choice one can set either or both $\Psi_3^{(0,1)}$ and  $\Psi_3^{(1,0)}$  to zero, which would lead to $\mathcal{K}^{(1,1)}=0$ in this gauge.  
   
   The last term appearing in \eqref{Teukeq11} that we would like to comment on is the ``matter source''   $\mathcal{T}^{(1,1)}$, which can be written similarly to equation \eqref{T00op} as operators acting on the components of the stress-energy tensor
 \begin{align}
\mathcal{T}^{(1,1)}=\mathfrak{T}^{(1,1)}( T^{(0,1)}_{ab}, T^{(1,0)}_{ab}) +\mathfrak{T}^{(0,0)ab} T^{(1,1)}_{ab} \,,
   \end{align} 
where the operator $\mathfrak{T}^{(0,0)}$ was introduced in \eqref{T00op} while the longer expression $\mathfrak{T}^{(1,1)}( T^{(0,1)}_{ab}, T^{(1,0)}_{ab}) $ can be decomposed as 
\begin{align}
\begin{split}
    \mathfrak{T}^{(1,1)}(T^{(0,1)}_{ab},T^{(1,0)}_{ab})=
        & \mathfrak{T}^{(1,0)ab}T^{(0,1)}_{ab}
        \\
        & + \mathfrak{T}^{(0,1)ab}T^{(1,0)}_{ab} 
        \\
        & +\mathfrak{T}^{(1,1)}_q( T^{(0,1)}_{ab}, T^{(1,0)}_{ab}) \,,
\end{split}
\end{align} 
where the form of the linear operator $\mathfrak{T}^{(1,0)}$ and the terms quadratic in the stress-energy tensor can be found in  Appendix \ref{app:modifTeuk}.

To summarize, we have derived a modified Teukolsky equation in a form similar to Refs \cite{Li_2023} and \cite{Cano_2023}. We have mostly adopted the notation of the former but with a difference in the definition of the operators $\mathcal{E_A}$ due to expanding the derivatives acting on $\Psi_2$ using Ricci identities. Another difference in Ref. \cite{Li_2023} is their choice of tetrad in which $\Psi^{(0,1)}_3=\Psi^{(1,1)}_3=0$. In our approach, we derived the modified Teukolsky equation in a gauge-independent form. As was already remarked by  \citet{Li_2023} and  \citet{wagle2023perturbationsspinningblackholes}, the variable $ \Psi_4^{(1,1)}$ itself is not gauge-independent; it is not invariant either with respect to coordinate transformations or with respect to tetrad rotations. One could construct a gauge-independent variable in a fashion similar to what was done for the second-order Teukolsky equation by \citet{Campanelli_1999}. Nevertheless, in our particular model, gauge invariance is not important, since the computation can be carried out in any gauge and the final GW fluxes computed from $ \Psi_4^{(1,1)}$ are invariant.
The full derivation of the modified Teukolsky equation is presented in the Mathematica notebook \texttt{ModifiedTeukolsky.nb} in the supplemental material \cite{MmaNB}.

%%%%%%%%%%%%%%%%%%%%%%%%%%%%%%%%%%%%%%%%%%%%%%%%%%%%%%%%%%%%%%%%%%%%%%%%%%%%%%%%%%%%%%%%%%%%%%%%%%%%%%%%%%%%%%%%%%
\section{Perturbing a black hole enclosed by a ring}  \label{sec:holeringpert}

This section translates our general modified Teukolsky formalism into a concrete, tractable model for EMRIs in environmental matter. First, in Section \ref{sec:pole-dipole-model}, we introduce our physical model: a thin matter ring surrounding the central black hole, characterized by its leading gravitational multipoles (monopole and dipole). Next, in Section \ref{backgroundNP}, we construct the complete Newman-Penrose description of this ring-hole spacetime, computing all the background curvature quantities that will source our modified equations. Finally, in Section \ref{subsec:decomposition}, we exploit this structure to decompose the modified Teukolsky equation into manageable pieces: a smooth part that can be solved by matching standard Teukolsky solutions at the ring radius, and a singular part that encodes all the environmental complexity in distributional sources localized at the ring position. This decomposition is the key technical achievement that makes the environmental problem computationally feasible.

%%%%%%%%%%%%%%%%%%%%%%%%%%%%%%%%%%%%%%%%%%%%%%%%%%%%%%%%%%%%%%%%%%%%%%%%%%%%%%%%%%%%%%%%%%%%%%%%%%%%%%%%%%%%%%%%%
\subsection{The pole-dipole ring-hole model}
\label{sec:pole-dipole-model}
%%%%%%%%%%%%%%%%%%%%%%%%%%%%%%%%%%%%%%%%%%%%%%%%%%%%%%%%%%%%%%%%%%%%%%%%%%%%%%%%%%%%%%%%%%%%%%%%%%%%%%%%%%%%%%%%%

%%%%%%%%%%%%%%%%%%%%%%%%%%%%%%%%%%%%%%%%%%%%%%%%%%%%%%%%%%%%%%%%%%%%%%%%%%%%%%%%%%%%%%%%%%%%%%%%%%%%%%%%%%%%%%%%%%
 \begin{figure}
    \begin{center}
        \includegraphics[clip, trim=0.1cm 2.25cm 0.1cm 1.5cm,width=0.49\textwidth]{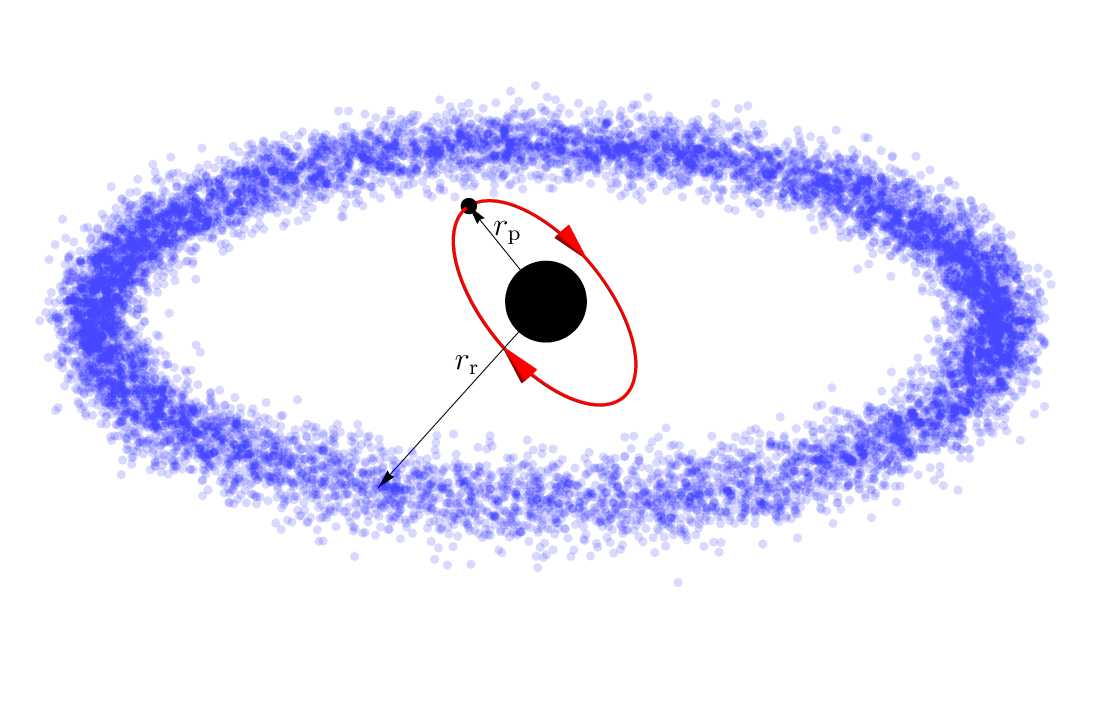}
    \caption{An illustration of the physical setup investigated in this paper. The ring-like matter cloud (blue) is located at a characteristic radius $r_{\rm r}$ that is assumed to be much larger than the radius of the orbit  $r_{\rm p}$ (red) of the secondary inspiraling into the black hole. As a result, the ring is characterized only by leading tidal multipoles.}   
    \label{fig:ringEMRI}
    \end{center}
\end{figure}

%%%%%%%%%%%%%%%%%%%%%%%%%%%%%%%%%%%%%%%%%%%%%%%%%%%%%%%%%%%%%%%%%%%%%%%%%%%%%%%%%%%%%%%%%%%%%%%%%%%%%%%%%%%%%%%%%

Our background space-time for the EMRI will be a Schwarzschild black hole of mass $M$ surrounded by a circumnuclear cloud of matter of mass $m_r$ at a distance $r_{\rm r}$ from the center, which is assumed to be much larger than the characteristic radius of the EMRI orbit $r_{\rm p}$. The setup is illustrated in Fig. \ref{fig:ringEMRI}. We approximate the toroidal cloud by its leading-order tidal multipoles to make the problem tractable while capturing the essential physics. Specifically, these multipoles are the monopole (mass) and dipole (angular momentum). This approximation is justified by the fact that the $l$-th tidal multipole falls off as $(r_{\rm p}/r_{\rm r})^l$ when $r_{\rm p}\ll r_{\rm r}$, making higher multipoles progressively weaker. 

Another simplifying assumption that we make is that the circumnuclear cloud has a small cross-section compared to the local curvature scale and the wavelength of the GWs passing through it. One concrete matter source for which all of the above-mentioned assumptions are fulfilled is a thin ring of almost pressureless gas or dust moving on circular geodesics at a large distance from the central black hole. The details and implications of these assumptions are discussed in Appendix \ref{app:background}. The sum of the assumptions allows us to write the metric in a simple and technically advantageous form. 

Specifically, we obtain our background metric by matching two linearized Kerr spacetimes at $r = r_{\rm r}$ using Israel junction conditions, which is carried out in Appendix \ref{app:israel}. We start with a background metric of the form
\begin{align}
\label{background}
\mathrm{d}s^2_{\rm BG} = \left(g^{(0,0)}_{\mu\nu} + \zeta h^{(1,0)}_{\mu\nu}\right)\mathrm{d}x^\mu \mathrm{d}x^\nu,
\end{align}
where $g^{(0,0)}_{\mu\nu}$ is the Schwarzschild metric and $\zeta$ is our perturbation parameter (of order $m_{\rm r}/M$). The Kerr matching Ansatz implies that the perturbation $h^{(1,0)}_{\mu\nu}$ describes a spacetime that is piecewise type D vacuum: inside the ring radius ($r < r_{\rm r}$), observers see only the central black hole mass $M$, while outside ($r > r_{\rm r}$), they measure total mass $M + m_{\rm r}$ and total angular momentum $Ma_{\rm r}$. Formally, this leads to a ``matter shell'' with non-zero mass density on the entire sphere $r=r_{\rm r}$ when interpreting the metric as an exact solution. However, we stress that we are motivated by the thin ring picture.

The matching introduces a coordinate transformation between inner and outer regions characterized by
\begin{align}
\label{coord-transform}
t^{\rm (in)} = (1 - z_{\rm r})t, \quad \phi^{\rm (in)} = \phi - \Omega_{\rm r}t,
\end{align}
where $z_{\rm r} = m_{\rm r}/(r_{\rm r} - 2M)$ is the gravitational redshift and $\Omega_{\rm r} = 2Ma_{\rm r}/r_{\rm r}^3$ is the frame-dragging angular velocity. As already discussed in Ref. \cite{Polcar_2022}, these transformations show that static observers near the black hole are both redshifted and rigidly rotating relative to asymptotic observers.

The nonzero components of the $(1,0)$ metric are
\begin{align}
\label{h10-components}
h_{tt}^{(1,0)} &= \frac{2m_{\rm r}(r - 2M - (r - r_{\rm r})\Theta(r - r_{\rm r}))}{r(r_{\rm r} - 2M)},\\
h_{rr}^{(1,0)} &= \frac{2m_{\rm r}r\Theta(r - r_{\rm r})}{(r - 2M)^2},\\
h_{t\phi}^{(1,0)} &= \frac{2Ma_{\rm r}\left((r^3 - r_{\rm r}^3)\Theta(r - r_{\rm r}) - r^3\right)\sin^2{\theta}}{rr_{\rm r}^3},
\end{align}
where $\Theta$ is the Heaviside step function. The ring parameters satisfy
\begin{align}
\label{ring-angular-momentum}
J_{\rm r} = Ma_{\rm r} = \frac{m_{\rm r}\sqrt{M}r_{\rm r}^{3/2}}{r_{\rm r} - 2M},
\end{align}
where the last equality is obtained by assuming that the angular momentum of the matter is that of a ring of matter that moves on circular geodesics in the $(0,0)$ space-time.

This thin-shell formulation is crucial for our analysis because it allows us to formulate standard Teukolsky equations in each vacuum region while confining all nontrivial modifications to delta-function sources at $r = r_{\rm r}$. 

Note that each step of the computation to follow also easily generalizes to central black holes with small spin. Technically, this spin would be added into the $h^{(1,0)}$ metric and would lead to an independent flux correction. We chose not to do so for simplicity.

%%%%%%%%%%%%%%%%%%%%%%%%%%%%%%%%%%%%%%%%%%%%%%%%%%%%%%%%%%%%%%%%%%%%%%%%%%%%%%%%%%%%%%%%%%%%%%%%%%%%%%%%%%%%%%%%%
\subsection{NP quantities in hole-ring background}\label{backgroundNP}

 Before formulating our strategy in detail, let us start by constructing the NP quantities corresponding to the background metric \eqref{background} introduced in the previous section. This is the first step which will provide us with insight into the individual terms present in the effective source $\mathcal{S}^{(1,1)}_{\rm eff}$; these terms can then be split into two groups: those living on the matter shell (proportional to $\delta(r-r_{\rm r})$) and those which are nonzero in at least one of the two regions of the spacetime.

In order to calculate the Newman-Penrose quantities, we first have to choose a suitable tetrad. As was discussed in section \ref{modifTueeqderivation}, our tetrad is expanded in our two perturbation parameters (see eq. \eqref{tetradexpansion}). For the background alone we have up to the order linear in $\zeta$
 \begin{align}
  \label{tetradBG}
e^{({\rm BG})\mu}_{b}=(e^{\mu}_{b})^{(0,0)}+\zeta (e^{\mu}_{b})^{(1,0)}.
 \end{align} 
Since the background spacetime is described by the Schwarzschild solution for $r<r_{\rm r}$, following Teukolsky, we can use the well-known Kinnersley tetrad $e^{({\rm K})\mu}_{b} (M,a)$  used in the Kerr spacetime with mass $M$ and spin $a$. This means that inside the shell $r<r_{\rm r}$ we have just the Schwarzschild-Kinnersley tetrad \cite{Kinnersley:1969zza} $e^{(0,0)\mu}_{a}=e^{({\rm K})\mu}_{b} (M,0)$. 

 However, the components of this inner tetrad $e^{(0,0)\mu^{({\rm in})}}_{b}(x^{({\rm in})})$ are expressed in the inner coordinates $x^{({\rm in})}$ introduced in \eqref{coord-transform} which is not desirable, as we decided that the coordinates of the outer metric \eqref{appoutermetric} will be used globally. For this reason, we have to find the components of the inner tetrad at the coordinates $x$, which means that we transform them as

 \begin{align}
  \label{innertetradtransf}
 e^{(0,0)\mu^{({\rm in})}}_{b}(x^{({\rm in})})\rightarrow e^{({\rm BG})\mu}_{b(r<r_{\rm r})}(x)=\frac{\partial x^{\mu} }{\partial x^{\mathrm{(in)}\nu}} e^{(0,0)\nu^{({\rm in})}}_{b}(x^{({\rm in})}).
 \end{align} 
Obviously, since we work in perturbation theory, we have to linearize the transformation in $\zeta$ (in $m_{\rm r}$ and $a_{\rm r}$) so the inner tetrad can then be written as

\begin{align}
  \label{innertetradtransflin}
e^{({\rm BG})\mu}_{a(r<r_{\rm r})}(x)&=e^{(0,0)\mu}_{a}(x)+\mathcal{L}_{\xi^{(t\phi)}} e^{(0,0)\mu}_{a}(x)+\mathcal{O}(\zeta^2) \, , \nonumber \\ \xi^{(t\phi)\mu}& =( z_{\rm r} t,0,0, \Omega_{\rm r} t)\,,
 \end{align} 
where $\mathcal{L}$ is the Lie derivative. For $r>r_{\rm r}$ the spacetime is described by a Kerr metric with mass $M+m_{\rm r} $ and spin $a_{\rm r} $ which is linearized in the pair $(m_{\rm r} , a_{\rm r} )$, similarly we can linearize the Kinnersley tetrad as

 \begin{align}
  \label{outer tetrad}
e^{({\rm BG})\mu}_{b(r>r_{\rm r})}= e^{(0,0)\mu}_{b}+m_{\rm r} \partial_1 e^{({\rm K})\mu}_{b} (M,0)+a_{\rm r} \partial_2 e^{({\rm K})\mu}_{b} (M,0).
 \end{align} 
 
 Putting the inner and outer tetrad together we obtain the tetrad covering the entire spacetime.
Unlike the $(0,0)$ tetrad, the $(1,0)$  tetrad is not continuous and can thus be written using the Heaviside step function, which is given in eq. \eqref{app10tetrad} in Appendix \ref{app:background NP quantities}.

Having chosen our background tetrad, we can calculate the spin coefficients and Weyl and Ricci scalars of the order $(0,0)$ and $(1,0)$, which are the building blocks of the modified Teukolsky equation. Since the NP derivatives in our formalism act only on scalar quantities, the NP derivatives can be written just using the coordinate derivatives, for example $\Delta^{(1,0)}=n^{\mu(1,0)}\partial_\mu$.  All NP scalars in the $(0,0)$ and $(1,0)$ order are given in Appendix \ref{app:background NP quantities}; here we shall only present the most relevant ones. 

We shall start with the Weyl scalars. In the $(0,0)$ order (Schwarzschild), the only nonzero scalar is $\Psi_2^{(0,0)}=-\frac{M}{r^3}$, which is in accordance with it being of Petrov type D. In the $(1,0)$ order, the nonzero Weyl scalars are

 \begin{align}
       \label{Weyl10}
\Psi_1^{(1,0)}&=-\frac{3 \mathrm{i} a_{\rm r}  M \sin (\theta) \delta (r-r_{\rm r})}{2 \sqrt{2} r_{\rm r}^3} \,,\\
\Psi_2^{(1,0)}&=-\frac{ (m_{\rm r} r+3 \mathrm{i} a_{\rm r}  M \cos (\theta))\Theta (r-r_{\rm r})}{r^4} \\
\nonumber&+\frac{m_{\rm r} (2 r_{\rm r}-3 M) \delta (r-r_{\rm r})}{6 r_{\rm r}^2 (r_{\rm r}-2 M)}\,,\\
\Psi_3^{(1,0)}&=\frac{3 \mathrm{i} a_{\rm r}  M \sin (\theta) (r_{\rm r}-2 M) \delta (r-r_{\rm r})}{4 \sqrt{2} r_{\rm r}^4}\,.
    \end{align} 
At first glance, it may be surprising that $\Psi_0^{(1,0)}$ and $\Psi_4^{(1,0)}$ vanish; however, these scalars have spin weight $\pm 2$, which means that they cannot describe monopole or dipole perturbations. The leading contributions to these scalars have a quadrupolar character, as is evident in their expansion into spin-weighted spherical harmonics in the following sections.  Similarly, the dominant contributions to the scalars $\Psi_1^{(1,0)}$ and $\Psi_3^{(1,0)}$ have a dipole character, which is why it is not surprising that they are proportional to the spin parameter $a_{\rm r}$. What is more important, however, is the fact that both scalars are proportional to $\delta (r-r_{\rm r})$, thus living only on the matter shell. This is a consequence of our choice of the tetrad $e^{({\rm BG})\mu}_{a}$, which is adapted to the symmetries of the two matched linearized Kerr spacetimes.

At this point one may ask if we can go one step further by eliminating $\Psi_1^{(1,0)}$ and $\Psi_3^{(1,0)}$ using tetrad rotation. This is indeed possible; we can apply type I and type II rotations (see Appendix \ref{app:NPGHP} and~\ref{app:background}) so that the only nonzero Weyl scalar is $\Psi_2^{(1,0)}$ thus proving that the background spacetime is of Petrov type D. 

This may come as a surprise, as we introduced the modified Teukolsky equation \eqref{Teukeq11source} to calculate an adiabatic inspiral in the background in which the standard approach is not feasible. However, while we can make all Weyl scalars apart from $\Psi_2^{(1,0)}$ vanish, some perturbed spin coefficients like $\kappa^{(1,0)}$ and $\nu^{(1,0)}$ remain nonzero. These spin coefficients (together with $\sigma$ and $\lambda$) have to vanish to obtain the standard Teukolsky equation.%
\footnotetext{More precisely, the scalars $S_3^{(0)}$ and $S_4^{(0)}$, constructed from spin coefficients and Ricci scalars, must be zero, otherwise the source would include terms like $\mathcal{E}_4^{(1)} S_4^{(0)}$ and $\mathcal{E}_3^{(1)} S_3^{(0)}$ that are not determined by $T^{(1)}_{\mu\nu}$.} 
In \textit{vacuum} Petrov type D spacetimes, these four scalars vanish automatically due to the Goldberg-Sachs theorem \cite{Stephani:2003tm}. Our spacetime, however, is not vacuum, and the matter at the ring location is sufficient to break the assumptions of the theorem.

 Returning to the Weyl scalars \eqref{Weyl10}, we see that even if $\Psi_2^{(1,0)}$ is the only nonzero Weyl scalar, one still has to use the modified Teukolsky equation to construct the source $\mathcal{T}^{(1,1)}$, which would involve metric reconstruction and other procedures discussed in the following sections. In addition to that, the components of the new (rotated) tetrad would include terms proportional to $ \delta (r-r_{\rm r})$, which would make it less regular than the tetrad \eqref{app10tetrad}.  For these reasons, we concluded that there is no great benefit in rotating the tetrad to eliminate $\Psi_1^{(1,0)}$ and $\Psi_3^{(1,0)}$, and we shall then from now on use the tetrad \eqref{app10tetrad}. 

 Apart from the Weyl scalars, we also have to calculate the $(1,0)$ spin coefficients, with the most important ones being 
\begin{align}
    \label{spincoef10}
    \nu^{(1,0)}&=  \frac{\mathrm{i} a_{\rm r}  \sin (\theta ) (r_{\rm r}-2 M)^2 \delta (r-r_{\rm r})}{4 \sqrt{2} r_{\rm r}^3} \,,\\
    \kappa^{(1,0)}&= -\frac{\mathrm{i} a_{\rm r}  \sin (\theta) \delta (r-r_{\rm r})}{\sqrt{2} r_{\rm r}}\,,
    \\\sigma^{(1,0)}& = 0 \,, \quad  \lambda^{(1,0)}= 0.
\end{align}
These are the four spin coefficients which are zero in type D vacuum zero; here we can see that two of them are nonzero on the matter shell. Similarly, we can calculate $\Phi_{AB}^{(1,0)}$ and $\Lambda^{(1,0)}$  from the components of the stress-energy tensor $  T_{\mu\nu}^{(1,0)\rm shell}$ \eqref{app10stressenergycomponents} using the definitions \eqref{app:Ricciscalars}. Naturally, all these scalars are proportional to $\delta (r-r_{\rm r})$. The full list of the $(0,0)$  $(1,0)$ NP quantities is given in Appendix \ref{app:background NP quantities}. 

Now that we have completely constructed the background ($(0,0)$ and  $(1,0)$)  Newman-Penrose quantities, we can proceed by formulating our approach to solving  the modified Teukolsky equation in this background. The calculations described in this section are summarized in the Mathematica notebook \texttt{Thebackgroundspacetime.nb} in our supplemental material  \cite{MmaNB}.

%%%%%%%%%%%%%%%%%%%%%%%%%%%%%%%%%%%%%%%%%%%%%%%%%%%%%%%%%%%%%%%%%%%%%%%%%%%%%%%%%%%%%%%%%%%%%%%%%%%%%%%%%%%%%%%%%
\subsection{Decomposition of the Teukolsky equation in the hole-ring spacetime}  \label{subsec:decomposition}   

The key to solving the modified Teukolsky equation lies in constructing the effective source $\mathcal{S}^{(1,1)}_{\rm eff}$. Once it is done, we can make use of the complete separability of $\mathcal{O}^{(0,0)}$ in the Schwarzschild background and mode decomposition to completely solve the problem. Let us recall that the effective source has the following structure
           \begin{align}
\mathcal{S}^{(1,1)}_{\rm eff}=&\mathcal{T}^{(1,1)}-\mathcal{O}^{(1,0)} \Psi_4^{(0,1)}- \mathcal{O}^{(0,1)} \Psi_4^{(1,0)}\\\nonumber&-\mathcal{K}^{(1,1)}(\Psi_3^{(0,1)},\Psi_3^{(1,0)}).
      \end{align}  

As was already stated above, the main simplification comes from the fact that our background can be thought of as two Petrov type D vacuum spacetimes separated by the shell $r=r_{\rm r}$, which means that one could formulate a standard Teukolsky equation on each side of the shell while we are also left with some terms in the modified Teukolsky equation that live only on the matter shell. 

We will now go through the parts of the effective source and identify the terms which are nonzero on each side of the boundary and those which are proportional to the delta function $\delta(r-r_{\rm r})$ and its derivatives.

We shall start with the curvature part of the effective source. Since $\Psi_4^{(1,0)}=0$ everywhere, the only nontrivial part with $\Psi_4$ in $\mathcal{S}^{(1,1)}_{\rm eff}$ is $\mathcal{O}^{(1,0)} \Psi_4^{(0,1)}$. The operator $\mathcal{O}^{(1,0)}$ can be decomposed (section \ref{modifTeukterms}) into a type D vacuum part and a correction $\mathcal{O}^{(1,0)}_{\mathrm{corr}}$. It is not surprising that $\mathcal{O}^{(1,0)}_{\mathrm{corr}}$ lives only on the matter shell as its parts are proportional either to $\Psi_3^{(1,0)}$ or $S_A^{(1,0)}$  (constructed from $\Phi^{(1,0)}_{AB}$). The type D vacuum part can be written as
\begin{align}
\begin{split}
    \mathcal{O}^{(1,0)}_\mathrm{D\mbox{-}vac} = 
    & \mathcal{O}^{(1,0)(-)}_\mathrm{D\mbox{-}vac} +\mathcal{O}^{(1,0)(r_{\rm r})}_\mathrm{D\mbox{-}vac}
    \\
    & + \Theta(r-r_{\rm r})\left(\mathcal{O}^{(1,0)(+)}_\mathrm{D\mbox{-}vac}-\mathcal{O}^{(1,0)(-)}_\mathrm{D\mbox{-}vac}\right)\,,
\end{split}
\end{align} 
   where $\mathcal{O}^{(1,0)(-)}_\mathrm{D\mbox{-}vac}$ is nonzero in the inner region, $\mathcal{O}^{(1,0)(+)}_\mathrm{D\mbox{-}vac}$, in the outer one and $\mathcal{O}^{(1,0)(r_{\rm r})}_\mathrm{D\mbox{-}vac}$ is the part containing delta functions (for example $\Psi_2^{(1,0)}$ contains a term with $\delta(r-r_{\rm r})$, see Appendix \ref{app:background NP quantities}).
   
  The other curvature part is  $\mathcal{K}^{(1,1)}(\Psi_3^{(0,1)},\Psi_3^{(1,0)})$ which is restricted only to the matter shell. Its first part  $\mathcal{K}^{(1,1)}_1(\Psi_3^{(0,1)},\Psi_3^{(1,0)})$  is similar to $\mathcal{O}^{(1,0)}_{\mathrm{corr}}$, all terms contain either $\Psi_3^{(1,0)}$ or $S_A^{(1,0)}$. Why, then, is the other part $\mathcal{K}^{(1,1)}_2(\Psi_3^{(0,1)},\Psi_3^{(1,0)})$ only a distribution on the matter shell? This becomes evident when using the re-expression of $\mathcal{K}^{(1,1)}_2$ using the $\tilde{\mathcal{K}}_2$ operators introduced in eq. \eqref{K2tilde} (see also Appendix \ref{app:multipolesofthering}). From this, it becomes clear that the operator vanishes by virtue of the piecewise vacuum Petrov type D character of the $(1,0)$ space-time.
  
  The last part is the "matter source" $\mathcal{T}^{(1,1)}$ which is only nonzero either at the location of the particle or on the shell, meaning it contains terms with $\delta(r-r_{\rm p}(\tau))$ and $\delta(r-r_{\rm r})$ where $r_{\rm p}(\tau)$ is the radial coordinate of a geodesic on which the particle moves. Since we assumed that the particle moves only in the inner region,  $r_{\rm p}(\tau)<r_{\rm r}$,which means that in the inner region the source $\mathcal{T}^{(1,1)}$ reduces to $\mathfrak{T}^{(1,0)ab}T^{(0,1)}_{ab}$ and $\mathfrak{T}^{(0,0)ab}T^{(1,1)}_{ab(\rm particle)}$. On the matter shell, we have complementary terms $\mathfrak{T}^{(0,1)ab}T^{(1,0)}_{ab}$ and $\mathfrak{T}^{(0,0)ab}T^{(1,1)}_{ab(\rm matter)}$, which requires solving  the $(1,1)$ equations of motion of the surrounding matter. The quadratic part of the matter source $\mathfrak{T}^{(1,1)}_q( T^{(0,1)}_{ab}, T^{(1,0)}_{ab}) $ vanishes everywhere as it contains products of the type  $\delta(r-r_{\rm p}(\tau))\delta(r-r_{\rm r})$ which are zero by our assumption. Thus in the outer region, all parts of $\mathcal{T}^{(1,1)}$ vanish.  The summary of all these terms is given in Table \ref{table:Teukolskyeqterms}.

\begin{table*}[]
 \caption{ This table shows the parts  of the effective source building blocks  sorted into three columns according to the regions in which they are nonvanishing. The last column on the other hand shows parts which vanish identically on our background. }
 \label{table:Teukolskyeqterms}
\centering
\begin{tabular}{|c| c| c| c||c|} 
 \hline
 Part of $\mathcal{S}^{(1,1)}_{\rm eff}$ nonzero at  & $r<r_{\rm r}$ & $r=r_{\rm r}$ & $r>r_{\rm r}$ & $0$ everywhere\\ 
 \hline
$ \mathcal{O}^{(1,0)} \Psi_4^{(0,1)}+ \mathcal{O}^{(0,1)} \Psi_4^{(1,0)}$   & $ \mathcal{O}^{(1,0)(-)}_\mathrm{D\mbox{-}vac}\Psi_4^{(0,1)}$ & $ \left(\mathcal{O}^{(1,0)(r_{\rm r})}_\mathrm{D\mbox{-}vac}+\mathcal{O}^{(1,0)}_{\mathrm{corr}}\right) \Psi_4^{(0,1)}$ & $ \mathcal{O}^{(1,0)(+)}_\mathrm{D\mbox{-}vac}\Psi_4^{(0,1)}$& $\mathcal{O}^{(0,1)} \Psi_4^{(1,0)}$ \\ 
 \hline
 $\mathcal{K}^{(1,1)}(\Psi_3^{(0,1)},\Psi_3^{(1,0)})$ & 0 & $\mathcal{K}^{(1,1)}(\Psi_3^{(0,1)},\Psi_3^{(1,0)})$ & $0$& - \\
 \hline
 $\mathcal{T}^{(1,1)}$  &\shortstack[l]{
$\mathfrak{T}^{(1,0)ab}T^{(0,1)}_{ab} $ \\
$+\mathfrak{T}^{(0,0)ab} T^{(1,1)}_{ab(\rm particle)}$
} & \shortstack[l]{
$\mathfrak{T}^{(0,1)ab}T^{(1,0)}_{ab} $ \\
$+\mathfrak{T}^{(0,0)ab} T^{(1,1)}_{ab(\rm matter)}$
}
 & $0$& $\mathfrak{T}^{(1,1)}_q( T^{(0,1)}_{ab}, T^{(1,0)}_{ab}) $ \\
 \hline
\end{tabular}
\end{table*}    

We can now separate the effective source as $\mathcal{S}^{(1,1)}_{\rm eff}=\mathcal{S}^{(1,1)}_{\rm ext}+\mathcal{S}^{(1,1)}_{r_{\rm r}}$  where 
\begin{align}
\label{Srssource}
    \begin{split}
    \mathcal{S}^{(1,1)}_{r_{\rm r}}= 
        &\mathfrak{T}^{(0,1)ab}T^{(1,0)}_{ab}+\mathfrak{T}^{(0,0)ab} T^{(1,1)}_{ab(\rm matter)}
        \\
        &-\Bigg[\left(\mathcal{O}^{(1,0)(r_{\rm r})}_\mathrm{D\mbox{-}vac}+\mathcal{O}^{(1,0)}_{\mathrm{corr}}\right)\Psi_4^{(0,1)}
        \\
        & +\mathcal{K}^{(1,1)}(\Psi_3^{(0,1)},\Psi_3^{(1,0)})\Bigg]\,,
    \end{split}
\end{align}  
 is the part of the source that lies only on the matter shell. This means that it can be written as      
      \begin{align}
       \label{Sourcedeltafunctions}
\mathcal{S}^{(1,1)}_{r_{\rm r}}=\displaystyle\sum_{i=0}^{3}\mathcal{S}^{(1,1)}_{r_{\rm r}(i)} \delta^{(i)}(r-r_{\rm r}).
      \end{align}     
This part of the source thus has a compact nature similarly to the point-particle source in the standard Teukolsky equation (the reason why the highest derivative of the delta function is $3$ will be shown in Section \ref{sec:equationsformatter}).  From this structure, it is easy to see that one needs to reconstruct the metric  $h^{(0,1)}_{\mu\nu}$   only on $r=r_{\rm r}$.

The other part of the source has an extended part and a compact one given by $\delta(r-r_{\rm p}(\tau))$ and its derivatives        
\begin{align}
    \begin{split}
    & \mathcal{S}^{(1,1)}_{\rm ext}= 
         \mathfrak{T}^{(1,0)ab}T^{(0,1)}_{ab}+\mathfrak{T}^{(0,0)ab} T^{(1,1)}_{ab(\rm particle)}
        \\
        & -\Bigg[\mathcal{O}^{(1,0)(-)}_\mathrm{D\mbox{-}vac}
         + \Theta(r-r_{\rm r})\left(\mathcal{O}^{(1,0)(+)}_\mathrm{D\mbox{-}vac}-\mathcal{O}^{(1,0)(-)}_\mathrm{D\mbox{-}vac}\right)\Bigg] \Psi_4^{(0,1)} .
    \end{split}
\end{align} 
 Similarly to the effective source one can also split the perturbation to the Weyl scalar as $\Psi_4^{(1,1)}=\Psi_{4(\mathrm{sm})}^{(1,1)}+\Psi_{4(r_{\rm r})}^{(1,1)}$, then instead of the original modified Teukolsky equation
    \begin{align}
\mathcal{O}^{(0,0)}\Psi_4^{(1,1)}=\mathcal{S}^{(1,1)}_{\rm eff},
      \end{align}        
we have two independent equations   
      \begin{align}
      \label{2Teukequations}
\mathcal{O}^{(0,0)}\Psi_{4(\mathrm{sm})}^{(1,1)}=\mathcal{S}^{(1,1)}_{\rm ext},\hspace{10pt} \mathcal{O}^{(0,0)}\Psi_{4(r_{\rm r})}^{(1,1)}=\mathcal{S}^{(1,1)}_{r_{\rm r}}.
      \end{align} 
The solution $\Psi_{4(\mathrm{sm})}^{(1,1)}$ to the first equation is smooth at $r=r_{\rm r}$. This smoothness arises because the source $\mathcal{S}^{(1,1)}_{\rm ext}$ contains only step functions $\Theta(r-r_{\rm r})$ but no delta functions. Since $\mathcal{O}^{(0,0)}$ is a second-order differential operator, step-function sources ensure that both $\Psi_{4 {\rm (sm)}}^{(1,1)}$ and its first derivative remain continuous across $r=r_{\rm r}$.
On the other hand  $\Psi_{4(r_{\rm r})}^{(1,1)}$ is not smooth or even continuous on the shell as it is sourced by delta functions and their derivatives. 
  
  So we are now left with two separate equations to solve. To find $\Psi_{4(r_{\rm r})}^{(1,1)}$ , one has to construct the source $\mathcal{S}^{(1,1)}_{r_{\rm r}}$, which is challenging as it requires metric reconstruction,  and also solving  the $(1,1)$ equations for the matter, this will be discussed in detail discussed in sections \ref{sec:sourceontheshell} and \ref{sec:equationsformatter}.

 The equation for $\Psi_{4 {\rm (sm)}}^{(1,1)}$  can be on the other hand recast as a completely different problem. If we now define
\begin{align}
    \mathcal{O}_{(\pm)}
        & =\mathcal{O}^{(0,0)}+\zeta\mathcal{O}^{(1,0)}_\mathrm{D\mbox{-}vac(\pm)} \,,
    \\
    \mathcal{T}_\mathrm{D\mbox{-}vac}
        &=\mathfrak{T}^{(0,0)ab}T^{(0,1)}_{ab}+\zeta\mathfrak{T}^{(1,0)ab}T^{(0,1)}_{ab}
        \\
        &\phantom{=} +\zeta\mathfrak{T}^{(0,0)ab} T^{(1,1)}_{ab(\rm particle)} \,,
    \\
    \Psi_{4}^{(\pm)}
        &=\Psi_{4}^{(0,1)}+\zeta \Psi_{4(\mathrm{sm})}^{(1,1)(\pm)}\,,
\end{align}  
where $\zeta$ is just a bookkeeping parameter as in other parts of the text. So instead of solving the equation for $\Psi_{4(\mathrm{sm})}^{(1,1)}$ we solve the pair of equations 

    \begin{align}
    \label{MatchedTeukek}
      \begin{split}
    \mathcal{O}_{(-)}\Psi_{4}^{(-)}&=\mathcal{T}_\mathrm{D\mbox{-}vac} \enspace \text{for} \enspace  r<r_{\rm r} \,,\\
\mathcal{O}_{(+)}\Psi_{4}^{(+)}&=0 \enspace  \text{for}\enspace  r>r_{\rm r} \,,
  \end{split} 
      \end{align} 
together with the smoothness conditions on $r_{\rm r}$        
   \begin{align}
  & \Psi_{4}^{(-)}\vert_{r=r_{\rm r}} =\Psi_{4}^{(+)}\vert_{r=r_{\rm r}}\,, 
  \\
  & \partial_\mu \Psi_{4}^{(-)}\vert_{r=r_{\rm r}} =\partial_\mu \Psi_{4}^{(+)}\vert_{r=r_{\rm r}} \,.  
      \end{align}        
The equations  \eqref{MatchedTeukek}  are just standard Teukolsky equations in linearized Kerr spacetime, more specifically if we denote the Teukolsky operator in Kerr spacetime as $\mathcal{O}^{({\rm K})}(M,a)$, we can write

\begin{align}
    \mathcal{O}_{(-)}=&\mathcal{O}^{({\rm K})}(M,0),  \\
    \begin{split}
    \mathcal{O}_{(+)} =
        &\mathcal{O}^{({\rm K})}(M,0)+m_{\rm r}\partial_{m} \mathcal{O}^{({\rm K})}(M,0)
        \\
        & +a_{\rm r} \partial_a \mathcal{O}^{({\rm K})}(M,0).
    \end{split}
\end{align}      
Thus, $\mathcal{O}_{(-)}$ is just a Schwarzschild Teukolsky operator $\mathcal{O}^{(0,0)}$, but written in the outer coordinates $x$ which are used globally, while the operator in the outer region corresponds to a slowly rotating Kerr with a mass perturbation $m_{\rm r}$. A similar argument can be made for the source $\mathcal{T}_\mathrm{D\mbox{-}vac}$, which is just the source of the Schwarzschild Teukolsky equation. All of this is not surprising at all since these Teukolsky equations correspond to their respective spacetimes \eqref{appinnermetric} and \eqref{appoutermetric}  on each side of the matter shell. 

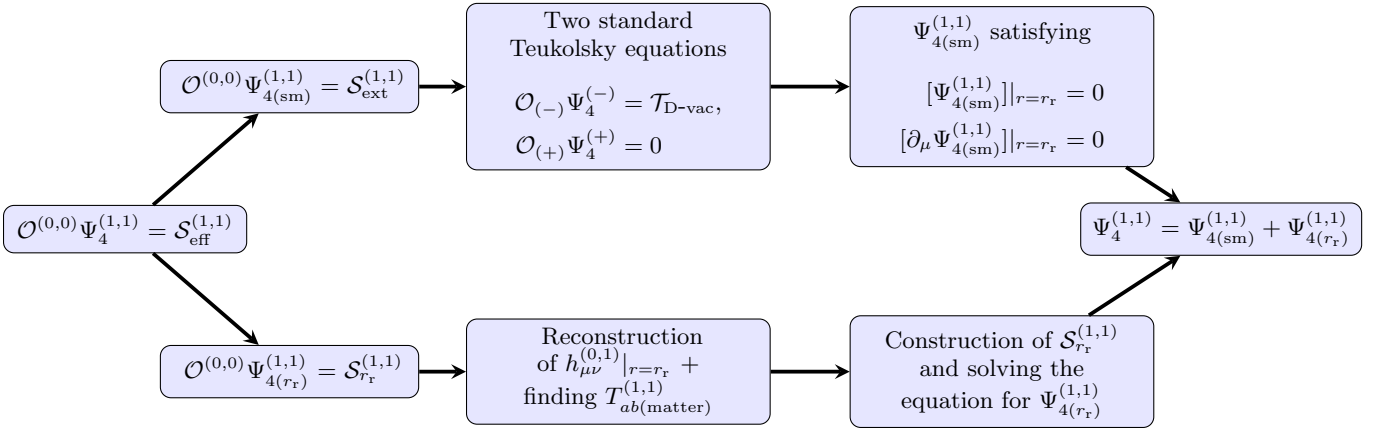
\begin{figure*}
\begin{tikzpicture}[scale=1.45]

    \tikzstyle{box} = [rectangle, draw, fill=blue!10, text centered, rounded corners]

    \node[box, text width=3.0 cm, minimum height=0.5cm] (box1)  at (0,0) {{$ \mathcal{O}^{(0,0)}\Psi_4^{(1,1)}=\mathcal{S}^{(1,1)}_{\rm eff}$  }    };
    \node[box, text width=3.2cm, minimum height=0.5cm ] (box2a)  at (1.5,1.3) {{ $\mathcal{O}^{(0,0)}\Psi_{4 {\rm (sm)}}^{(1,1)}=\mathcal{S}^{(1,1)}_{\rm ext}$}};
    \node[box, text width=3.2cm, minimum height=0.5cm] (box2b)  at (1.5,-1.3) {{ $\mathcal{O}^{(0,0)}\Psi_{4(r_{\rm r})}^{(1,1)}=\mathcal{S}^{(1,1)}_{r_{\rm r}}$}};
    \node[box, text width=3.8cm, minimum height=0.5cm]  (box3a)  at (4.5,1.3) {{Two standard Teukolsky equations 
     \begin{align}
     \nonumber\mathcal{O}_{(-)}\Psi_{4}^{(-)}&=\mathcal{T}_\mathrm{D\mbox{-}vac}, \\\nonumber\mathcal{O}_{(+)}\Psi_{4}^{(+)}&=0
      \end{align}
    }};
    \node[box, text width=3.8cm, minimum height=0.5cm] (box3b) at (4.5,-1.3) {{ Reconstruction of $ h^{(0,1)}_{\mu\nu}\vert_{r=r_{\rm r}}$ + finding $T^{(1,1)}_{ab(\rm matter)}$ }};
    \node[box, text width=3.8cm, minimum height=0.5cm] (box4a) at (8,1.3) {{$\Psi_{4 {\rm (sm)}}^{(1,1)}$ satisfying 
     \begin{align}
     \nonumber [\Psi_{4 {\rm (sm)}}^{(1,1)}]\vert_{r=r_{\rm r}}=0\\\nonumber [\partial_\mu\Psi_{4 {\rm (sm)}}^{(1,1)}]\vert_{r=r_{\rm r}}=0
      \end{align}
    }};
    \node[box, text width=3.8cm, minimum height=0.5cm] (box4b) at (8,-1.3) {{Construction of $\mathcal{S}^{(1,1)}_{r_{\rm r}}$ and solving the equation for $\Psi_{4(r_{\rm r})}^{(1,1)}$}};
 \node[box, text width=3.5cm, minimum height=0.5cm] (box5) at (10,0) {{$\Psi_4^{(1,1)}=\Psi_{4 {\rm (sm)}}^{(1,1)}+\Psi_{4(r_{\rm r})}^{(1,1)}$}};

    \draw[->, thick, >=stealth, line width=0.5mm] (box1) -- (box2a);
   \draw[->, thick, >=stealth, line width=0.5mm] (box1) -- (box2b);
  \draw[->, thick, >=stealth, line width=0.5mm] (box2a) -- (box3a);
    \draw[->, thick, >=stealth, line width=0.5mm] (box3a) -- (box4a);
            \draw[->, thick, >=stealth, line width=0.5mm] (box4a) -- (box5);
  \draw[->, thick, >=stealth, line width=0.5mm] (box2b) -- (box3b);
    \draw[->, thick, >=stealth, line width=0.5mm] (box3b) -- (box4b);
            \draw[->, thick, >=stealth, line width=0.5mm] (box4b) -- (box5);

\end{tikzpicture}
\caption{This flowchart shows how the modified Teukolsky equation is reduced to two independent  problems due to the symmetries of the background metric \eqref{background}. The key insight is that our piecewise spacetime allows us to solve standard Teukolsky equations in each region, with all environmental complexity confined to matching conditions at the matter shell. One part of the problem involves a solution to two standard Teukolsky equation matched on the shell $r=r_{\rm r}$ while the bottom part of the flowchart describes a construction of a complicated source living on the matter shell which is essential for finding $\Psi_{4(r_{\rm r})}^{(1,1)}$.  The prediction for the environmental contribution to the GW flux is obtained from the sum of the outputs of these two independent calculations.}
 \label{flowchart2psi4}
 \end{figure*}

Now that we have formulated our strategy, which is summarized in the flowchart \ref{flowchart2psi4}, we can use the complete separability of the operators $\mathcal{O}^{(0,0)}$, $\mathcal{O}_{(-)}$ and $\mathcal{O}_{(+)}$ to solve the equation for $\Psi_{4(r_{\rm r})}^{(1,1)}$ and find the smooth part $\Psi_{4(\rm sm)}^{(1,1)}$  by matching the solutions of the equations \eqref{MatchedTeukek} on the matter shell. To use the separability we have to expand all the quantities into individual modes.

%%%%%%%%%%%%%%%%%%%%%%%%%%%%%%%%%%%%%%%%%%%%%%%%%%%%%%%%%%%%%%%%%%%%%%%%%%%%%%%%%%%%%%%%%%%%%%%%%%%%%%%%%%%%%%%
\section{Solution of the modified Teukolsky equation}
\label{sec:solteuk}
%%%%%%%%%%%%%%%%%%%%%%%%%%%%%%%%%%%%%%%%%%%%%%%%%%%%%%%%%%%%%%%%%%%%%%%%%%%%%%%%%%%%%%%%%%%%%%%%%%%%%%%%%%%%%%%

Having established the theoretical framework and decomposed the problem in the space-time of a black hole surrounded by the pole-dipole ring, we now describe the solution of the final modified Teukolsky equation in order to obtain corrections to GW emission. We break the problem into five interconnected computational tasks. First, in Section \ref{Smoothmatching} we present the solution of the ``smooth part'', which involves matching standard Teukolsky solutions across the formal matter shell, accounting for frequency shifts and mode coupling introduced by the ring's mass and angular momentum. Second, in Section \ref{sec:reconstruction}, we tackle metric reconstruction: converting the Weyl scalar $\Psi_4^{(0,1)}$ sourced by the inspiraling particle back into metric perturbations using the CCK method, but only at the ring location where it contributes to environmental coupling. The heart of the calculation appears in the third Section \ref{sec:sourceontheshell}, where we assemble the effective source terms on the shell. A separate challenging aspect, addressed in Section \ref{sec:equationsformatter}, involves solving the coupled equations of motion for the ring matter itself, which oscillates under gravitational-wave driving and contributes additional emission. We close in Section \ref{subsec:finalsolution} by assembling these components into the complete solution, yielding the environmental corrections to  the Weyl scalar $\Psi_4^{(1,1)}$.

%%%%%%%%%%%%%%%%%%%%%%%%%%%%%%%%%%%%%%%%%%%%%%%%%%%%%%%%%%%%%%%%%%%%%%%%%%%%%%%%%%%%%%%%%%%%%%%%%%%%%%%%%%%%%%%
\subsection{The smooth part of the solution  }    
\label{Smoothmatching}
%%%%%%%%%%%%%%%%%%%%%%%%%%%%%%%%%%%%%%%%%%%%%%%%%%%%%%%%%%%%%%%%%%%%%%%%%%%%%%%%%%%%%%%%%%%%%%%%%%%%%%%%%%%%%%%
  
Let us now introduce our procedure for smoothly matching the solutions of the two Teukolsky equations $\Psi_4^{(-)}$ and $\Psi_4^{(+)}$ on the shell $r=r_{\rm r}$. In this section, we will sometimes write  $\Psi_4^{(-)}$ and   $\Psi_4^{(+)}$  in an unperturbed (full Kerr) form, but keep in mind that we consider only terms which are at most linear in $m_{\rm r}$ and $a_{\rm r}$ (or $\zeta$). With that said, we can mode-decompose these scalars as,
  
 \begin{align} 
  \label{twopsi4parts} 
   \begin{split}
\Psi_4^{(-)}=\frac{1}{r^4}\displaystyle\sum_{lm\omega^{\mathrm{(in)}}} R^{(-)}_{lm\omega^{\mathrm{(in)}}}(r)Y_{lm}(\theta)e^{\mathrm{i}(m\phi^{\mathrm{(in)}}-\omega^{\mathrm{(in)}} t^{\mathrm{(in)}})} ,
\\
\Psi_4^{(+)}=\frac{1}{(r-i a_{\rm r}\cos\theta)^4}\displaystyle\sum_{lm\omega} R^{(+)}_{lm\omega}(r) S_{lm\omega}(\theta)e^{\mathrm{i}(m\phi-\omega t)} ,
 \end{split} 
    \end{align} 
where we sum over the discrete frequencies $\omega_{mnk}$. We can see that $\Psi_4^{(-)}$ is expressed in the inner (Schwarzschild) coordinates, this affects also the ``inner frequency'' $\omega^{(i)}$ which can be written as 
 
  \begin{align}
\omega^{(\rm in)}_{mnk}=\frac{\mathrm{d}}{\mathrm{d}t^{\mathrm{(in)}}}(n\psi^r+k\psi^{\theta}+m\psi^{\mathrm{(in)}\phi}).
   \end{align}      
where $(\psi^r,\psi^\theta,\psi^{\mathrm{(in)}\phi})$ are the angle coordinates (part of action-angle coordinates in Kerr spacetime). Since the geodesic motion is separable and $r$ and $\theta$  coordinates are global, the angles $\psi^r$ and $\psi^\theta$ do not transform. As for  $\psi^{\mathrm{(in)}\phi}$, this can be separated into three parts, $\psi^{\mathrm{(in)}\phi}=\phi^{\mathrm{(in)}}-\Delta\phi_r-\Delta\phi_\theta$  \cite{van_de_Meenttransport}, where $\Delta\phi_r$   and  $\Delta\phi_\theta$  are functions of $\psi^r$ and $\psi^\theta$, respectively. We can now use the transformation relations \eqref{coord-transform} to express $\omega^{(i)}$ using $\omega$
         
  \begin{align}
  \label{infrequency}
\omega^{(\rm in)}_{mnk}=(1+z_{\rm r})\omega_{mnk}-m \Omega_{\rm r}.
   \end{align}       
The presence of the matter shell thus rescales and shifts the frequency by parameters   $z_{\rm r} \propto m_{\rm r}$ and $\Omega_{\rm r} \propto a_{\rm r}$.  The  $t$-$\phi$ part in the solutions \eqref{twopsi4parts} is then automatically matched   
    \begin{align}
m\phi^{\mathrm{(in)}}-\omega^{\mathrm{(in)}} t^{\mathrm{(in)}}=m\phi-\omega t+\mathcal{O}(\zeta^2).
   \end{align}    
The $r$-$\theta$ part is more difficult; we have to smoothly match the solutions and simultaneously satisfy the boundary conditions at infinity and the horizon. In the absence of the matter shell, the solution \eqref{TeukolskyRadialsolution} with functions $C^{(A)}$ given by \eqref{TeukolskyRadialvariationofconstants} satisfies the boundary conditions. We shall denote this solution as  $R^{(\rm part)}$ as it is a particular solution to the inhomogeneous equation \eqref{TeukolskyKerrRadial}.  When the matter shell is present we shall consider the following ansatz \begin{align}
\begin{split}
   R^{(-)}_{lm\omega^{\mathrm{(in)}}}(r)&=R^{(\rm part)}_{lm\omega^{\mathrm{(in)}}}(r)+C^{(\infty)(\rm corr)}_{lm\omega^{\mathrm{(in)}}}R^{(H)}_{lm\omega^{\mathrm{(in)}}}(r) ,\\
       R^{(+)}_{lm\omega}(r)&=C^{(H)}_{lm\omega} R^{(\infty)}_{lm\omega}(r), 
\end{split} 
\end{align} 
where $C^{(\infty)(\rm corr)}$    and $C^{(H)}$ are so far undetermined constants.   In general, we would have two constants on each side of the boundary $r=r_{\rm r}$ since a general solution can be written as a particular solution plus a general homogeneous solution, but the boundary conditions leave us only with  $C^{(\infty)(\rm corr)}$  and $C^{(H)}$. At infinity and on the horizon, we then have the desired result 
\begin{align}
\begin{split}
   R^{(-)}_{lm\omega^{\mathrm{(in)}}} (r\rightarrow 2M)&\sim  \left(C^{(\infty)(\rm part)}_{lm\omega^{\mathrm{(in)}}}+C^{(\infty)(\rm corr)}_{lm\omega^{\mathrm{(in)}}}\right) \Delta^{2}e^{-\mathrm{i}p r_{*}}, \\
    R^{(+)}_{lm\omega} (r\rightarrow \infty)&\sim C^{(H)}_{lm\omega}r^{3}e^{\mathrm{i}\omega r_{*}}. 
    \end{split}
     \end{align}  
 Now, to find the constants $C^{(\infty)(\rm corr)}$ and $C^{(H)}$ by matching, we have to expand all the quantities to the order $\mathcal{O}(\zeta)$. For the radial function this means
   \begin{align}
   \label{radialfuncexpansion}
   \begin{split}
R^{(\pm)}_l(M+m_{\rm r},a_{\rm r})\approx R_{({\rm Schw})l}\\ +m_{\rm r}\partial_m R_{l}^{(\pm)}+a_{\rm r}\partial_a R_{l}^{(\pm)} ,
\end{split}
   \end{align} 
where the leading term $R_{({\rm Schw})l}$ corresponds to Schwarzschild space-time. From now on, we shall omit all indices apart from $l$ for notational simplicity. 
Let us note that $R^{(-)}_l$ depends on $m_{\rm r}$ and $a_{\rm r}$ through the frequency  $\omega^{(\rm in)}$ as was shown in eq.  \eqref{infrequency}. 

There is yet another problem in the fact that due to spin parameter $a_{\rm r}$, the angular part of  $\Psi_4^{(+)}$ depends on spin-weighted spheroidal harmonics rather than spin-weighted  spherical harmonics $Y_{l}$ (we omit the spin  weight $-2$ here).   Since $m_{\rm r}$ and $a_{\rm r}$ are in principle independent, we can treat the terms in the expansion \eqref{radialfuncexpansion} separately. Starting with the terms proportional to $m_{\rm r}$, we know that for $a_{\rm r}=0$ the angular part just reduces to $Y_{l}$. We can thus formulate the smoothness matching condition for the radial function and its derivative as
      \begin{align}
         \label{massmatching}
         \begin{split}
\partial_m R_{l}^{(-)}(r_{\rm r})&=\partial_m R_{l}^{(+)}(r_{\rm r}), \\
\partial_m (R')_{l}^{(-)}(r_{\rm r})&=\partial_m (R')_{l}^{(+)}(r_{\rm r}) ,
    \end{split}
   \end{align}   
where we further expand the constants  $C^{(\infty)(\rm corr)}$    and $C^{(H)}$ as well as the homogeneous solutions in the same fashion as   \eqref{radialfuncexpansion} to get the contributions linear in $m_{\rm r}$.

We can now look at the more complicated part, the terms proportional to $a_{\rm r}$. As was already stated, the angular part of $\Psi_4^{(+)}$ is given by $S_{l}(\theta)$ depending on $a_{\rm r}$ while there is also a prefactor $\xi^{-4}$  depending on $a_{\rm r}$ and $\theta$ which itself is not separable. Fortunately one can expand in $a_{\rm r}$ so that one can have both  $\Psi_4^{(-)}$ and $\Psi_4^{(+)}$   in the form  
 \begin{align}   
  \label{newradialmodes}
\Psi_4^{(\pm)}=\frac{1}{r^4}\displaystyle\sum_{lm\omega}\mathfrak{R}^{(\pm)}_{lm\omega}(r)Y_{lm}(\theta)e^{\mathrm{i}(m\phi-\omega t)} ,
    \end{align} 
where all the dependence on $a_{\rm r}$ is absorbed into  the new radial functions $\mathfrak{R}^{(\pm)}$. To pass from $R^{(\pm)}$ to $\mathfrak{R}^{(\pm)}$ we perform the slow-rotation expansion of spin-weighted spheroidal harmonics \cite{ShahWhiting}
  
  \begin{align}
  \begin{split}
S_{l}(\theta)=& Y_{l}(\theta)+a_{\rm r}\left(b_{l-1}Y_{l-1}(\theta)+b_{l+1}Y_{l+1}(\theta) \right)
\\ & +\mathcal{O}(a_{\rm r}^2) ,
\end{split}
   \end{align}     
 and we  also expand  the factor $\xi^{-4}$  
      \begin{align}
\frac{1}{(r-\mathrm{i}a_{\rm r}\cos(\theta) )^4}\approx \frac{1}{r^4}+\frac{4 i a_{\rm r} \cos (\theta )}{r^5}.
  \end{align}
To eliminate the angular dependence in $\cos(\theta)$ we use the identity \cite{seibert2018spin}
 \begin{align}
\cos (\theta )Y_{l}(\theta)=\sum_{j=l-1}^{l+1}c_j Y_{j}(\theta) \,.
  \end{align}  
  Putting it together, we can  write
 \begin{align}
 \begin{split}
 \label{angularexpansion}
\xi^{-4} S_{l}(\theta)=& \frac{1}{r^4}\left(Y_{l}(\theta)+a_{\rm r}\sum_{j=l-1}^{l+1}d_j(r) Y_{j}(\theta)\right)
\\ & +\mathcal{O}(a_{\rm r}^2) ,
\end{split}
  \end{align}    
where the coefficients $b_j$, $c_j$ and $d_j$ can be found in Appendix \eqref{app:new angularn expansion}.  By comparing the spherical harmonics expansion  \eqref{angularexpansion} with the mode decomposition  \eqref{newradialmodes}, we can find the form of $\mathfrak{R}^{(\pm)}$, which can be expanded in the same way as  \eqref{radialfuncexpansion}. On the Schwarzschild level we have $\mathfrak{R}_{(\rm Schw)l}^{(\pm)}(r)=R_{(\rm Schw)l}^{}(r)$ while for the terms proportional to $a_{\rm r}$ the result is

 \begin{align} 
     \begin{split}
 \partial_a\mathfrak{R}_{l}^{(-)}(r)&=\partial_a R_{l}^{(-)}(r)\,,
 \\
\partial_a\mathfrak{R}_{l}^{(+)}(r)&=\partial_a R_{l}^{(+)}(r)+\sum_{j=l-1}^{j=l+1}d_j(r)R_{(\rm Schw)j}(r) \,.
    \end{split}
    \end{align}  
 We can clearly see that there is a mode mixing in $\partial_a\mathfrak{R}_{l}^{(+)}(r)$. Conceptually, this is because even though the Teukolsky equation is individually separable in each of the regions $r<r_{\rm r}, r>r_{\rm r}$, they separate into different bases. Specifically, the spheroidal harmonics depend on the spheroidicity parameter $a\omega$, which is different on each side of the shell. Effectively, the problem can then be viewed from either side of the shell as solving the Teukolsky equation with non-separable boundary conditions. From these considerations, it is then apparent  that  $\mathfrak{R}^{(+)}$ is not an individual mode solution to the radial Teukolsky equation   \eqref{TeukolskyKerrRadial} but rather a sum of modes. The matching condition then reads
  \begin{align}
   \label{spinmatching}
        \begin{split}
 \partial_a\mathfrak{R}_{l}^{(-)}(r_{\rm r})&= \partial_a\mathfrak{R}_{l}^{(+)}(r_{\rm r}) \,,\\
  \partial_a\mathfrak{(R')}_{l}^{(-)}(r_{\rm r})&= \partial_a\mathfrak{(R')}_{l}^{(+)}(r_{\rm r}) \,.
      \end{split}
     \end{align} 
Unlike in the matching condition     \eqref{massmatching}  the matched constants in $\partial_a\mathfrak{R}_{l}^{(\pm)}(r)$ now depend on the solutions to the Schwarzschild Teukolsky inhomogeneous equation  $R_{(\rm Schw)j}(r)$ where $j=l-1,l,l+1$.

Solving the matching conditions   \eqref{massmatching} and \eqref{spinmatching} provides the smooth part of the solution to the modified Teukolsky equation  $\Psi_{4(\rm sm)}^{(1,1)(\pm)}$, which can be decomposed as in  \eqref{twopsi4parts}   with the radial part 
   \begin{align}
R_{l(\rm sm)}^{(1,1)(\pm)}=m_{\rm r}\partial_m R_{l}^{(\pm)}+a_{\rm r}\partial_a R_{l}^{(\pm)}
     \end{align} 
determined by the matching conditions. When calculating the GW  fluxes to infinity and through the horizon, the required  quantities are the two constants  $C^{(H)}_{lm\omega}$ and $C^{(\infty)}_{lm\omega^{\mathrm{(in)}}}=C^{(\infty)(\rm part)}_{lm\omega^{\mathrm{(in)}}}+C^{(\infty)(\rm corr)}_{lm\omega^{\mathrm{(in)}}}$ found through the matching conditions   \eqref{massmatching} and \eqref{spinmatching} and expanded to the linear order  in $\zeta$.

%%%%%%%%%%%%%%%%%%%%%%%%%%%%%%%%%%%%%%%%%%%%%%%%%%%%%%%%%%%%%%%%%%%%%%%%%%%%%%%%%%%%%%%%%%%%%%%%%%%%%%%%%%%%%%%
\subsection{Reconstruction of the metric \texorpdfstring{$h^{(0,1)}_{\mu\nu}$}{h(0,1)munu}} 
\label{sec:reconstruction}
%%%%%%%%%%%%%%%%%%%%%%%%%%%%%%%%%%%%%%%%%%%%%%%%%%%%%%%%%%%%%%%%%%%%%%%%%%%%%%%%%%%%%%%%%%%%%%%%%%%%%%%%%%%%%%%
The curvature scalar $\Psi_4^{(0,1)}$ captures only the perturbations to the Weyl curvature tensor, which is directly related to gravitational radiation emerging from the EMRI at large distances. However, in our case, we also need to understand the local interaction of the radiation from the EMRI with the matter environment, which requires reconstructing the metric perturbation $h^{(0,1)}_{\mu\nu}$ from the $\Psi_4^{(0,1)}$ field.

To obtain the effective source $\mathcal{S}^{(1,1)}_{r_{\rm r}}$  one has to first reconstruct the metric perturbation because most  terms in $\mathcal{S}^{(1,1)}_{r_{\rm r}}$ contain $(0,1)$ NP quantities/operators. The crucial simplification of our model lies in the fact that one has to reconstruct the metric and its derivatives only at $r=r_{\rm r}$. This means we have to essentially find a metric satisfying homogeneous linearized Einstein equations $\delta G_{\mu\nu}[h^{(0,1)}]=0$ on the Schwarzschild background. Thanks to this special setup, we can use the CCK metric reconstruction approach introduced in Refs \cite{Chrzanowski:1975wv,Kegeles:1979an}. This involves obtaining the metric from a single scalar called the Hertz potential \cite{Pound_2021}. The metric is then obtained from the potential in the radiation gauge, in which one requires the metric to be traceless and to have vanishing components with respect to projections into a null vector. There are two different radiation gauges, the ingoing radiation gauge (IRG) and the outgoing radiation gauge (ORG) along with their respective Hertz potentials. The IRG and ORG differ in the choice of the null direction in the gauge condition; here we will be working in the ORG, which yields asymptotically flat metric perturbations \cite{van_de_MeentShah}.

  The CCK approach is closely tied to the standard Teukolsky formalism and can be done in any type D vacuum spacetime, so we will introduce it in that way. The CCK metric reconstruction relies on the operator identity \cite{Wald}
  \begin{align} 
  \label{Waldoperator} 
\mathscr{\tilde{S}}^{(0)}_4\mathcal{E}^{(0)}=\mathcal{O}^{(0)}\mathscr{\tilde {T}}^{(0)}_4 ,
  \end{align} 
where $\mathcal{O}^{(0)}$ is the operator in the standard Teukolsky equation, while the other operators are defined by the relations
 \begin{align}  
\mathscr{\tilde{S}}^{(0)}_4 T^{(1)}&=\mathcal{T}^{(1)},\enspace \mathscr{\tilde {T}}^{(0)}_4 h^{(1)}=\Psi_4^{(1)},\\ 
\mathcal{E}^{(0)}h^{(1)}&=\delta G_{\mu\nu}[h^{(1)}].
  \end{align} 
Applying the identity   \eqref{Waldoperator}  on a metric perturbation $h_{\mu\nu}^{(1)}$ will give us the standard Teukolsky equation for $\Psi_4^{(1)}$
  \begin{align}  
  \label{StandardTeukolsky2}
\mathcal{O}^{(0)}\Psi_4^{(1)}=\mathcal{T}^{(1)}.
  \end{align} 
However, if we now take the adjoint (see \cite{Wald}) of the operator identity , we arrive at%
\footnote{
We would like to note that there are several different conventions for the Hertz potential and the operators, differing by some rescaling factors.  One of them is presented, for example, by \citet{Pound_2021}, while others can be found in \citet{van_de_MeentShah} or the original paper of \citet{Chrzanowski:1975wv}. In this work we use the operators $\mathscr{S}^{(0)}_4=2\mathscr{\tilde{S}}^{(0)}_4$ and $\mathscr{T}^{(0)}_4=2\mathscr{\tilde{T}}^{(0)}_4$, which does not change the identity \eqref{Waldoperator} and its adjoint form \eqref{Waldoperatoradj}.
}
  \begin{align}  
    \label{Waldoperatoradj} 
\mathcal{E}^{(0)}\mathscr{\tilde{S}}^{(0)\dagger}_4=\mathscr{\tilde{T}}^{(0)\dagger}_4\mathcal{O}^{(0)\dagger},
  \end{align} 
where  we made use of the fact that the Einstein operator  $\mathcal{E}^{(0)}$  is self-adjoint.

 If we now define the metric perturbation using the Hertz potential  $\Psi_{\rm Hz}^{(1)}$ as 
  \begin{align}  
      \label{metricfromHZpot} 
h_{\mu\nu}^{(1)}=\left(\mathscr{S}^{(0)\dagger}_4\Psi_{\rm Hz}^{(1)}\right)_{\mu\nu},
  \end{align} 
then this metric satisfies the  linearized vacuum Einstein equation by the identity  \eqref{Waldoperatoradj} provided that  $\mathcal{O}^{(0)\dagger}\Psi_{\rm Hz}^{(1)}=0$. In summary, we have  
  \begin{align}  
        \label{adjTeukolsky} 
\mathcal{O}^{(0)\dagger}\Psi_{\rm Hz}^{(1)}=0 \Rightarrow  \mathcal{E}^{(0)}h^{(1)}=0.
  \end{align}
 By construction, the metric \eqref{metricfromHZpot} is in the ORG, satisfying
 \begin{align}  
 n^{\mu}  h_{\mu\nu}^{(1)}=0,\enspace  h^{(1)\mu}_{\mu}=0 .
  \end{align}  
  where the tetrad vector    $n^{\mu} $ corresponds to the background type D metric as does the trace. Since we use the NP/GHP formalism, we are only interested in the tetrad components $ h_{ab}^{(1)}$. We can then split the operator $\mathscr{S}^{(0)\dagger}_4$ into three parts%
  \footnote{Note that here the lower index denotes NP tetrad components (that is, the indices are \textit{not} related to the multipole expansion).} 
    \begin{align}  
     \begin{split}  
\mathscr{S}^{(0)\dagger}_{4ll}=&-\left(\ethm'-\tau'\right)\left(\ethm'+3\tau'\right) \,,
\\
\mathscr{S}^{(0)\dagger}_{4lm}=&-\frac{1}{2}\left(\text{\textthorn}'-\rho'+\bar{\rho}'\right)\left(\ethm'+3\tau'\right) \,,
\\
& -\frac{1}{2}\left(\ethm'-\tau'+\bar{\tau}\right)\left(\text{\textthorn}'+3\rho'\right) \,,
\\
\mathscr{S}^{(0)\dagger}_{4mm}=&-\left(\text{\textthorn}'-\rho'\right)(\text{\textthorn}'+3\rho') \,.
     \end{split}  
   \end{align}  
  Using this, we can explicitly write the nonzero tetrad components of the metric perturbation as 
  \begin{subequations}          \label{recmetriccomponents}  
     \begin{align}  
 h_{ll}^{(1)}&=\mathscr{S}^{(0)\dagger}_{4ll}\Psi_{\rm Hz}^{(1)}+\overline{\mathscr{S}^{(0)\dagger}_{4ll}\Psi_{\rm Hz}^{(1)}} \,,
 \\
  h_{lm}^{(1)}&=\mathscr{S}^{(0)\dagger}_{4lm}\Psi_{\rm Hz}^{(1)}=\overline{h_{l\bar{m}}^{(1)}} \,,
  \\
   h_{mm}^{(1)}&=\mathscr{S}^{(0)\dagger}_{4mm}\Psi_{\rm Hz}^{(1)}=\overline{h_{\bar{m}\bar{m}}^{(1)}} \,.
   \end{align} 
    \end{subequations}  
  Returning to the condition for the Hertz potential \eqref{adjTeukolsky}, one can show that the adjoint Teukolsky operator is related to the operator from the Teukolsky master equation with spin weight $s=2$ by $\mathcal{O}^{(0)\dagger}\Psi_{\rm Hz}^{(1)}\propto {}_{2}\mathcal{O}^{(0)}\left(\xi^{4}\Psi_{\rm Hz}^{(1)}\right)$ \cite{LoustoWhiting,Pound_2021}. This means that we can use the same separation ansatz as for $\Psi_{0}^{(1)}$. The Hertz potential must satisfy yet another condition; by applying $ \mathscr{\tilde {T}}^{(0)}_4$ on the metric \eqref{metricfromHZpot} 
 we obtain the Weyl scalar perturbation $\Psi_{4}^{(1)}$  in terms of the Hertz potential. This Weyl scalar is then a solution of the previously solved standard Teukolsky equation \eqref{StandardTeukolsky2}. Thus, combining the two conditions, we have to satisfy%
 \footnote{ Our Hertz potential is related to the one in Pound and Wardell \cite{Pound_2021} by $\Psi_{\rm Hz}^{(1)}=\frac{1}{2}\Psi_{\rm Hz}^{(\mathrm{P-W} )(1)}$, while in van de Meent and Shah \cite{van_de_MeentShah} the form used is $\Psi_{\rm Hz}^{(1)}=\xi^{-4}\Psi_{\rm Hz}^{(\mathrm{vdM-S} )(1)}$. }
      \begin{align}
      \label{Hzpot conditions}  
 {}_{2}\mathcal{O}^{(0)}\left(\xi^{4}\Psi_{\rm Hz}^{(1)}\right)=0,\enspace   \frac{1}{2}(\text{\textthorn}')^4\overline{\Psi_{\rm Hz}^{(1)}}=\Psi_{4}^{(1)}.
   \end{align}
  
  We can now translate this formalism to our particular model, where we construct the $(0,1)$  metric perturbation on the $(0,0)$ Schwarzschild background, meaning that 
  \begin{align}  
    \begin{split} 
h^{(1)}_{\mu\nu} &\rightarrow h^{(0,1)}_{\mu\nu},\enspace  \Psi_4^{(1)} \rightarrow \Psi_4^{(0,1)},\\
 \Psi_{\rm Hz}^{(1)}&\rightarrow \Psi_{\rm Hz}^{(0,1)}  ,\enspace \mathcal{O}^{(0)} \rightarrow \mathcal{O} ^{(0,0)}.
     \end{split} 
  \end{align}   
    We can now satisfy the two equations  \eqref{Hzpot conditions}  using the mode decomposition introduced in the previous section.  The Weyl scalar $\Psi_4^{(0,1)}$ can be expanded into spin-weighted spherical harmonics with $s=-2$
 \begin{align}  
 \label{psi4Schwepasnion}
\Psi_4^{(0,1)}=\frac{1}{r^4}\displaystyle\sum_{lm\omega}{}_{-2} R^{(0,1)}_{lm\omega}(r){}_{-2}Y_{lm}(\theta)e^{i(m\phi-\omega t)}.
    \end{align}   
 Likewise, the function $R^{(0,1)}_{lm\omega}$ is a solution to the Teukolsky radial equation \eqref{TeukolskyKerrRadial} with $s=-2$. The Hertz potential, on the other hand, also solves the (homogeneous) Teukolsky equation but with $s=2$, which means that we can write down an analogical decomposition for $\Psi_{\rm Hz}^{(0,1)}$ as well
   \begin{align} 
       \label{modesHertz} 
\Psi_{\rm Hz}^{(0,1)}=\frac{1}{r^4}\displaystyle\sum_{lm\omega}{}_{2} R^{\mathrm{Hz}(0,1)}_{lm\omega}(r){}_{2}Y_{lm}(\theta)e^{i(m\phi-\omega t)} .
    \end{align}
We are still left with the other condition in \eqref{Hzpot conditions}, which is a fourth-order differential  equation  relating $\Psi_{\rm Hz}^{(0,1)}$  directly to $\Psi_{4}^{(0,1)}$.  We can use the fact that we need to reconstruct the metric on $r=r_{\rm r}$  where the Teukolsky equation is homogeneous ( $r_{\rm r}>r_{\rm max}>r(t)$), and then the respective radial functions read 
     \begin{align} 
        \begin{split} 
{}_{-2}R^{(0,1)}_{lm\omega}(r>r_{\rm max})&={}_{-2}C^{(\rm H)}_{(\mathrm {Schw})lm\omega}{}_{-2} R^{(\infty)}_{(\mathrm{Schw})lm\omega}(r)\\
{}_{2}R^{\mathrm{Hz}(0,1)}_{lm\omega}(r>r_{\rm max})&=C^{\rm Hz}_{lm\omega}{}_{2} R^{(\infty)}_{(\mathrm{Schw})lm\omega}(r).
     \end{split} 
    \end{align}   
Thus, all we have to do is express the amplitude $C^{\rm Hz}$ using ${}_{-2}C^{(\rm H)}$. Fortunately, this problem was already solved in Ref. \cite{van_de_MeentShah} and there is a simple algebraic relation, which can be used in our particular model \footnote{\citet{van_de_MeentShah} considered only eccentric equatorial orbits, which is the reason why the index $k$ corresponding to an inclined motion is missing in their formula.}
    
   \begin{align} 
   \label{amplitudesHertz} 
C^{\rm Hz}_{lmnk}=(-1)^{l+m+k} \frac{2}{\omega_{mnk}^4}{}_{-2}C^{(\rm H)}_{(\mathrm {Schw})lmnk}.
    \end{align}      
The procedure for obtaining the Hertz potential $\Psi_{\rm Hz}^{(0,1)}$ is therefore straightforward. First, we solve the $(0,1)$ (Schwarzschild) Teukolsky equation  \eqref{Teuk01} , then use relation \eqref{amplitudesHertz} to obtain the amplitudes of the Hertz potential whose individual modes  \eqref{modesHertz}  are given by the solutions to the $s=2$ Teukolsky master equation.

From the metric components   \eqref{recmetriccomponents}, one can derive all the  $(0,1)$ NP quantities forming the building blocks of the effective source (see terms in Table \ref{table:Teukolskyeqterms}), starting from the tetrad for which we use the following choice 
    \begin{align}  
    \label{01tetrad}
        \begin{split}  
 l^{\mu(0,1)} &=\frac{1}{2} h^{(0,1)}_{ll} n^{\mu(0,0)} \,,
 \\
  n^{\mu(0,1)} &=\frac{1}{2}h^{(0,1)}_{nn}l^{\mu(0,0)}+h^{(0,1)}_{ln}n^{\mu(0,0)} \,,
  \\
 m^{\mu(0,1)} &=h^{(0,1)}_{nm}l^{\mu(0,0)}-\frac{1}{2}h^{(0,1)}_{m\bar{m}}m^{\mu(0,0)} 
 \\&-\frac{1}{2}h^{(0,1)}_{mm}\bar{m}^{\mu(0,0)}+h^{(0,1)}_{lm}n^{\mu(0,0)} \,.
   \end{split}  
   \end{align}  
This choice allows one to define the perturbed tetrad for a metric perturbation in any gauge, and one can easily obtain different choices under tetrad rotations. We adopt the choice in eq. \eqref{01tetrad} because $n^{\mu(0,1)}$ then vanishes in the ORG. This choice of tetrad was also adopted in Ref. \cite{spiersmoxon}, where the perturbed Weyl scalars and spin coefficients are also expressed in terms of $h^{(0,1)}_{ab}$. We refer the readers to the Mathematica notebook accompanying Ref. \cite{spiersmoxon} for the formulas for perturbed NP quantities, as we will not include them in this paper. 

Let us make several remarks before concluding this section. The metric as expressed in \eqref{metricfromHZpot} is missing the so-called completion terms \cite{van_de_MeentShah,Merlin} describing the $l=0,1$ multipoles of the gravitational field sourced by the inspiraling particle. These terms correspond to the mass and angular momentum of the secondary body and are obviously not included in the Teukolsky equation for $\Psi_4$ or $\Psi_0$ as these Weyl scalars have spin weight $\pm2$. However, they can be easily added as mass and angular momentum perturbations to the Kerr parameters of the metric. As a result, they simply shift the equilibrium radius and rotation rate of the ring. However, such a shift of the ring does not contribute either to GW fluxes, nor to the waveform at $\mathcal{O}(\zeta \varepsilon)$, so we will ignore them.

The CCK metric reconstruction naturally produces a metric in the radiation gauge; however, as analyzed by \citet{PoundMerlin}, this gauge is not regular in the whole spacetime. It may contain singular string(s) emanating from the particle or produce discontinuities on a surface containing the particle at a given time.  However, this is not of concern to us as we reconstruct the metric outside of the source; this corresponds to the regular half of a  half-string solution  as demonstrated in Ref. \cite{PoundMerlin}.

Since we work on the Schwarzschild background, we could have in principle used the Regge-Wheeler-Zerilli formalism, which solves the linearized Einstein equations directly in the metric components \cite{Regge:1957td,Zerilli}. We chose the CCK reconstruction as it is based on the Teukolsky/NP/GHP formalism already employed in this work, while the $(0,1)$  metric is represented by a single scalar potential, which simplifies the expressions in the effective source. 

%%%%%%%%%%%%%%%%%%%%%%%%%%%%%%%%%%%%%%%%%%%%%%%%%%%%%%%%%%%%%%%%%%%%%%%%%%%%%%%%%%%%%%%%%%%%%%%%%%%%%%%%%%%%%%%

\subsection{The source on the shell }
\label{sec:sourceontheshell}
%%%%%%%%%%%%%%%%%%%%%%%%%%%%%%%%%%%%%%%%%%%%%%%%%%%%%%%%%%%%%%%%%%%%%%%%%%%%%%%%%%%%%%%%%%%%%%%%%%%%%%%%%%%%%%%

With the metric reconstruction method established, we turn to constructing the source terms that drive the environmental corrections. Specifically, we proceed with constructing the source  $\mathcal{S}^{(1,1)}_{r_{\rm r}}$ from eq. \eqref{Srssource}, which is nonzero only on the shell $r=r_{\rm r}$. The general strategy here is to use the definitions of the terms of the modified Teukolsky equation comprising $\mathcal{S}^{(1,1)}_{r_{\rm r}}$ and express them using the NP quantities, which we now know how to calculate. The next step is then to project the source parts onto spin-weighted spherical harmonics so that we can end up with a radial Teukolsky equation for $\Psi_{4(r_{\rm r})}^{(1,1)}$ (eq.  \eqref{2Teukequations}).

The source can be separated into two parts based on whether they require metric reconstruction or not
  \begin{align} 
\label{Srs2parts}
   \mathcal{S}^{(1,1)}_{r_{\rm r}}= -\left(\mathcal{O}^{(1,0)(r_{\rm r})}_\mathrm{D\mbox{-}vac}+\mathcal{O}^{(1,0)}_{\mathrm{corr}}\right)\Psi_4^{(0,1)}+\mathcal{S}^{(1,1)}_{\rm reconst}.
    \end{align}
The Weyl scalar  $\Psi_4^{(0,1)}$  is the solution to the $(0,1)$ Teukolsky equation, while the operator $\mathcal{O}^{(1,0)(r_{\rm r})}=\mathcal{O}^{(1,0)(r_{\rm r})}_\mathrm{D\mbox{-}vac}+\mathcal{O}^{(1,0)}_{\mathrm{corr}}$ is constructed from the $(1,0)$ NP quantities, which we know explicitly. The remaining terms denoted as $\mathcal{S}^{(1,1)}_{\rm reconst}$ all need metric reconstruction, which is done using the CCK procedure described in the previous section.

We shall start with the term $\mathcal{O}^{(1,0)(r_{\rm r})}\Psi_4^{(0,1)}$.  We first take the $(1,0)$ NP quantities from Section \ref{backgroundNP}  and Appendix \ref{app:background NP quantities}, use the definition of the operator $\mathcal{O}^{(1,0)}$   \eqref{O10operdef}, and apply it to $\Psi_4^{(0,1)}$, which we expand into spin-weighted spherical harmonics as in \eqref{psi4Schwepasnion}. We  would now like to achieve  the following mode decomposition 
   \begin{align} 
      \begin{split} 
&\mathcal{O}^{(1,0)(r_{\rm r})}\Psi_4^{(0,1)} =\\ &\sum_{lm\omega} \mathcal{R}\left[\mathcal{O}^{(1,0)(r_{\rm r})}\Psi_4^{(0,1)}\right]_{lm\omega}  {}_{-2}Y_{lm}(\theta)e^{\mathrm{i}(m\phi-\omega t)},
  \end{split} 
  \end{align}  
 where the symbol $\mathcal{R}$ denotes the radial part of the mode which depends only on $r$.   When analyzing the individual terms in $\mathcal{O}^{(1,0)(r_{\rm r})}\Psi_4^{(0,1)}$ we can see that while some of them are already proportional to ${}_{-2}Y_{lm}(\theta,\phi)$, other terms contain angular dependence in the form $\cos(\theta){}_{-2}Y_{lm}(\theta,\phi)$ or even derivatives of spin-weighted spherical harmonics. In order to find the radial part we can use various identities for the spherical harmonics outlined in Appendix \ref{app:SWSHs}. The most complicated terms are those containing derivatives, which we can group together to form an expression in the form  
 \begin{align} 
   \begin{split} 
 \label{deriivativesindecomposition}
&\left[\partial_\phi-\mathrm{i}\sin(\theta)\partial_\theta\right]{}_{-2}Y_{lm}(\theta,\phi)=\\&-\mathrm{i}\sin(\theta)\sqrt{2}r\ethm( {}_{-2}Y_{lm}(\theta,\phi))+2\mathrm{i}\cos(\theta){}_{-2}Y_{lm}(\theta,\phi) \,,
  \end{split} 
  \end{align}
  where we used the definition of the GHP derivative $\ethm$ \eqref{GHPangularderivatives}, which acts as a raising operator on   $ {}_{-2}Y_{lm}(\theta,\phi)$.  Now, to simplify the right-hand side of  \eqref{deriivativesindecomposition} we can use the fact that 
 \begin{align} 
\cos(\theta)\sim {}_{0}Y_{10}(\theta)\,, \enspace \sin(\theta)\sim {}_{-1}Y_{10}(\theta) \,.
  \end{align}
  This enables us to  use the decomposition \eqref{Yproduct} using coefficients $C^{slm}_{ s_1 l_1 m_1 s_2 l_2 m_2} $ to fully express the angular terms using only ${}_{-2}Y_{lm}(\theta,\phi))$. Alternatively, one could use identities for $\cos(\theta){}_{s}Y_{lm}(\theta,\phi)$ (Appendix \ref{app:new angularn expansion}) and $\sin(\theta){}_{s}Y_{lm}(\theta,\phi)$ can be found in Ref. \cite{seibert2018spin}. After applying the identities, we find that for a fixed $l$ in the expansion of $\Psi_4^{(0,1)}$ we encounter terms with ${}_{-2}Y_{l-1m}(\theta,\phi)$, ${}_{-2}Y_{lm}(\theta,\phi)$, and ${}_{-2}Y_{l+1m}(\theta,\phi)$. Consequently,    the radial part of $\mathcal{O}^{(1,0)(r_{\rm r})}\Psi_4^{(0,1)}$ can be written as  
     \begin{align} 
     \label{RadOPsi4}
        \begin{split} 
 \mathcal{R}\left[\mathcal{O}^{(1,0)(r_{\rm r})}\Psi_4^{(0,1)}\right]_{lm\omega} &=\sum_{j=l-1}^{l+1}\bigg( \mathcal{RO}^{(1,0)}_{(0)jm\omega}(r){}_{-2}R^{(0,1)}_{jm\omega}(r)\\&+\mathcal{RO}^{(1,0)}_{(1)jm\omega}(r)({}_{-2}R')^{(0,1)}_{jm\omega}(r)\bigg),
    \end{split} 
    \end{align}  
 where $\mathcal{RO}^{(1,0)}_{(A)jm\omega}(r)$ are just functions of $r$.  We can notice that the presence of the angular momentum of the surrounding matter ($a_{\rm r}$) introduces mode mixing into the picture, which we have already encountered in Section \ref{Smoothmatching} where we smoothly  matched $\Psi_{4\rm(sm)}^{(1,1)}$  at $r=r_{\rm r}$. Since $\mathcal{O}^{(1,0)(r_{\rm r})}$ is a first-order differential operator, the resulting function \eqref{RadOPsi4} contains both the radial function ${}_{-2}R^{(0,1)}(r)$ and its derivative $({}_{-2}R')^{(0,1)}(r)$. As we have already discussed, the effective source $\mathcal{S}^{(1,1)}_{r_{\rm r}}$ is nonzero only at $r=r_{\rm r}$,  which is why it is convenient to express \eqref{RadOPsi4} using the delta function and its derivative (as was sketched in eq. \eqref{Sourcedeltafunctions})
     \begin{align} 
       \begin{split} 
 \mathcal{R}\left[\mathcal{O}^{(1,0)(r_{\rm r})}\Psi_4^{(0,1)}\right]_{lm\omega} =&\mathcal{R}\left[\mathcal{O}^{(r_{\rm r})}\Psi_4\right]^{(1,1)}_{(0)lm\omega}\delta(r-r_{\rm r})\\+&\mathcal{R}\left[\mathcal{O}^{(r_{\rm r})}\Psi_4\right]^{(1,1)}_{(1)lm\omega}\delta'(r-r_{\rm r}).
   \end{split} 
    \end{align}      
    
We can now turn our attention to the other part of $\mathcal{S}^{(1,1)}_{r_{\rm r}}$ 
\begin{align}
     \label{reconstsource}
    \begin{split}
    \mathcal{S}^{(1,1)}_{\rm reconst}&= 
        \mathfrak{T}^{(0,1)ab}T^{(1,0)}_{ab}+\mathfrak{T}^{(0,0)ab} T^{(1,1)}_{ab(\rm matter)}
        \\
        & -\mathcal{K}^{(1,1)}(\Psi_3^{(0,1)},\Psi_3^{(1,0)})\,.
    \end{split}
\end{align}  
To obtain $\mathcal{S}^{(1,1)}_{\rm reconst}$ we have to first reconstruct the $(0,1)$ metric; there is, however, one term which requires also solving the $(1,1)$ equations for the matter. If we now expand the tetrad components $T^{(1,1)}_{ab(\rm matter)}$ of the stress-energy tensor we find that
\begin{align}
 \begin{split}
T^{(1,1)}_{ab(\rm matter)}&=2(e^{\mu}_{a})^{(0,1)} (e^{\nu}_{b})^{(0,0)}T_{\mu\nu}^{(1,0)\rm matter} \\&+ (e^{\mu}_{a})^{(0,0)} (e^{\nu}_{b})^{(0,0)}T_{\mu\nu}^{(1,1)\rm matter}.
 \end{split}
\end{align}  
We can now identify that it is the last term  arising from the interaction of $h_{\mu\nu}^{(0,1)}$ and $T_{\mu\nu}^{(1,0)}$ which has to be constructed from the solution of the  $(1,1)$ equations of motion. This term which we denote as 
  \begin{align} 
  \label{Tdynamical}
\mathfrak{T}^{(1,1)}_{\rm dynamical}:=\mathfrak{T}^{(0,0)ab}(e^{\mu}_{a})^{(0,0)} (e^{\nu}_{b})^{(0,0)}T_{\mu\nu}^{(1,1)\rm matter},
\end{align}  
will be discussed in detail in the following section.  The other terms can then be expressed directly using the Hertz potential. First we take all the $(0,1)$ NP quantities in \eqref{reconstsource}, which means the tetrad $(e^{\mu}_{a})^{(0,1)}$   \eqref{01tetrad} the spin coefficients $\gamma_{abc}^{(0,1)}$ and Weyl scalars $\Psi_A^{(0,1)}$ (see Ref. \cite{spiersmoxon}), and write them using the $(0,0)$  tetrad  components of the $(0,1)$ metric $h_{ab}^{(0,1)}$. Then we can use formulas \eqref{recmetriccomponents} relating these tetrad components and the Hertz potential $\Psi_{\rm Hz}^{(0,1)}$. Finally, we transform the lengthy NP expression into a more compact GHP form  and we can write the reconstructed part as
  \begin{align} 
   \label{Sreconst}
\mathcal{S}^{(1,1)}_{\rm reconst}=\Sigma^{(1,0)}\overline{\Psi_{\rm Hz}^{(0,1)}} +\mathfrak{T}^{(1,1)}_{\rm dynamical}.
  \end{align}  
Here $\Sigma^{(1,0)}$ is a fourth-order differential operator constructed from $(1,0)$ GHP scalars and also $(0,0)$ (Schwarzschild) scalars and GHP derivatives; this can be summarized as
 \begin{align} 
\Sigma^{(1,0)}=\Sigma^{(1,0)}\left(T^{(1,0)}_{ab},\Psi^{(1,0)}_{3},\nu^{(1,0)},\gamma_{abc},\Psi_2,\text{\textthorn},\text{\textthorn}',\ethm,\ethm'\right),
  \end{align}  
where we have omitted the $(0,0)$ index for Schwarzschild scalars/derivatives for brevity.  It is interesting to note that there is no term with an operator acting on $\Psi_{\rm Hz}^{(0,1)}$  in \eqref{Sreconst} but we should realize that $\mathcal{S}^{(1,1)}_{\rm reconst}$ is not a real quantity, so we cannot expect a symmetric term $\overline{\Sigma^{(1,0)}}\Psi_{\rm Hz}^{(0,1)}$. However, we could still have some terms with $\Psi_{\rm Hz}^{(0,1)}$. This is indeed the case in the term $\mathcal{O}^{(0,1)}\Psi_4^{(1,0)}$, which is zero in our case and thus does not contribute to $\mathcal{S}^{(1,1)}_{\rm reconst}$.

To simplify $\Sigma^{(1,0)}$, we use the GHP commutation relations  \eqref{app:GHPcommutation}
to put all derivatives in the ordered form $\text{\textthorn}\text{\textthorn}'\ethm\ethm'$. This will put together the terms with the same combination of GHP derivatives; additionally, we also use the Bianchi and Ricci identities in vacuum type D spacetimes to eliminate some derivatives acting on Schwarzschild Weyl scalars/spin coefficients.

Let us remark that throughout the entire calculation of the source term $\Sigma^{(1,0)}\Psi_{\rm Hz}^{(0,1)}$ we keep track of its GHP weight  which is   $\lbrace-4,0\rbrace$ (the same as $\Psi_4$) since this is a useful consistency check. At this point, we have just an abstract GHP form of $\Sigma^{(1,0)}\Psi_{\rm Hz}^{(0,1)}$, before substituting the coordinate forms of the expression, we expand all quantities in spin-weighted spherical harmonics. For a scalar $\mathcal{A}$ with a spin weight $s$ and a boost weight $b$ we have
 \begin{align}   
 \label{generalswshexpansion}
\mathcal{A}=\displaystyle\sum_{lm}{}_{b}\mathcal{A}_{lm}(t,r) {}_{s}Y_{lm}(\theta,\phi),
    \end{align} 
where we assign the spin weight $s=0$ to the radial part ${}_{b}\mathcal{A}_{lm}$ and $b=0$ to the angular part ${}_{s}Y_{lm}$ meaning we can also assign to them a GHP weight. The radial part is obviously time independent for $(1,0)$ scalars, while the Hertz potential can be expanded into frequencies.  Since we have commuted the angular GHP derivative $\ethm$ and $\ethm'$ to the right, we can let them act on ${}_{s}Y_{lm}$ as raising and lowering operators \eqref{GHPraisinglowering}, thus eliminating these derivatives from the picture. We can also use the formula \eqref{Yconjugate} to remove complex conjugate spherical harmonics ${}_{s}\bar{Y}_{lm}$.   

By applying the relation \eqref{Yproduct}, we then express products of two spin-weighted spherical harmonics ${}_{s_1}Y_{l_1m_1}{}_{s_2}Y_{l_2 m_2}$ using a single function ${}_{-2}Y_{lm}$. Since the surrounding matter is approximated only by the monopole and dipole term, the index $l$ in the expansion of $(0,1)$ quantities takes only values $l=0,1$. This means that similarly to \eqref{RadOPsi4}, the radial part of $\Sigma^{(1,0)}\Psi_{\rm Hz}^{(0,1)}$ will include $R^{\mathrm{Hz}(0,1)}_{l-1\omega}$, $R^{\mathrm{Hz}(0,1)}_{l\omega}$, and $R^{\mathrm{Hz}(0,1)}_{l+1\omega}$.

 Subsequently, we substitute in the coordinate form of the NP scalars and expand the Hertz potential in frequencies. The derivatives $\text{\textthorn}$ and $\text{\textthorn}'$ then give rise to derivatives with respect to $t$ (which are trivial) and $r$. As we mentioned,  $\Sigma^{(1,0)}$ is a fourth-order operator, but the higher derivatives with respect to $r$ can be eliminated as we can use the Teukolsky radial equation ${}_{2}\mathfrak{D}_{lm\omega}{}_{2}R^{\mathrm{Hz}(0,1)}_{lm\omega}=0$  or rather its complex conjugate  form. The complete result for the radial part of $\Sigma^{(1,0)}\overline{\Psi_{\rm Hz}^{(0,1)}}$ can then be written in full analogy with \eqref{RadOPsi4} as
  \begin{align} 
     \label{Sreconstradial}
       \begin{split}
 \mathcal{R}\left[\Sigma^{(1,0)}\overline{\Psi_{\rm Hz}^{(0,1)}}\right]_{lm\omega} &=\sum_{j=l-1}^{l+1}\bigg( \mathcal{R}\Sigma^{(1,0)}_{(0)jm\omega}(r)\overline{{}_{2}R^{\mathrm{Hz}(0,1)}_{j-m\omega}}(r) \\&+    \mathcal{R}\Sigma^{(1,0)}_{(1)jm\omega}(r)\overline{({}_{2}R')^{\mathrm{Hz}(0,1)}_{j-m\omega}}(r) \bigg),
       \end{split} 
    \end{align}  
where we shifted the index $m\rightarrow-m$ due to the presence of  the conjugate functions ${}_{s}\bar{Y}_{lm}$ in the expansion of  $\overline{\Psi_{\rm Hz}^{(0,1)}}$. The convenient form of the source is then written using the delta function and its first two derivatives
  \begin{align} 
    \begin{split}
 &\mathcal{R}\left[\Sigma^{(1,0)}\overline{\Psi_{\rm Hz}^{(0,1)}}\right]_{lm\omega} = \\&\displaystyle\sum_{i=0}^{2} \mathcal{R}\left[\Sigma^{(1,0)}\overline{\Psi_{\rm Hz}^{(0,1)}}\right]^{(1,1)}_{(i)lm\omega}\delta^{(i)}(r-r_{\rm r}).
    \end{split} 
    \end{align}  
The entire procedure for constructing the source $\Sigma^{(1,0)}\overline{\Psi_{\rm Hz}^{(0,1)}}$ and decomposing it into ${}_{-2}Y_{l_2 m_2}$ is done primarily in the notebook \texttt{SourceinNPGHPformalism.nb} while the final (coordinate) form is derived in another notebook \texttt{Sourceincoordinates.nb}  (supplemental material \cite{MmaNB}).

% %%%%%%%%%%%%%%%%%%%%%%%%%%%%%%%%%%%%%%%%%%%%%%%%%%%%%%%%%%%%%%%%%%%%%%%%%%%%%%%%%%%%%%%%%%%%%%%%%%%%

 \subsection{Solving the \texorpdfstring{$(1,1)$}{(1,1)} equations for matter }
 \label{sec:equationsformatter}
 %%%%%%%%%%%%%%%%%%%%%%%%%%%%%%%%%%%%%%%%%%%%%%%%%%%%%%%%%%%%%%%%%%%%%%%%%%%%%%%%%%%%%%%%%%%%%%%%%%%%%%%%%%%%%%%
 
The last but also the most complicated part of the source $\mathcal{S}^{(1,1)}_{r_{\rm r}}$ is  $\mathfrak{T}^{(1,1)}_{\rm dynamical}$. This term is constructed from $T_{\mu\nu}^{(1,1)\rm matter}$ as per eq. \eqref{Tdynamical}. 

Let us first examine how to obtain the radial modes of $\mathfrak{T}^{(1,1)}_{\rm dynamical}$ from eq. \eqref{Tdynamical}. The procedure is very similar to the way this was done for $\mathcal{R}\left[\Sigma^{(1,0)}\overline{\Psi_{\rm Hz}^{(0,1)}}\right]_{lm\omega}$ in the previous section. We start by defining the components of the $(1,1)$ stress-energy tensor with respect to the $(0,0)$ tetrad
 \begin{align} 
\mathsf{T}_{ab}^{(1,1)\rm matter}:=(e^{\mu}_{a})^{(0,0)}(e^{\nu}_{b})^{(0,0)}T_{\mu\nu}^{(1,1)\rm matter}.
\end{align} 
We can now act with the operator $\mathfrak{T}^{(0,0)ab}$ on the components $\mathsf{T}_{ab}^{(1,1)\rm matter}$ and then expand all the quantities into spin-weighted spherical harmonics according to their spin weight (eq. \eqref{generalswshexpansion}). Next, we apply the angular GHP derivatives  $\ethm$ and  $\ethm'$ on  $ {}_{s}Y_{lm} $. This directly yields the radial part, as the second-order differential operator $\mathfrak{T}^{(0,0)ab}$ maintains spherical symmetry. We then express the result in terms of functions $\mathcal{R}\mathfrak{T}^{(0,0)ab}_{(i)lm\omega}(r)$
 \begin{align} 
 \label{T11radial1}
 \mathcal{R}\left[\mathfrak{T}^{(1,1)}_{\rm dynamical}\right]_{lm\omega}=\displaystyle\sum_{i=0}^{2}  \mathcal{R}\mathfrak{T}^{(0,0)ab}_{(i)lm\omega}(r) \frac{\mathrm{d}^i}{\mathrm{d}r^i}\mathcal{R}\mathsf{T}_{ab\enspace lm\omega}^{(1,1)\rm matter }(r) .
\end{align} 
This result is valid for any $(1,1)$ stress-energy tensor. Now, the only remaining task is to determine the $(1,1)$ stress-energy tensor in our specific model.  

As already discussed, the interaction between $T_{\mu\nu}^{(1,0)\rm matter}$ and $h^{(1,0)}_{\mu\nu}$ gives rise to the stress-energy tensor $T_{\mu\nu}^{(1,1)\rm matter}$ whose 10 independent components can only be found if we know the degrees of freedom of the matter. In other words, the four equations $\nabla_\mu T^{\mu\nu}=0$ must be supplemented by some set of constitutive relations. Here, we resolve this issue by dropping the pole-dipole approximation given by $T_{\mu\nu}^{(1,0)\rm shell}$ as derived from the Israel junction conditions and instead using the full stress-energy tensor of the ring  
 \begin{align}
 \label{ring stress energy}
 T^{\mu\nu}_{\rm ring}=\rho_{\rm ring} u^{\mu} u^{\nu}.
 \end{align}
Assuming stress-energy conservation leads to the usual equations of motion for dust %
 \begin{align}
 \label{ring EOM}
a^{\nu}=u^{\mu}\nabla_\mu u^{\nu}=0, \enspace \nabla_\mu\left( \rho_{\rm ring} u^{\mu}\right)=0\,.
 \end{align} 
 Here, the first equation is the geodesic equation, and the second is the continuity equation. At the $(1,0)$ order, these equations are satisfied by
\begin{align}
 \label{densityvelocity}
& \rho_\mathrm{ring}^{(1,0)}=\frac{m_{\rm r}}{2\uppi r_{\rm r}^2}\frac{r_{\rm r}-3M}{r_{\rm r}-2M}\delta\left(r-r_{\rm r}\right)\delta\left(\theta-\frac{\uppi}{2}\right) \,,\\
& u^{(0,0)\mu}=\left(\sqrt{\frac{r_{\rm r}}{r_{\rm r}-3M}},0,0,\frac{1}{r_{\rm r}}\sqrt{\frac{M}{r_{\rm r}-3M}}\right).
 \end{align}
where $u^{(0,0)\mu}$ is the four-velocity of the Schwarzschild equatorial circular orbit located at the radius $r=r_{\rm r}$.  We can now continue with the $(1,1)$ order, starting with the geodesic equation.  Collecting the $(1,1)$ terms in $\nabla_\mu T^{\mu\nu}_{\rm ring}=0$, we find that the term with the geodesic equation has the form  $\rho_\mathrm{ring}^{(1,0)}a^{\nu(0,1)}$. This means that the four-acceleration $a^{\nu(0,1)}$ is evaluated at $r=r_{\rm r}$ and $\theta=\frac{\uppi}{2}$ due to the fact that $\rho_\mathrm{ring}^{(1,0)}$ is given by delta functions. The perturbed geodesic equation then reads
\begin{align}
 \label{(0,1)geod}
 \begin{split}
a^{\nu(0,1)}=&u^{\mu(0,0)}\nabla^{(0,0)}_\mu u^{\nu(0,1)}+u^{\mu(0,1)}\nabla^{(0,0)}_\mu u^{\nu(0,0)}\\&+\Gamma^{\nu(0,1)}_{\mu\lambda}u^{\mu(0,0)}u^{\lambda(0,0)}=0.
 \end{split}
 \end{align}
The unknown variable is $u^{\nu(0,1)}$, which is better reflected if we rewrite the equation as
\begin{align}
 \label{(0,1)geodrewrittern}
  \begin{split}
&u^{t(0,0)}\partial_t u^{\nu(0,1)}+u^{\phi(0,0)}\partial_\phi u^{\nu(0,1)}\\&+\tensor{\mathcal{U}}{^{\nu(0,0)}_\mu} u^{\mu(0,1)}=\mathcal{F}^{\nu(0,1)}.
  \end{split}
 \end{align}
 This is a linear differential equation with a source   $\mathcal{F}^{\nu(0,1)}=-\Gamma^{\nu(0,1)}_{\mu\lambda}u^{\mu(0,0)}u^{\lambda(0,0)}$ constructed from the metric $h_{\mu\nu}^{(0,1)}$. In accordance with the rest of this paper, we express equation \eqref{(0,1)geodrewrittern} in the $(0,0)$ NP tetrad components. We then define the perturbed four velocity projection as%
 \footnote{This should not be confused with  $u^{a(0,1)}=(e^{a}_{\mu})^{(0,0)}u^{(0,1)\mu}+(e^{a}_{\mu})^{(0,1)}u^{(0,0)\mu} $.}
 $\mathsf{u}^{a(0,1)}:=(e^{a}_{\mu})^{(0,0)}u^{(0,1)\mu} $. So far, we did not have to distinguish between upper and lower tetrad indices. To avoid confusion, we will label the tetrad components with the respective $(0,0)$ tetrad vectors, regardless of the index position. For example, $\mathsf{u}^{n(0,1)}:=n_{\mu}^{(0,0)}u^{\mu(0,1)}$ is equivalent to $\mathsf{u}^{(0,1)}_n:=n_{\mu}^{(0,0)}u^{\mu(0,1)}$.  Although $\mathsf{u}^{\bar{m}(0,1)}$ is the complex conjugate of $\mathsf{u}^{m(0,1)}$, we will treat these components as independent. 
 
The left-hand side of  equation  \eqref{(0,1)geodrewrittern} contains only derivatives with respect to the Killing coordinates $(t,\phi)$. It is then natural to expand the four-velocity components as
\begin{align}
 \label{fourvelocity   expansion}
 \mathsf{u}^{a(0,1)}(t,\phi)=\displaystyle\sum_{m\omega} \mathsf{u}^{a(0,1)}_{m\omega} e^{\mathrm{i}(m\phi-\omega t)}\,.
 \end{align}
The forcing term  $\mathcal{F}^{\nu(0,1)}$ is processed using the familiar method from the previous sections. We express it using the GHP formalism and expand the Hertz potential as well as other quantities into  $ {}_{s}Y_{lm}$. Next, we apply the derivatives $\ethm$ and $\ethm'$ to these angular functions. What is different this time is that we do not decompose the source $\mathcal{F}^{\nu(0,1)}$ into spin-weighted spherical harmonics. Instead, we evaluate the source at $r=r_{\rm r}$ and $\theta=\frac{\uppi}{2}$. This approach leaves us with the dependence on $t$  and $\phi$ only
\begin{align}
 \label{Fexpansion}
 \begin{split}
\mathcal{F}^{a(0,1)}(t,\phi)&=\displaystyle\sum_{lm\omega} \mathcal{F}^{(+)a(0,1)}_{lm\omega}e^{\mathrm{i}(m\phi-\omega t)}\\&+\displaystyle\sum_{lm\omega}\mathcal{F}^{(-)a(0,1)}_{lm\omega}e^{-\mathrm{i}(m\phi-\omega t)}.
 \end{split}
 \end{align}
 The sources $\mathcal{F}^{a(0,1)}$ are real only for the $a=l^\mu,n^\mu$ components and generally complex for $a=m^\mu,\bar{m}^\mu$ NP tetrad components.
 The functions $\mathcal{F}^{(\pm)a(0,1)}_{lm\omega}$ are given in terms of the $lm\omega$ mode decomposition of the Hertz potential 
 \begin{subequations}
\begin{align}
 \label{Fmodes}
  \begin{split}
\mathcal{F}^{(+)a(0,1)}_{lm\omega}&=f^{(+)a(0,0)}_{(0)lm\omega}{}_{2}R^{\mathrm{Hz}(0,1)}_{lm\omega}(r_{\rm r})\\&+f^{(+)a(0,0)}_{(1)lm\omega}({}_{2}R')^{\mathrm{Hz}(0,1)}_{lm\omega}(r_{\rm r})\,,
 \end{split}
\\
 \begin{split}
\mathcal{F}^{(-)a(0,1)}_{lm\omega}&=f^{(-)a(0,0)}_{(0)lm\omega}\overline{{}_{2}R^{\mathrm{Hz}(0,1)}_{lm\omega}}(r_{\rm r})\\&+f^{(-)a(0,0)}_{(1)lm\omega}\overline{({}_{2}R')^{\mathrm{Hz}(0,1)}_{lm\omega}}(r_{\rm r}).
 \end{split}
 \end{align}
 \end{subequations}
 As in the previous instances, we used the radial Teukolsky equation ${}_{2}\mathfrak{D}{}_{2}R^{\mathrm{Hz}(0,1)}=0$ to eliminate the second and higher derivatives of the Hertz potential, which is evaluated at $r=r_{\rm r}$. It is interesting to note that this source depends both on the Hertz potential and its complex conjugate. This is different from the term $\Sigma^{(1,0)}\overline{\Psi_{\rm Hz}^{(0,1)}}$ discussed in the previous section.  
 If we are to expand both sides of equation \eqref{(0,1)geodrewrittern} into $ e^{\mathrm{i}(m\phi-\omega t)}$, we define
\begin{align}
 \label{Fmodesres}
\mathcal{F}^{a(0,1)}_{m\omega}:=\displaystyle\sum_{l}\left(\mathcal{F}^{(+)a(0,1)}_{lm\omega}+\mathcal{F}^{(-)a(0,1)}_{l-m-\omega}\right) ,
 \end{align}
which is the desired counterpart to $ \mathsf{u}^{a(0,1)}_{m\omega}$. The original differential equations  \eqref{(0,1)geodrewrittern} are then reduced to a  system of linear algebraic equations
\begin{align}
 \label{(0,1)geodmodes}
\left(\tensor{\mathcal{L}}{^{a(0,0)}_{b}}\right)_{m\omega}\mathsf{u}^{b(0,1)}_{m\omega}=\mathcal{F}^{a(0,1)}_{m\omega}.
 \end{align}
These equations can be solved just by inverting the matrix $\tensor{\mathcal{L}}{^{a(0,0)}_{b}}$, giving us the four-velocity modes
\begin{align}
 \label{(0,1)geodmodessolved}
\mathsf{u}^{a(0,1)}_{m\omega}=\left(\tensor{\mathcal{L}}{^{a(0,0)}_{b}}\right)^{-1}_{m\omega}\mathcal{F}^{b(0,1)}_{m\omega}.
 \end{align}
The path to solving the perturbed geodesic equation is then straightforward.  We first evaluate the radial part of the Hertz potential (and its derivative) at $r=r_{\rm r}$, then sum over the index $l$ to obtain the source \eqref{Fmodesres}. The modes of the four-velocity are then explicitly given by \eqref{(0,1)geodmodessolved}.  It should be kept in mind that all the functions are evaluated at the  location of the (unperturbed) ring meaning that $f^{(\pm)a(0,0)}_{(A)lm\omega}=f^{(\pm)a(0,0)}_{(A)lm\omega}(r=r_{\rm r},\theta=\frac{\uppi}{2})$ and $\tensor{\mathcal{L}}{^{a(0,0)}_{b}}=\tensor{\mathcal{L}}{^{a(0,0)}_{b}}(r=r_{\rm r},\theta=\frac{\uppi}{2})$. This is also true for the perturbed four-velocity $u^{\mu(0,1)}$ which, unlike $u^{\mu(0,0)}$, generally has all its components non-zero.

A useful consistency check for the result \eqref{(0,1)geodmodessolved} is the normalization of the four-velocity  $u^{\mu}$. Since at the $(0,0)$ order we have $g^{(0,0)}_{\mu\nu}u^{\mu(0,0)}u^{\nu(0,0)}=-1$, we expect at the $(0,1)$ to have
\begin{align}
 \label{fourvelocitynormalisation}
2g^{(0,0)}_{\mu\nu}u^{\mu(0,1)}u^{\nu(0,0)} +h^{(0,1)}_{\mu\nu}u^{\mu(0,0)}u^{\nu(0,0)}=0.
 \end{align}
If we insert the perturbed four-velocity  $u^{\nu(0,1)}$ and the reconstructed metric $h^{(0,1)}_{\mu\nu}$ into this equation, which we rewrite in a manner similar to eq. \eqref{(0,1)geodmodes}, we find that the normalization condition is satisfied.

With the geodesic equation solved, we can now proceed to the $(1,1)$ continuity equation   
\begin{align}
 \label{(1,1)continuityeq}
 \begin{split}
&u^{\mu(0,0)}\partial_{\mu}\rho_{\rm ring}^{(1,1)}+\rho_{\rm ring}^{(1,1)}\nabla^{(0,0)}_\mu u^{\mu(0,0)}+u^{\mu(0,1)}\partial_{\mu}\rho_{\rm ring}^{(1,0)}\\
&+\rho_{\rm ring}^{(1,0)}\nabla^{(0,0)}_\mu u^{\mu(0,1)}+\rho_{\rm ring}^{(1,0)}\Gamma^{\mu(0,1)}_{\mu\lambda}u^{\lambda(0,0)}=0.
 \end{split}
 \end{align} 
Similarly to the geodesic equation, we can recast the continuity equation as 
\begin{align}
 \label{(1,1)continuityeqreqritten}
 u^{\mu(0,0)}\partial_{\mu}\rho_{\rm ring}^{(1,1)}+\rho_{\rm ring}^{(1,1)}\nabla^{(0,0)}_\mu u^{\mu(0,0)}=\mathcal{F}^{(1,1)},
 \end{align}  
where  the sourcing term is given by $\mathcal{F}^{(1,1)}=-u^{\mu(0,1)}\partial_{\mu}\rho_{\rm ring}^{(1,0)}-\rho_{\rm ring}^{(1,0)}\nabla^{(0,0)}_\mu u^{\mu(0,1)}$.

When formulating the equation in the NP formalism, we find that $\Gamma^{\mu(0,1)}_{\mu\lambda}u^{\lambda(0,0)}$ is given by $h^{(0,1)}_{ln}$, $h^{(0,1)}_{m\bar{m}}$ and their derivatives. However, in the radiation gauge, these metric components are zero. This is why the term with the Christoffel symbols vanishes and is not included in the source $\mathcal{F}^{(1,1)}$ .

Now, if we examine the first term in the source $\mathcal{F}^{(1,1)}$,  $u^{\mu(0,1)}\partial_{\mu}\rho_{\rm ring}^{(1,0)}$, we find that it generates derivatives of delta functions with respect to $r$ and $\theta$. In contrast, the other term retains the delta functions in their original form in $\rho_{\rm ring}^{(1,0)}$. Therefore, we adopt an ansatz for $ \rho_{ \rm ring}^{(1,1)}$ that aligns with the structure of the source $\mathcal{F}^{(1,1)}$ 
\begin{align}
 \label{densitydeltafunctions}
  \begin{split}
  \rho_{ \rm ring}^{(1,1)}&=\delta\rho_{\rm r}^{(1,1)}\delta\left(r-r_{\rm r}\right)\delta\left(\theta-\frac{\uppi}{2}\right)\\
  &-\rho_{\rm r}^{(1,0)}\delta'\left(r-r_{\rm r}\right)\delta\left(\theta-\frac{\uppi}{2}\right)\delta r^{(0,1)}\\
   &-\rho_{\rm r}^{(1,0)}\delta\left(r-r_{\rm r}\right)\delta'\left(\theta-\frac{\uppi}{2}\right)\delta \theta^{(0,1)}.
   \end{split}
  \end{align} 
The three unknown functions $\delta\rho_{\rm r}^{(1,1)}$ , $\delta r^{(0,1)}$, and $\delta \theta^{(0,1)}$ can be better understood by expressing the total density along with its perturbative expansion as follows:
  \begin{align}
 \label{densitycompact}
  \begin{split}
  &\rho_{ \rm ring}=\rho_{\rm r}\delta\left(r-r_{\rm ring}\right)\delta\left(\theta-\theta_{\rm ring}\right) ,\\
  & \rho_{ \rm ring}=\zeta \rho_{\rm ring}^{(1,0)}+\zeta\varepsilon\rho_{\rm ring}^{(1,1)}+... \ .
   \end{split}
  \end{align}
 By comparing this with  \eqref{densitydeltafunctions} we can see that the   density prefactor and the position of the ring    at the leading orders are given by
   \begin{align}
 \label{densityperturbations}
  \begin{split}
 & \rho_{\rm r}\approx\zeta\left(\rho_{\rm r}^{(1,0)}+\varepsilon\delta\rho_{\rm r}^{(1,1)}(t,\phi)\right),
 \\
 & r_{\rm ring}\approx r_{\rm r}+\varepsilon\delta r^{(0,1)}(t,\phi),
 \\
 & \theta_{\rm ring} \approx\frac{\uppi}{2}+\varepsilon\delta \theta^{(0,1)}(t,\phi).
   \end{split}
  \end{align} 
   The ring is then displaced from its original coordinate position and deformed by the GWs described by the metric $h_{\mu\nu}^{(0,1)}$.

    \begin{figure}
    \begin{center}
        \includegraphics[width=0.48\textwidth]{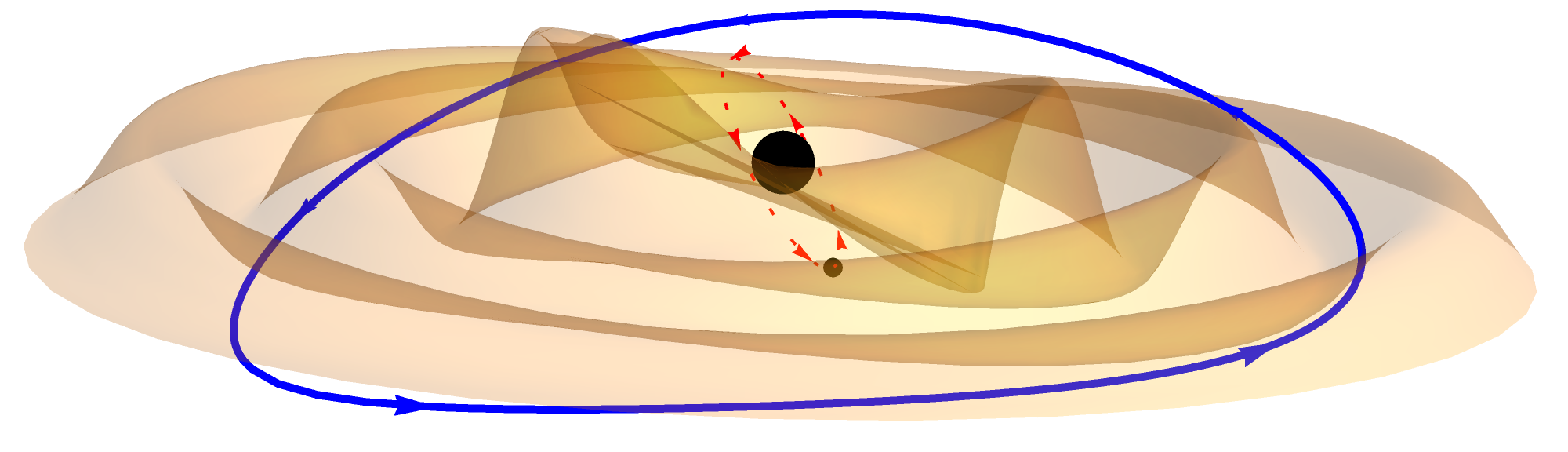}
    \caption{Deformation of the ring due to GWs $h^{(0,1)}_{\mu\nu}$ sourced by a particle on a circular inclined orbit. The waves in the figure depict the polarization  $h^{(0,1)}_{+}=\Re(h^{(0,1)}_{mm})$ as evaluated on the equatorial plane. The ring oscillates both in the radial and polar directions, and this leads to the $\mathfrak{T}^{(1,1)}_{\rm dynamical}$ source term in the modified Teukolsky equations.}     
    \label{fig:deformedring}
    \end{center}
\end{figure}

\begin{figure*}[ht]
\begin{center}
 {\subfloat{\includegraphics[width=0.48\textwidth]{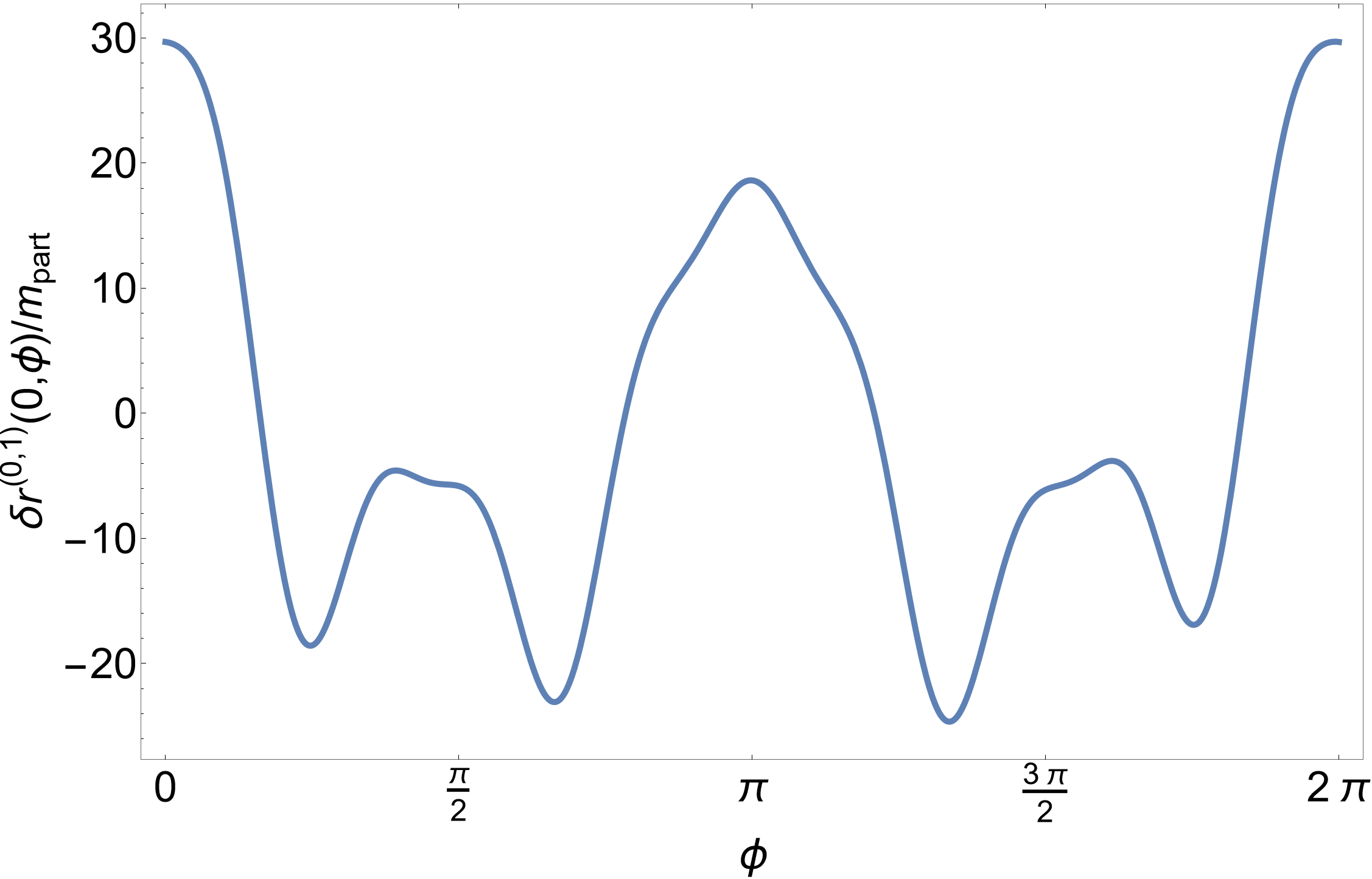}}
{  \subfloat{\includegraphics[width=0.48\textwidth]{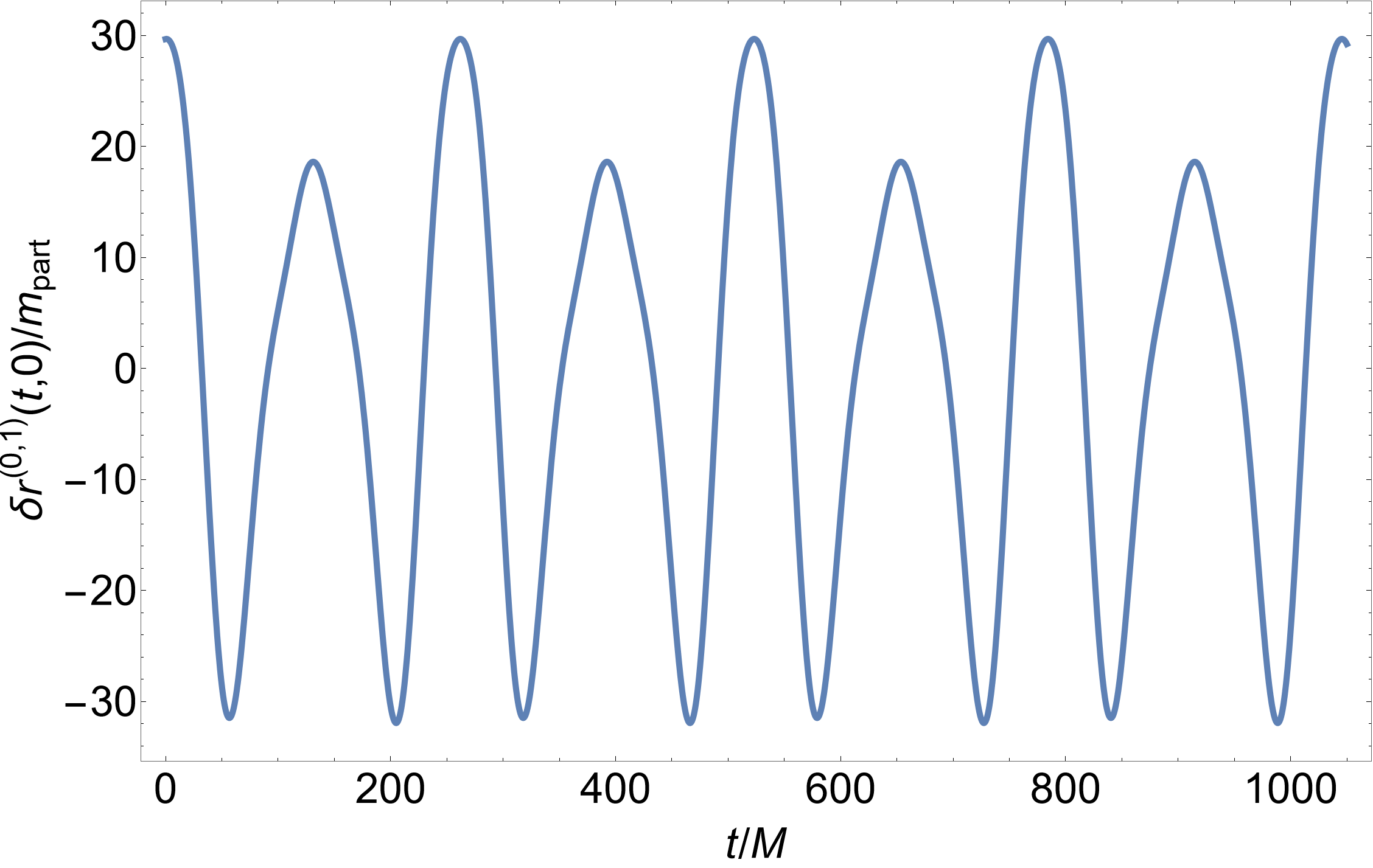}}}
}   
\end{center}
   \caption{The radial displacement $\delta r^{(0,1)}$ of the matter ring caused by the gravity of an inspiraling object of mass $m_{\rm part}$. The inspiraling secondary is on an inclined circular orbit of radius $r_{\rm p}=12M$  and an inclination of $60^\circ$ with respect to the plane of the ring; the unperturbed ring radius is $r_{\rm r}=20 M$. (The location of the ring is placed relatively close to the orbit to obtain more pronounced effects.)  The radial displacement at $t=0$  is depicted on the left while the oscillation of  element $\phi=0 $ is shown in the right figure. The oscillations are anharmonic, and the structural deformation has a non-trivial structure.}
        \label{fig:ringoscilations}
\end{figure*}

   To solve the continuity equation, we expand the three functions into $ e^{\mathrm{i}(m\phi-\omega t)}$ as we did with the four-velocity $u^{\mu(0,1)}$, which we found from the geodesic equation. 
\begin{align}
 \label{densityexpansions}
    \begin{split}
 \delta\rho_{\rm r}^{(1,1)}&\rightarrow (\delta\rho_{\rm r}^{(1,1)})_{m\omega},
 \\\enspace \delta r^{(0,1)}&\rightarrow \delta r^{(0,1)}_{m\omega},\hspace{10pt}  \delta \theta^{(0,1)}\rightarrow \delta \theta^{(0,1)}_{m\omega}.
    \end{split}
  \end{align}  
 As stated above, the source $\mathcal{F}^{(1,1)}$ does not depend explicitly on $h_{\mu\nu}^{(0,1)}$. Instead, it is determined by the perturbed four-velocity. By eliminating the derivatives, we can express the solution in terms of linear relationships between the mode-decomposed functions $\lbrace \delta\rho_{\rm r}^{(1,1)},\delta r^{(0,1)},\delta \theta^{(0,1)} \rbrace$ and $\mathsf{u}^{\mu(0,1)}$
 \begin{align}
 \label{densitysolutions}
     \begin{split}
(\delta\rho_{\rm r}^{(1,1)})_{m\omega}&=\left(\tensor{\mathcal{C}}{^{(1,0)(\rho)}_{a}}\right)_{m\omega}\mathsf{u}^{a(0,1)},\\
\delta r^{(0,1)}_{m\omega}&=\left(\tensor{\mathcal{C}}{^{(0,0)(r)}_{a}}\right)_{m\omega}\mathsf{u}^{a(0,1)},\\
\delta \theta^{(0,1)}_{m\omega}&=\left(\tensor{\mathcal{C}}{^{(0,0)(\theta)}_{a}}\right)_{m\omega}\mathsf{u}^{a(0,1)}.
    \end{split}
  \end{align} 
In Figures \ref{fig:deformedring} and \ref{fig:ringoscilations}, we present examples of the deformations of the ring obtained from a numerical implementation of the procedure described above. To compute the amplitudes of $\Psi_4^{(0,1)}$, we used the Black Hole Perturbation Toolkit \cite{BHPToolkit}.
  
Finally, having solved both the geodesic and continuity equations, we can construct $\mathsf{T}_{ab}^{(1,1)\rm matter}$ from $\mathsf{u}^{a(0,1)}$ and $\rho_{ \rm ring}^{(1,1)}$. The projection of $\mathsf{T}_{ab}^{(1,1)\rm matter}$ onto spin-weighted spherical harmonics is straightforward. This is because the $\theta$-dependence is only found in $\delta\left(\theta-\frac{\uppi}{2}\right)$ and $\delta'\left(\theta-\frac{\uppi}{2}\right)$. This leads to the radial functions  $\mathcal{R}\mathsf{T}_{ab\enspace lm\omega}^{(1,1)\rm matter }$ shown in eq. \eqref{T11radial1}, which can now be written as
 \begin{align} 
 \mathcal{R}\left[\mathfrak{T}^{(1,1)}_{\rm dynamical}\right]_{lm\omega}= \left(\mathcal{R}\mathfrak{t}^{(1,0)}_{a}(r) \right)_{lm\omega} \mathsf{u}^{a(0,1)}_{m\omega}\,,
\end{align} 
where we used the fact that the density perturbation can be expressed using   $\mathsf{u}^{a(0,1)}_{m\omega}$. As in the previous section, we can expand  it using the delta function and its derivatives to obtain the following result
\begin{align} 
 \mathcal{R}\left[\mathfrak{T}^{(1,1)}_{\rm dynamical}\right]_{lm\omega}= \displaystyle\sum_{i=0}^{3} \mathcal{R}\left[\mathfrak{T}^{(1,1)}_{\rm dynamical}\right]_{(i)lm\omega}\delta^{(i)}(r-r_{\rm r}),
\end{align}  
which contains the third derivative of the delta function $\delta(r-r_{\rm r})$. This is evident from the fact that already in   $\mathsf{T}_{ab}^{(1,1)\rm matter}$  we have a term with $\delta'\left(r-r_{\rm r}\right)$ (eq.  \eqref{densitydeltafunctions}) on which we act with a second-order differential operator (Eq. \eqref{T11radial1}).

Having completed the calculation of  $\mathfrak{T}^{(1,1)}_{\rm dynamical}$ (see supplementary notebook \texttt{Sourceincoordinates.nb} \cite{MmaNB}), we can now interpret this term as resulting from the ring absorbing part of the radiation $h_{\mu\nu}^{(0,1)}$, which causes it to oscillate (see Fig. \ref{fig:ringoscilations}). The oscillating ring is then a source of GWs, which manifests itself in the $(1,1)$ Teukolsky equation as the term $\mathfrak{T}^{(1,1)}_{\rm dynamical}$.

Are there other modeling assumptions we could have taken to obtain $\mathfrak{T}^{(1,1)}_{\rm dynamical}$? As already mentioned, the fundamental issue is that $\nabla_\mu T^{\mu\nu}=0$ constrains at most 4 matter degrees of freedom. Here, we chose those degrees of freedom to be $\rho, u^\mu$ (only 3 components of $u^\mu$ are independent due to $u^\mu_\mu = -1$). The only way to self-consistently obtain $\mathfrak{T}^{(1,1)}_{\rm dynamical}$ was then through the ring dynamics discussed above. Interestingly, this means that the dynamics had to be modeled without any multipole truncation. This was necessary because the source of the radiation was assumed to be relativistic and thus with a generic multipole structure. When deciding on this approach, we had also attempted to formulate consistent matter dynamics for the ``pole-dipole shell." However, isolating four degrees of freedom of the shell in a natural way is challenging. Accomplishing this would require extensive parameterizations of the dynamics without a clear physical interpretation. In conclusion, the approach taken here is the most frugal and physically clear approach we could conceive.

%%%%%%%%%%%%%%%%%%%%%%%%%%%%%%%%%%%%%%%%%%%%%%%%%%%%%%%%%%%%%%%%%%%%%%%%%%%%%%%%%%%%%%%%%%%%%%%%%%%%%%%%%%%%%%%
  \subsection{The complete solution} \label{subsec:finalsolution}
%%%%%%%%%%%%%%%%%%%%%%%%%%%%%%%%%%%%%%%%%%%%%%%%%%%%%%%%%%%%%%%%%%%%%%%%%%%%%%%%%%%%%%%%%%%%%%%%%%%%%%%%%%%%%%%

Now that we have assembled all the components of the effective source at $r_{\rm r}$, we can finally formulate the equation for $\Psi_{4(r_{\rm r})}^{(1,1)}$. The effective source consists of a sum of distributional sources
      \begin{align}
\mathcal{S}^{(1,1)}_{r_{\rm r}}=\displaystyle\sum_{i=0}^{3}\mathcal{S}^{(1,1)}_{r_{\rm r}(i)} \delta^{(i)}(r-r_{\rm r}).
      \end{align}   
Since we have decomposed the source term into spin-weighted spherical harmonics and frequencies, we can represent it using the radial functions
     \begin{align}
\mathcal{R}\left[\mathcal{S}^{(1,1)}_{r_{\rm r}}\right]_{lm\omega}(r)=\displaystyle\sum_{i=0}^{3}\mathcal{R}\left[\mathcal{S}^{(1,1)}_{r_{\rm r}(i)} \right]_{lm\omega}(r)\delta^{(i)}(r-r_{\rm r}).
      \end{align}  
The $r$-dependent functions that appear in front of the delta function and its derivatives can be described more explicitly as
\begin{widetext}
 \begin{align}
  \begin{split}
\mathcal{R}\left[\mathcal{S}^{(1,1)}_{r_{\rm r}(i)} \right]_{lm\omega}(r)&=\sum_{j=l-1}^{l+1}\bigg( \mathcal{RO}^{(1,0)}_{(i)(0)jm\omega}(r){}_{-2}R^{(0,1)}_{jm\omega}(r)+\mathcal{RO}^{(1,0)}_{(i)(1)jm\omega}(r)({}_{-2}R')^{(0,1)}_{jm\omega}(r)\bigg)\\
&+\sum_{j=l-1}^{l+1}\bigg( \mathcal{R}\Sigma^{(1,0)}_{(i)(0)jm\omega}(r)\overline{{}_{2}R^{\mathrm{Hz}(0,1)}_{j-m\omega}}(r)+  \mathcal{R}\Sigma^{(1,0)}_{(i)(1)jm\omega}(r)\overline{({}_{2}R')^{\mathrm{Hz}(0,1)}_{j-m\omega}}(r) \bigg) \\
&+\left(\mathcal{R}\mathfrak{t}^{(1,0)}_{(i)a}(r) \right)_{lm\omega} \mathsf{u}^{a(0,1)}_{m\omega}.
  \end{split}
\end{align} 
The structure of these terms was discussed in previous sections. In summary, we notice three distinct parts. First, the terms containing the radial parts of $\Psi_4^{(0,1)}$ (the function ${}_{-2}R^{(0,1)}_{jm\omega}(r)$) were obtained without the need for metric reconstruction. Next, we have the terms that explicitly depend on the Hertz potential ${}_{2}R^{\mathrm{Hz}(0,1)}_{lm\omega}(r)$. Finally, the last term contains the perturbed four-velocity coefficients $\mathsf{u}^{a(0,1)}_{m\omega}$, which arise from the oscillatory motion of the ring. Remember that this is the only term calculated for all multipoles of the ring simultaneously.

The resulting source is then fully determined by the functions
 \begin{align}
\left \lbrace\mathcal{RO}^{(1,0)}_{(i)(k)lm\omega}(r),\mathcal{R}\Sigma^{(1,0)}_{(i)(k)lm\omega}(r),\left(\mathcal{R}\mathfrak{t}^{(1,0)}_{(i)a}(r) \right)_{lm\omega}\right\rbrace
\end{align}
which are listed in the supplemental material \cite{MmaNB}  (Mathematica notebook  \texttt{Sourceincoordinates.nb} ). By decomposing $\Psi_{4(r_{\rm r})}^{(1,1)}$ as
 \begin{align}  
\Psi_{4(r_{\rm r})}^{(1,1)}=\frac{1}{r^4}\displaystyle\sum_{lm\omega}{}_{-2} R^{(1,1)}_{(r_{\mathrm{r}})lm\omega}(r){}_{-2}Y_{lm}(\theta)e^{i(m\phi-\omega t)},
    \end{align}   
we can formulate an ordinary differential equation for its radial part ${}_{-2} R^{(1,1)}_{(r_{\mathrm{r}})lm\omega}(r)$
 \begin{align} 
 \label{(1,1) Teukolsky radial} 
 {}_{-2}\mathfrak{D}_{lm\omega}\left( {}_{-2}R^{(1,1)}_{(r_{\mathrm{r}})lm\omega}(r)\right)=-2r^6\displaystyle\sum_{i=0}^{3}\mathcal{R}\left[\mathcal{S}^{(1,1)}_{r_{\rm r}(i)} \right]_{lm\omega}(r)\delta^{(i)}(r-r_{\rm r}).
    \end{align} 
    \end{widetext}
This is just a radial Teukolsky equation on the Schwarzschild background with the radial part of the source $\mathcal{S}^{(1,1)}_{r_{\rm r}}$. Notice the factor $-2r^6$ on the right-hand side. We have already introduced this factor in section \ref{Teukolsky Kerr} where we discussed the Teukolsky equation in Kerr spacetime, denoting it as $\mathfrak{G}(a)$. In the Schwarzschild case, this expression reduces to $\mathfrak{G}(0)=-2r^6$.

Equation \eqref{(1,1) Teukolsky radial} is now ready to be solved by using variation of parameters. This involves applying the formula \eqref{TeukolskyRadialvariationofconstants} which satisfies the boundary conditions at infinity and on the horizon. Since the source consists of delta functions, we can compute the integrals over $r$ in \eqref{TeukolskyRadialvariationofconstants}  analytically.

Once equation \eqref{(1,1) Teukolsky radial} is solved, we can put $\Psi_{4(r_{\rm r})}^{(1,1)} $ together with the part smoothly matched at $r=r_{\rm r}$ (section \ref{Smoothmatching}) to get
 \begin{align} 
  \label{The complete solution}  
\Psi_{4}^{(1,1)}=\Psi_{4(\mathrm{sm})}^{(1,1)}+\Psi_{4(r_{\rm r})}^{(1,1)} .
    \end{align}   
 This is then the complete solution of the $(1,1)$ modified Teukolsky equation in our model where the ring is represented by its monopole and dipole. Equivalently, we can say that all the information is encoded in the partial amplitudes present in the mode decomposed $\Psi_{4}^{(1,1)}$
 \begin{align} 
   \label{The complete amplitudes}   
\Psi_{4}^{(1,1)}\leftrightarrow  \left \lbrace C^{(1,1)(H)}_{lm\omega},C^{(1,1)(\infty)}_{lm\omega} \right\rbrace.
    \end{align}    
Their calculation will be the focus of our future work. In this paper, we have laid all the groundwork by deriving the necessary equations which can then be solved in a straightforward manner.

%%%%%%%%%%%%%%%%%%%%%%%%%%%%%%%%%%%%%%%%%%%%%%%%%%%%%%%%%%%%%%%%%%%%%%
%%%%%%%%%%%%%%%%%%%%%%%%%%%%%%%%%%%%%%%%%%%%%%%%%%%%%%%%%%%%%%%%%%%%%%
\section{Summary} \label{sec:summary}
%%%%%%%%%%%%%%%%%%%%%%%%%%%%%%%%%%%%%%%%%%%%%%%%%%%%%%%%%%%%%%%%%%%%%%
%%%%%%%%%%%%%%%%%%%%%%%%%%%%%%%%%%%%%%%%%%%%%%%%%%%%%%%%%%%%%%%%%%%%%%

The present work develops a fully relativistic framework for gravitational perturbations of slowly rotating black holes surrounded by axisymmetric stationary matter distributions, with a specific application to EMRIs evolving within such environments. We employ a two-parameter perturbation scheme, which expands in both the mass ratio $\varepsilon =m_{\rm part}/M$ and a parameter $\zeta = m_{\rm r}/M$ characterizing the deviation from vacuum Kerr spacetime due to the surrounding matter cloud or ring of mass $m_{\rm r}$. This approach enables us to derive a modified Teukolsky equation at order $\mathcal{O}(\varepsilon\zeta)$ that captures the leading-order environmental effects on EMRI dynamics.

Our formulation focuses on the pole-dipole approximation for the matter distribution. We achieve this through a thin shell of matter at radius $r_{\rm r}$ surrounding the central black hole. This configuration, illustrated in Fig. \ref{fig:ringEMRI}, is motivated by various astrophysical scenarios. These scenarios include remnant accretion structures, nuclear stellar distributions, and idealized models of dark matter profiles. However, the thinness of the shell implies that it should primarily describe ``cold'' structures without significant internal pressure. This approximation allows for analytical progress while capturing the essential physics of environmental gravitational effects. Since the effect of the shell is consistently linearized, one can later construct a model for matter of finite thickness by superposing such shells at different radii.

A key step lies in our use of Israel junction conditions on the matter shell \cite{Israel:1966rt,Poisson_2004} to match the interior vacuum region (within which the EMRI evolves) with the exterior region. This matching procedure results in a piecewise Petrov type D spacetime. It features carefully controlled junction behavior, enabling the application of Teukolsky-based methods in each region. This approach also ensures that we properly account for the non-trivial matching at $r = r_{\rm r}$. Our analysis shows that local inertial observers near the black hole are redshifted. They are also rigidly rotating compared to observers at infinity. Specifically, the redshift is $m_{\rm r}/(r_{\rm r} - 2M)$, and the angular frequency of the rigid rotation is $2Ma_{\rm r}/r_{\rm r}^3$, where $a_{\rm r}$ is the angular momentum of the ring. This picture highlights the importance of global analysis of the spacetime, rather than solely focusing on the local neighborhood of the central black hole. 

The modified Teukolsky equation we derive has the structure
\begin{equation}
	\mathcal{O}^{(0,0)}\Psi_4^{(1,1)} = \mathcal{S}^{(1,1)}_{\rm eff},
\end{equation}
where $\mathcal{O}^{(0,0)}$ is the standard Teukolsky operator on the Schwarzschild background. The effective source $\mathcal{S}^{(1,1)}_{\rm eff}$ includes contributions from both particle motion and the matter-modified background geometry. The resulting solution naturally breaks down into smooth and singular components. The singular components are confined to the location of the matter shell. This decomposition allows for efficient solution strategies that take advantage of the piecewise structure of the background. The effective source for the smooth part includes step functions, while the singular part is sourced by delta functions at $r = r_{\rm r}$.

For the reconstruction of the metric perturbation sourced by the particle, we use the CCK formalism in the outgoing radiation gauge (ORG). This approach builds on recent developments \cite{van_de_MeentShah} and adapts them to our non-vacuum context. The reconstruction procedure requires careful treatment of the particle perturbations on the matter shell. Here, the standard Hertz potential method must be enhanced with solutions to transport equations for the perturbed Newman-Penrose quantities. We provide explicit formulas for the reconstructed metric components using derivatives of a Hertz potential decomposed into spherical harmonic modes. 

A particularly challenging aspect concerns the equations for the perturbations of environmental matter by the particle. Unlike in the vacuum case, these equations cannot be solved mode-by-mode in the multipole expansion. Instead, the perturbed stress-energy conservation and continuity equations must be solved for the full matter distribution. We show that the matter ring oscillates under the influence of GWs from the inspiraling particle, causing oscillations of the ring and additional GW emissions.

Our framework yields several concrete results for EMRIs surrounded by the pole-dipole ring. We demonstrate that the solution of the modified Teukolsky equation decomposes naturally into two components: a smooth part $\Psi_{4(\rm sm)}^{(1,1)}$ obtained by matching standard Teukolsky solutions across the matter shell, and a singular part $\Psi_{4(r_{\rm r})}^{(1,1)}$ sourced by delta functions at $r = r_{\rm r}$. The effective source contains three distinct physical contributions: direct coupling terms $\mathcal{O}^{(1,0)}\Psi_4^{(0,1)}$ that require no metric reconstruction, terms involving the reconstructed Hertz potential $\Sigma^{(1,0)}\overline{\Psi_{\rm Hz}^{(0,1)}}$, and a dynamical component $\mathfrak{T}^{(1,1)}_{\rm dynamical}$ arising from ring oscillations under GW driving. The angular momentum of the surrounding matter ($a_{\rm r}$) introduces mode-mixing between adjacent $l$-modes, coupling solutions with indices $l-1$, $l$, and $l+1$. We show that the ring oscillation amplitude scales as $\varepsilon(r_{\rm p}/r_{\rm r})^{3/2}$ in the adiabatic regime, where the ring absorbs GW energy and re-radiates it, contributing additional terms to the accumulated flux at order $\mathcal{O}(\varepsilon \zeta)$. These theoretical developments will be key for building full inspiral waveforms and statements on the detectability of the environment in EMRIs.

%%%%%%%%%%%%%%%%%%%%%%%%%%%%%%%%%%%%%%%%%%%%%%%%%%%%%%%%%%%%%%%%%%%%%%%%%%%%%%%%%%%%%%%%%%%%%%%%%%%%%%%%%%%%%%%%%
 \section{Discussion and outlook}   
 \label{sec:discussion}
%%%%%%%%%%%%%%%%%%%%%%%%%%%%%%%%%%%%%%%%%%%%%%%%%%%%%%%%%%%%%%%%%%%%%%%%%%%%%%%%%%%%%%%%%%%%%%%%%%%%%%%%%%%%%%%%%

%%%%%%%%%%%%%%%%%%%%%%%%%%%%%%%%%%%%%%%%%%%%%%%%%%%%%%%%%%%%%%%%%%%%%%%%%%%%%%%%%%%%%%%%%%%%%%%%%%%%%%%%%%%%%%%%%
\subsection{Generating inspiral waveforms}

Our final goal is to compute the inspiral and the corresponding waveform, a task for which we now have all ingredients. We can now comment on how this final task will be carried out in a paper that will follow this work. 

To drive the inspirals, we only need the GW fluxes and a suitable flux-balance law for the decay of the orbital degrees of freedom. Since the spacetime is vacuum and type D everywhere apart from the shell,  we can use the standard formulas for fluxes at infinity and on the horizon ($\mathcal{F}^{(\infty)}$ and $\mathcal{F}^{(\rm H)}$).

However, these fluxes alone do not provide enough information to compute the evolution since the secondary body can exchange energy and angular momentum with the ring. Therefore, we must also compute the fluxes $\mathcal{F}^{(\mathcal{S})}$ through a sphere $\mathcal{S}$ of radius $r = r_{\mathcal{S}}$ located between the inspiraling particle and the ring ($r_{\rm max} < r_{\mathcal{S}} < r_{\rm r}$). Since this sphere lies within a vacuum type D region, we can reconstruct the metric perturbation $h_{\mu\nu}^{(1,1)}$ from $\Psi_4^{(1,1)}$ using the CCK method at $r = r_{\mathcal{S}}$, and from this obtain $\mathcal{F}^{(\mathcal{S})}$.

As a result, we will adiabatically evolve the parameters of the orbit of the secondary body $I^{\rm (p)}_A$ by using a modified flux-balance law
\begin{align} 
\begin{split} 
\label{fluxesevolution}  
\frac{\mathrm{d}I^{\rm (p)}_A}{\mathrm{d}t}=&
-\mathcal{F}^{(\rm H)}_A(\mathbf{I^{\rm (p)}},\mathbf{I^{\rm (ring)}})
\\ &
-\mathcal{F}^{(\mathcal{S})}_A(\mathbf{I^{\rm (p)}},\mathbf{I^{\rm (ring)}})\,,
\end{split} 
\end{align}   
where the fluxes will be expanded as   
$ \mathcal{F}_A=\varepsilon\mathcal{F}^{(0,1)}_A+\varepsilon\zeta\mathcal{F}^{(1,1)}_A $. Consequently, we will expand the evolutions $I^{\rm (p)}_A(t)$ into leading adiabatic pieces and environmental corrections.  This flux-balance law differs from the usual one by $\mathcal{F}^{(\mathcal{S})}_A - \mathcal{F}^{(\infty)}_A$, which is of order $\mathcal{O}(\epsilon\zeta)$ and corresponds to energy and angular momentum transferred into the ring. Even though the scaling of this new term with $\epsilon \zeta$ should hold, we suspect the difference to be numerically suppressed due to the $r_{\rm p} \ll r_{\rm r}$ assumption. We will leave the investigation of this question for the next paper in this series. Additionally, we could also formulate an adiabatic evolution equation for the ring parameters $ I^{\rm (ring)}_A$ , but their evolution would contribute to the EMRI phase only at subleading order in the $\epsilon \zeta$ expansion.

  For equatorial orbits, the motion is characterized only by energy and angular momentum. Then, the calculation of the fluxes in equation \eqref{fluxesevolution}  is straightforward, as the timelike and azimuthal Killing vectors exist globally and one can use the usual Teukolsky formulas \cite{Teukolsky:1974yv,Drasco_2006}. For inclined particle orbits, the calculations will additionally require evolving the Carter constant, for which we will have to slightly generalize the computation of \citet{Sago_2006}.

Similarly, we plan to calculate and expand the GW phase and produce the resulting waveform using methods outlined in Ref. \cite{Hughes:2021exa}. We want to stress again that while various authors (including us in Ref. \cite{Polcar_2022}) have computed similar results in inconsistent approximations and kludge toy models, fully self-consistent relativistic inspirals and waveforms in matter environments have not been computed to date.

%%%%%%%%%%%%%%%%%%%%%%%%%%%%%%%%%%%%%%%%%%%%%%%%%%%%%%%%%%%%%%%%%%%%%%%%%%%%%%%%%%%%%%%%%%%%%%%%%%%%%%%%%%%%%%%%%
\subsection{Technical simplifications?}
Is it possible that the computations we are carrying out could be done in a simpler way? While we have clearly made a number of advantageous assumptions, there are avenues for further technical simplifications. 

For example, instead of the full cross-term perturbation to the Weyl scalar $\Psi^{(1,1)}_4$, we could try to solve for an analogue of the ``reduced Teukolsky scalar'' as discussed by Refs \cite{Campanelli_1999,spiersmoxon}. The reduced scalar $\Psi^{(1,1)}_{4\mathrm{L}}$ is formally obtained from the cross-term metric perturbation $h^{(1,1)}$ as if it were a purely linear metric perturbation, or 
\begin{align}
    \Psi^{(1,1)}_{4\mathrm{L}}=\mathscr{\tilde {T}}^{(0,0)}_4 h^{(1,1)} \,,
\end{align}
where the operator $\mathscr{\tilde {T}}^{(0,0)}_4$  was discussed in Section \ref{sec:reconstruction}. Although the reduced scalar is somewhat unphysical, the analogue of this reduced scalar in second-order perturbation theory can be used to extract GW fluxes and its use leads to simplified computations \cite{Campanelli_1999,spiersmoxon}. However, we examined the equation for the resulting scalar in our set-up and found no particular advantage in using it; it is possible it could be useful in some of the extensions and generalizations discussed below.

Another point that could lead to simplifications is a deeper exploitation of the $r_{\rm r} \gg r_{\rm p}$ limit. Because the radiative metric perturbation by the central inspiral falls off only as $\sim 1/r_{\rm r}$, we did not assume that the interaction of the radiation with the matter ring was particularly simplified by the $r_{\rm r} \gg r_{\rm p}$ limit. However, it is conceivable that a consistent approximation could be obtained in a post-Minkowski expansion. In particular, the field of the ring, the black hole, and the GW field could possibly be treated as linearized perturbations on a flat background in the vicinity of $r_{\rm r}$. The various $(1,1)$ cross-terms could then be computed from second-order perturbation theory on a flat background. We leave an analysis of this idea to the future.

%%%%%%%%%%%%%%%%%%%%%%%%%%%%%%%%%%%%%%%%%%%%%%%%%%%%%%%%%%%%%%%%%%%%%%%%%%%%%%%%%%%%%%%%%%%%%%%%%%%%%%%%%%%%%%%%%
\subsection{Extensions and generalizations}

Let us now make several remarks regarding possible extensions of this calculation to higher tidal multipoles of the matter environments and possibly to more generic models. 

First, as was mentioned in Section \ref{sec:holeringpert}, the computation is fully linearized in the environmental perturbation. As such, our flux correction can be viewed as a ``Green's function'' from which one can easily obtain corrections corresponding to extended environmental matter configurations by weighted integration over the ring parameters. For example, a smooth superposition of many ring contributions of different masses, radii, and inclinations should serve as a good model for extended halos of cold dark matter. Second, as was sketched  in the flowchart in Fig. \ref{fig:Teukeqscheme}, the modified Teukolsky equation can be used in a large variety of models. The $(1,0)$ metric and stress-energy tensor do not have to be axially symmetric or stationary; they could represent any matter configurations including ``hot'' gases and fluids, eccentric disks, or even a distant orbiting body \cite{Bonga:2019ycj}.

  In fact, the derivation of the modified Teukolsky equation relies on the Bianchi and Ricci identities, which are independent of the field equations. As such, it can also be used in the context of alternative theories of gravity where $h^{(1,0)}_{\mu\nu}$ is a perturbation to the black hole spacetime due to the beyond-GR effects. This was already demonstrated at the level of quasinormal modes in modified gravities \cite{Li_2023,wagle2023perturbationsspinningblackholes,Cano_2023,Weller}. The main difference here is the relation between the Ricci tensor and the stress-energy tensor $R_{\mu\nu}\leftrightarrow T_{\mu\nu}$, which generally looks different in alternative theories of gravity and may also include other gravitational degrees of freedom. 

In any of the setups mentioned above, the calculation would be structurally similar to the calculation presented here. However, for a more generic model, there will be several key new difficulties. Most notably, for a general environmental perturbation, the Petrov-D property is violated everywhere, or $\Psi^{(1,0)}_4\neq 0$. Technically, this would imply that we have to reconstruct the metric on the particle worldline to get $\mathcal{O}^{(0,1)}\Psi^{(1,0)}_4$. Along the worldline, the purely vacuum CCK reconstruction method using the Hertz potential would no longer be applicable, and one would have to use a different approach such as the Green-Hollands-Zimmerman reconstruction method \cite{Green:2019nam}. Alternatively, in Schwarzschild space-time one could use the Regge-Wheeler-Zerilli formalism or work directly in the Lorenz gauge \cite{Regge:1957td,Zerilli:1970se,Barack:2005nr}.  Additionally, the effective source $\mathcal{S}^{(1,1)}_{\rm eff}$ would no longer be compact in the radial direction, which would mean that the integration over $r$ in the variation of constants \eqref{TeukolskyRadialvariationofconstants} would have to be done numerically, unlike in our setup.

Another problem which we would have to take into account in a more general environmental model is the perturbation of the geodesic of the inspiraling body, which would contribute to the effective source through $T^{(1,1)\mu\nu}_{\rm particle}$. In such a case, the geodesic motion would no longer be integrable but could be in principle treated as a nearly integrable Hamiltonian system. This problem was the focus of our previous paper \cite{Polcar_2022} where we employed canonical perturbation theory to obtain a new set of approximate integrals of motion which were then subject to adiabatic evolution in the inspiral.    

Finally, for a generic perturbed background, one should modify the formula for the flux-balance laws  \eqref{fluxesevolution}   since the established formulas were derived in the Kerr spacetime. This is perhaps the least clear of the future tasks, since in general there may be no Killing vectors and tensors available. Using the approximate action-angle variables mentioned in the previous paragraph, one could likely employ the formalisms of Refs \cite{Isoyama:2018sib,Grant:2024ivt} to derive flux-balance formulae away from orbital resonances. However, along orbital resonances, canonical perturbation theory fails \cite{arnold2006mathematical}. Finding flux-balance laws in such a case represents an interesting open problem on its own.

Another possible avenue for generalization is the computation of the Lindblad and corotation resonances, which we neglected due to the $r_{\rm r}\gg r_{\rm p}$ limit. In essence, our computation could in part follow that of \citet{Hirata:2010vn,Hirata:2010vn}. The computation of Hirata presents a heuristic balance argument to estimate the excitation of the environment by the perturbing particle and possibly also migration torques on the particle itself (see also Ref. \cite{Duque:2025yfm}). We plan to explore the relationship of the work of Hirata to our work in the second paper of this series. 

%%%%%%%%%%%%%%%%%%%%%%%%%%%%%%%%%%%%%%%%%%%%%%%%%%%%%%%%%%%%%%%%%%%%%%%%%%%%%%%%%%%%%%%%%%%%%%%%%%%%%%%%%%%%%%%%%
\begin{acknowledgments}
We would like to thank Georgios Lukes-Gerakopoulos and Samuel Upton for discussions on this topic. The authors were supported by the Charles U. \textit{Primus} Research Program 23/SCI/017.
\end{acknowledgments}

%%%%%%%%%%%%%%%%%%%%%%%%%%%%%%%%%%%%%%%%%%%%%%%%%%%%%%%%%%%%%%%%%%%%%%%%%%%%%%%%%%%%%%%%%%%%%%%%%%%%%%%%%%%%%%%%%
%%%%%%%%%%%%%%%%%%%%%%%%%%%%%%%%%%%%%%%%%%%%%%%%%%%%%%%%%%%%%%%%%%%%%%%%%%%%%%%%%%%%%%%%%%%%%%%%%%%%%%%%%%%%%%%%%
\appendix
%%%%%%%%%%%%%%%%%%%%%%%%%%%%%%%%%%%%%%%%%%%%%%%%%%%%%%%%%%%%%%%%%%%%%%%%%%%%%%%%%%%%%%%%%%%%%%%%%%%%%%%%%%%%%%%%%
%%%%%%%%%%%%%%%%%%%%%%%%%%%%%%%%%%%%%%%%%%%%%%%%%%%%%%%%%%%%%%%%%%%%%%%%%%%%%%%%%%%%%%%%%%%%%%%%%%%%%%%%%%%%%%%%%

%%%%%%%%%%%%%%%%%%%%%%%%%%%%%%%%%%%%%%%%%%%%%%%%%%%%%%%%%%%%%%%%%%%%%%%%%%%%%%%%%%%%%%%%%%%%%%%%%%%%%%%%%%%%%%%%%
\section{NP and GHP formalism}\label{app:NPGHP}
%%%%%%%%%%%%%%%%%%%%%%%%%%%%%%%%%%%%%%%%%%%%%%%%%%%%%%%%%%%%%%%%%%%%%%%%%%%%%%%%%%%%%%%%%%%%%%%%%%%%%%%%%%%%%%%%%
This part of the Appendix provides additional details  of  Newman-Penrose (NP) and Geroch-Held-Penrose (GHP) formalism relevant to the calculations in this paper.

The NP connection is represented by the Ricci rotation coefficients $\gamma_{cba}=e^{\nu}_{c} e^{\mu}_{b}  \nabla_\mu( e_{a\nu })$ which are antisymmetric in the first and third index   $\gamma_{cba}=-\gamma_{abc}$. In the NP formalism, it is convenient to instead use the spin coefficients which are defined as
\begin{subequations}
\begin{align}
    \kappa &= - m^\nu l^\mu \, \nabla_\mu l_\nu=-\gamma_{mll}, \\
    \sigma &= - m^\mu m^\nu \, \nabla_\nu l_\mu=-\gamma_{mml}, \\
    \lambda &= \bar{m}^\mu \bar{m}^\nu \, \nabla_\nu n_\mu=\gamma_{\bar{m}\bar{m}n}, \\
    \nu &=     \bar{m}^\mu    n^\nu \, \nabla_\nu n_\mu=\gamma_{\bar{m}nn}, \\
    \rho &= - m^\mu \bar{m}^\nu \, \nabla_\nu l_\mu=-\gamma_{m\bar{m}l}, \\
    \mu &=  \bar{m}^\mu  m^\nu\, \nabla_\nu n_\mu=\gamma_{\bar{m}mn}, \\
    \tau &= - m^\mu n^\nu \, \nabla_\nu l_\mu=-\gamma_{mnl}, \\
    \pi &= \bar{m}^\nu l^\nu \, \nabla_\nu n_\mu=\gamma_{\bar{m}ln}, \\
    \epsilon &=- \frac{1}{2} ( n^\mu l^\nu\, \nabla_\nu l_\mu - \bar{m}^\mu l^\nu \, \nabla_\nu m_\mu)= \frac{1}{2}(\gamma_{\bar{m}lm}-\gamma_{nll}), \\
    \alpha &=- \frac{1}{2} (n^\mu \bar{m}^\nu \, \nabla_\nu l_\mu - \bar{m}^\mu  \bar{m}^\nu \, \nabla_\nu m_\mu)=\frac{1}{2}(\gamma_{\bar{m}\bar{m}m}-\gamma_{n\bar{m}l}), \\
    \beta &= -\frac{1}{2} (n^\mu m^\nu \, \nabla_\nu l_\mu - \bar{m}^\mu m^\nu \, \nabla_\nu m_\mu)=\frac{1}{2}(\gamma_{\bar{m}mm}-\gamma_{nml}),\\
      \gamma &= -\frac{1}{2} (n^\mu n^\nu \, \nabla_\nu l_\mu - \bar{m}^\mu n^\nu \, \nabla_\nu m_\mu)= \frac{1}{2}(\gamma_{\bar{m}nm}-\gamma_{nnl}).
\end{align}
\end{subequations}
Using the definitions of the spin coefficients, one can then derive commutation relations for the NP derivatives.
The Riemann curvature tensor can be decomposed into the Weyl tensor $C_{\alpha \beta \gamma \delta} $ whose ten independent components are encoded in five complex Weyl scalars
\begin{subequations}
\begin{align}
    \Psi_0 &= C_{\alpha \beta \gamma \delta} l^\alpha m^\beta l^\gamma m^\delta, \\
    \Psi_1 &= C_{\alpha \beta \gamma \delta} l^\alpha n^\beta l^\gamma m^\delta, \\
    \Psi_2 &= C_{\alpha \beta \gamma \delta} l^\alpha m^\beta \bar{m}^\gamma n^\delta, \\
    \Psi_3 &= C_{\alpha \beta \gamma \delta} l^\alpha n^\beta \bar{m}^\gamma n^\delta, \\
    \Psi_4 &= C_{\alpha \beta \gamma \delta} n^\alpha \bar{m}^\beta n^\gamma \bar{m}^\delta
\end{align}
\end{subequations}

and the Ricci tensor $R_{\alpha \beta} $ whose traceless part is described by the $\Phi$-scalars (3 real and 3 complex) while its trace is represented by  $ \Lambda$
\begin{subequations}
\label{app:Ricciscalars}
\begin{align}
    \Phi_{00} &= \frac{1}{2} R_{\alpha \beta} l^\alpha l^\beta, \\
    \Phi_{01} &= \frac{1}{2} R_{\alpha \beta} l^\alpha m^\beta=\overline{\Phi_{10}}, \\
    \Phi_{02} &= \frac{1}{2} R_{\alpha \beta} m^\alpha m^\beta=\overline{\Phi_{20}}, \\
    \Phi_{11} &= \frac{1}{4} R_{\alpha \beta} (l^\alpha n^\beta + m^\alpha \bar{m}^\beta), \\
    \Phi_{12} &= \frac{1}{2} R_{\alpha \beta} n^\alpha m^\beta=\overline{\Phi_{21}}, \\
    \Phi_{22} &= \frac{1}{2} R_{\alpha \beta} n^\alpha n^\beta , \\
    \Lambda &=\frac{R}{24}.
\end{align}
\end{subequations}
In general relativity one could use the Einstein field equations $R_{\mu \nu}-\frac{1}{2}R g_{\mu \nu}=8\uppi T_{\mu \nu} $ to use $ T_{ab}$ instead of $ \Phi_{AB}$ and $\Lambda$, for example $\Phi_{11}=2\uppi(T_{ln}+T_{m\bar{m}}) $.

By projecting the Ricci identities 
\begin{align}
    \left(\nabla_\mu \nabla_\nu-\nabla_\nu\nabla_\mu\right)e^{\alpha}_{a} =\tensor{R}{_\mu_\nu^\alpha _\beta}e^{\beta}_{a} \,,
\end{align}
onto the Newman-Penrose tetrad we can obtain their   equivalents in terms of NP quantities, one such example being 
\begin{align}
    \begin{split}
    & \left(\Delta+\mu+\bar{\mu}+3\gamma-\bar{\gamma}\right)\lambda +\Psi_4
    \\
    &  -\left(\bar{\delta}+3\alpha+\bar{\beta}+\pi-\bar{\tau}\right)\nu=0.
    \end{split}
\end{align} 
Similarly, one can also project the  differential Bianchi identities
$$R_{\mu\nu\alpha\beta;\gamma}+R_{\mu\nu\beta\gamma;\alpha}+R_{\mu\nu\gamma\alpha;\beta}=0,$$
of which the most relevant to our calculations are
\begin{subequations}
\begin{align}
\left(\delta+4\beta-\tau\right)\Psi_4-\left(\Delta+ 2\gamma+4\mu\right)\Psi_3+3\nu\Psi_2=S_3\\
\left(D+4\varepsilon-\rho\right)\Psi_4-\left(\bar{\delta}+4\pi+2\alpha\right)\Psi_3+3\lambda\Psi_2=S_4\\
\left(\bar{\delta}+3\pi\right)\Psi_2-\left(D+2\varepsilon-2\rho\right)\Psi_3-2\lambda\Psi_1-\kappa\Psi_4=S_5\\
 \left(\Delta+3\mu\right)\Psi_2-\left(\delta+2\beta-2\tau\right)\Psi_3-2\nu\Psi_1-\sigma\Psi_4=S_6.
\end{align} 
\end{subequations}
where $S_A$ are explicitly given in terms of  Ricci  scalars $\Phi_{AB}$  and $\Lambda$ 
\begin{subequations}
\begin{align}
    \begin{split}
        S_3=
            &-\left(\Delta+2\bar{\mu}+2\gamma\right)\Phi_{21}+\left(\bar{\delta}-\bar{\tau}+2\alpha+2\bar{\beta}\right)\Phi_{22}
            \\
            &+2\nu\Phi_{11}+\bar{\nu}\Phi_{20}-2\lambda\Phi_{12}\,,
    \end{split}
    \\
    \begin{split}
        S_4=
            & -\left(\Delta+\bar{\mu}+2\gamma-2\bar{\gamma}\right)\Phi_{20}+(\bar{\delta}+2\alpha-2\bar{\tau})\Phi_{21}
            \\
            & +2\nu\Phi_{10} -2\lambda\Phi_{11}+\bar{\sigma}\Phi_{22}\,,
    \end{split}
    \\
    \begin{split}
        S_5=
            & \left(\delta-2\bar{\alpha}+2\beta+\bar{\pi}\right)\Phi_{20}-\left(D-2\bar{\rho} +2\epsilon\right)\Phi_{21}
            \\
            &-2\mu\Phi_{10}+2\pi\Phi_{11}-\bar{\kappa}\Phi_{22}-2\bar{\delta}\Lambda \,,
    \end{split}
    \\
    \begin{split}
        S_6=
            & \left(\delta+2\bar{\pi} +2\beta \right)\Phi_{21}-\left(D-\bar{\rho}+2\varepsilon+2\bar{\varepsilon}\right)\Phi_{22}
            \\
            &-2\mu\Phi_{11}-\bar{\lambda}\Phi_{20}+2\pi\Phi_{12}-2\Delta\Lambda\,.
    \end{split}
\end{align}
\end{subequations}

The Newman-Penrose tetrad is not unique, it can be rotated (Lorentz transformed) at each point of the spacetime. The 6-parametric Lorentz group can be represented by the following three types of rotations \cite{Chandrasekhar:1984siy}

 \begin{align}
 \label{app:tetradrotation}
\text{I}:\enspace & l^\mu\rightarrow l^\mu,\enspace m^\mu\rightarrow m^\mu+\upsilon l^\mu,\\
\nonumber & n^\mu\rightarrow  n^\mu+\bar{\upsilon} m^\mu+\upsilon \bar{m}^\mu+\vert\upsilon\vert^2 l^\mu\\
\text{II}:\enspace & n^\mu\rightarrow n^\mu,\enspace  m^\mu\rightarrow m^\mu+\varpi l^\mu,\\
\nonumber & l^\mu\rightarrow  l^\mu+\bar{\varpi} m^\mu+\varpi \bar{m}^\mu+\vert\varpi\vert^2 n^\mu\\
\text{III}:\enspace & l^\mu\rightarrow A l^\mu,\enspace  n^\mu\rightarrow A^{-1} n^\mu,\enspace   m^\mu \rightarrow e^{i\vartheta} m^\mu
   \end{align} 
where $\upsilon$ and $\varpi$ are complex  while $A$ and   $\vartheta$ are are real functions. \\
The type III transformations preserve the direction of the null vectors and it is the starting point to introduce the 
Geroch-Held-Penrose formalism. If we use a single complex function $\eta$ related to the two real functions by $\eta^2=A e^{i\vartheta}$, we can rewrite the type III transformation as 

\begin{align}
l^\mu\rightarrow  \eta\bar{\eta} l^\mu  \hspace{10pt}  n^\mu\rightarrow  \eta^{-1}\left( \bar{\eta}\right)^{-1} n^\mu \hspace{10pt} m^\mu\rightarrow  \eta\left(\bar{\eta}\right)^{-1} m^\mu.
\end{align}
 A general quantity constructed from the tetrad then transforms as
\begin{align}
\Theta\rightarrow\eta^p  \bar{\eta}^q\Theta.
\end{align}

The pair $\{p,q\}$ is then called the GHP weight of $\Theta$.
  The GHP formalism  also utilizes a transformation ' :$\zeta\rightarrow \zeta'$ defined by exchanging the tetrad vectors $l\leftrightarrow n$ and $m\leftrightarrow \bar{m}$. Using this transformation we can adopt the following notation
\begin{align}
\begin{split}
\kappa':=-\nu,\hspace{8pt} \sigma':=-\lambda,\hspace{8pt} \rho':=-\mu ,\\
 \tau':=-\pi,\hspace{8pt} \beta':=-\alpha,\hspace{8pt}\epsilon':=-\gamma .
 \end{split}
\end{align}

If we apply the exchange transformation ', then the GHP weight of $\Theta'$  is $\{-p,-q\}$   while a complex-conjugate quantity $\bar{\Theta}$  has the  weight $\{q,p\}$.

For some spin coefficients, it is possible to assign the GHP weight, these are

\begin{align}
\kappa: \{3,1\},\hspace{8pt}  \sigma: \{3,-1\},\hspace{8pt}  \rho: \{1,1\},\hspace{8pt}  \tau: \{1,-1\}, 
\end{align}
while the remaining spin coefficients can be combined with the NP derivatives, which themselves do not have a well-defined  GHP weight, to define the derivative operators

\begin{align}
\begin{split}
\text{\textthorn}=D-p\epsilon-q\bar{\epsilon},\hspace{15pt}  \text{\textthorn}'=\Delta+p\epsilon'+q\bar{\epsilon}', \\ \nonumber  \ethm=\delta-p\beta+q\bar{\beta'}, \hspace{15pt} \ethm'=\bar{\delta}+p\beta'-q\bar{\beta},
\end{split}
\end{align}

whose GHP weight is then
\begin{align}
    \begin{split}
    & \text{\textthorn}: \{1,1\},\hspace{10pt} \text{\textthorn}': \{-1,-1\},\hspace{10pt} 
    \\
    & \ethm: \{1,-1\},\hspace{10pt}  \ethm': \{-1,1\}.
    \end{split}
\end{align}
As for the Weyl scalars, we have
\begin{align}\begin{split}
& \Psi_0: \{4,0\},\hspace{4pt}  \Psi_1: \{2,0\},\hspace{4pt}  \Psi_2: \{0,0\},
\\ 
& \Psi_3: \{-2,0\} ,\hspace{4pt} \Psi_4: \{-4,0\}.
\end{split}
\end{align}
while for the Ricci scalars the weights read
\begin{align}\begin{split}
& \Phi_{00}: \{2,2\},\hspace{4pt}  \Phi_{11}: \{0,0\},\hspace{4pt} \Phi_{22}: \{-2,-2\},
\\ 
& \Phi_{01}: \{2,0\},\hspace{4pt}  \Phi_{02}: \{2,-2\},\hspace{4pt} \Phi_{12}: \{0,-2\},
\\ 
& \Lambda: \{0,0\}.
\end{split}
\end{align}

In a similar fashion, we can deduce  the GHP weight of other quantities.

The GHP derivatives satisfy commutation relations; for our calculations the most important ones are the following
\begin{subequations}
\label{app:GHPcommutation}
\begin{align}
    \begin{split}
    [\text{\textthorn},\ethm]= 
        & -\bar{\tau}'\text{\textthorn}-\kappa\text{\textthorn}'+\bar{\rho}\ethm+\sigma\ethm'+q\left(\bar{\rho}\bar{\tau}'-\bar{\kappa}\bar{\sigma}'\right)
        \\
        & + p\left(\sigma\tau'-\rho'\kappa-\Psi_1\right)  \,, 
    \end{split}
    \\
    \begin{split}
    [\text{\textthorn},\ethm']=
        & -\tau'\text{\textthorn}-\bar{\kappa}\text{\textthorn}'+\bar{\sigma}\ethm+\rho\ethm' + p\left(\rho\tau'-\kappa\sigma'\right)
        \\
        & +q\left(\bar{\sigma}\bar{\tau}'-\bar{\rho}'\bar{\kappa}-\bar{\Psi}_1\right) \,.
    \end{split}
\end{align}
\end{subequations}
In this part of the Appendix we have listed only the most important identities, for a complete list see the original papers \cite{Newman:1961qr} (NP formalism) and \cite{Geroch:1973am} (GHP formalism).

%%%%%%%%%%%%%%%%%%%%%%%%%%%%%%%%%%%%%%%%%%%%%%%%%%%%%%%%%%%%%%%%%%%%%%%%%%%%%%%%%%%%%%%%%%%%%%%%%%%%%%%%%%%%%%%%%
\section{Spin-weighted spherical harmonics}\label{app:SWSHs}
%%%%%%%%%%%%%%%%%%%%%%%%%%%%%%%%%%%%%%%%%%%%%%%%%%%%%%%%%%%%%%%%%%%%%%%%%%%%%%%%%%%%%%%%%%%%%%%%%%%%%%%%%%%%%%%%%
Here we outline  several identities which are useful when constructing the effective source $\mathcal{S}^{(1,1)}_{r_{\rm r}}$ in Section \ref{sec:sourceontheshell}.   

First, for complex conjugation we have

\begin{align} 
\label{Yconjugate}
{}_{s}\bar{Y}_{lm}=(-1)^{s+m}{}_{s}Y_{l-m}.
    \end{align} 
 For a fixed spin weight $s$ the spin-weighted spherical harmonics are a complete orthonormal basis on the sphere $S^2$    
    
\begin{align}   
\displaystyle\int_{S^2} {}_{s}\bar{Y}_{l_1m_1}(\theta,\phi) {}_{s}Y_{l_2 m_2}(\theta,\phi)\mathrm{d}\Omega= \delta_{l_1 l_2} \delta_{m_1 m_2}.
    \end{align}

 For spin weight $0$ we have the ordinary spherical harmonics which are given by associated Legendre polynomials $P_l^m$   
    \begin{align}  
{}_{0}Y_{lm}(\theta,\phi)=\mathcal{N}_{lm} P_l^m(\cos\theta)e^{\mathrm{i}m \phi}.
  \end{align} 
The spin-weighted spherical harmonics of different weights can be obtained by applying spin raising and lowering operators  on ${}_{0}Y_{lm}$. Fortunately, these operators are proportional to the GHP derivatives $\ethm$ and $\ethm'$ in our $(0,0)$ Schwarzschild background where they have an explicit form \cite{Pound_2021}

\begin{align}   
\label{GHPangularderivatives}
\begin{split}
\ethm&=\frac{1}{\sqrt{2} r}\left(\partial_\theta+\frac{\mathrm{i}}{\sin\theta}\partial_\phi-s\cot\theta\right) ,
\\
\ethm'&=\frac{1}{\sqrt{2} r}\left(\partial_\theta-\frac{\mathrm{i}}{\sin\theta}\partial_\phi+s\cot\theta\right) .
\end{split}
    \end{align} 
This means that they act only on the angular coordinates (on the sphere) and they raise and lower the spin weight as
\begin{align}  
\label{GHPraisinglowering}
\begin{split}
\sqrt{2} r\ethm {}_{s}Y_{lm}&=-\sqrt{l(l+1)-s(s+1)}\enspace {}_{s+1}Y_{lm},\\ 
\sqrt{2} r\ethm'{}_{s}Y_{lm}&=\sqrt{l(l+1)-s(s-1)}\enspace {}_{s-1}Y_{lm}.
\end{split}
    \end{align} 
    When constructing the source $\mathcal{S}^{(1,1)}_{r_{\rm r}}$     one tends    to encounter  a product ${}_{s_1}Y_{l_1m_1}{}_{s_2}Y_{l_2 m_2}$ which we will have to decompose into ${}_{s}Y_{lm}$ meaning we have to find  coefficients $C^{slm}_{ s_1 l_1 m_1 s_2 l_2 m_2} $ in the expansion 
    
    \begin{align}   
    \label{Yproduct}
{}_{s_1}Y_{l_1m_1}{}_{s_2}Y_{l_2 m_2}=\displaystyle\sum_{lm} C^{slm}_{ s_1 l_1 m_1 s_2 l_2 m_2} {}_{s}Y_{lm}.
    \end{align} 
    These can be obviously found by calculating integrals of the form

\begin{align} 
\begin{split}
&C^{slm}_{ s_1 l_1 m_1 s_2 l_2 m_2}=\\&\displaystyle\int_{S^2}  {}_{s}\bar{Y}_{lm}(\theta,\phi) {}_{s_1}Y_{l_1m_1}(\theta,\phi) {}_{s_2}Y_{l_2 m_2}(\theta,\phi)\mathrm{d}\Omega.
\end{split}   
    \end{align} 
Fortunately, there exists a  closed-form expression involving  a product of Wigner 3-j symbols  \cite{spiers2024analyticallyseparatingsourceteukolsky}
\begin{align}  
\begin{split}
C^{slm}_{ s_1 l_1 m_1 s_2 l_2 m_2}&=(-1)^{s+m}\sqrt{\frac{(2l+1)(2l_1+1)(2l_2+1)}{4\uppi}}\\
&\times  \begin{pmatrix} l & l_1 & l_2 \\ s & -s_1 & -s_2 \end{pmatrix}\begin{pmatrix} l & l_1 & l_2 \\ -m & m_1 & m_2 \end{pmatrix} .
\end{split}
    \end{align}
  The coefficients $C^{slm}_{ s_1 l_1 m_1 s_2 l_2 m_2} $ then inherit the  selection rules  of the 3-j symbols
    
\begin{align} 
\begin{split}
s&=s_1+s_2,\enspace  m=m_1+m_2,\\  \vert &l_1-l_2\vert\leq l\leq l_1+l_2 ,
\end{split}
    \end{align} 
which are the well-known rules for spin-addition in quantum mechanics.

The identities presented here will become relevant for our calculations in section \ref{sec:sourceontheshell} where we shall expand various quantities into ${}_{s}Y_{lm}$.

%%%%%%%%%%%%%%%%%%%%%%%%%%%%%%%%%%%%%%%%%%%%%%%%%%%%%%%%%%%%%%%%%%%%%%%%%%%%%%%%%%%%%%%%%%%%%%%%%%%%%%%%%%%%%%%%%
\section{Full modified Teukolsky equation}\label{app:modifTeuk}
%%%%%%%%%%%%%%%%%%%%%%%%%%%%%%%%%%%%%%%%%%%%%%%%%%%%%%%%%%%%%%%%%%%%%%%%%%%%%%%%%%%%%%%%%%%%%%%%%%%%%%%%%%%%%%%%%

In this section of the Appendix, we shall present the (modified) Teukolsky equation in detail by writing down all its individual parts. 

We define the following operators
\begin{subequations}
\begin{align}
\label{operatordefapp}
 & \mathcal{E}_3:=\bar{\delta}+3\alpha+\bar{\beta}+4\pi-\bar{\tau},
 \\ 
 & \mathcal{E}_4:=\Delta+4\mu+\bar{\mu}+3\gamma-\bar{\gamma}
 \\
 & F_3:=\delta+4\beta-\tau, \quad
 F_4:=D+4\varepsilon-\rho ,
 \\
 & J_3:=\Delta+ 2\gamma+4\mu, \quad
 J_4:=\bar{\delta}+4\pi+2\alpha ,
 \\
 & G_3:=D+2\varepsilon-2\rho, \quad
 G_4:=\delta+2\beta-2\tau,
\end{align}  
\end{subequations}
which are explicitly given in terms of NP quantities thus reducing the length of the NP expressions.  Using the expansion
 \begin{align}
 g_{\mu\nu}=g^{(0,0)}_{\mu\nu}+\zeta h^{(1,0)}_{\mu\nu}+\varepsilon h^{(0,1)}_{\mu\nu}+\varepsilon\zeta h^{(1,1)}_{\mu\nu}+...,
 \end{align}
we can derive a Teukolsky equation at each perturbative order. At the $\mathcal{O}(\varepsilon)$ order we have the standard Teukolsky equation in the form
 \begin{align}
 \mathcal{O}^{(0,0)} \Psi_4^{(0,1)}=\mathcal{T}^{(0,1)}.
\end{align}
The operator $ \mathcal{O}^{(0,0)} $  and the source $\mathcal{T}^{(0,1)}$ read
\begin{subequations}
\begin{align}
    \mathcal{O}^{(0,0)}=& \mathcal{E}_4^{(0,0)} F_4^{(0,0)}-\mathcal{E}_3^{(0,0)} F_3^{(0,0)} -3\Psi_2^{(0,0)} \,,
    \\
    \begin{split}\label{Sourceoper00app}
        \mathcal{T}^{(0,1)} 
            = & \mathfrak{T}^{(0,0)ab} T^{(0,1)}_{ab} 
            \\
            = &\left(\mathcal{E}_4^{(0,0)} \mathcal{S}_4^{(0,0)ab}-\mathcal{E}_3^{(0,0)} \mathcal{S}_3^{(0,0)ab}\right) T^{(0,1)}_{ab}.
    \end{split}
\end{align}
\end{subequations}

At the  $\mathcal{O}(\varepsilon\zeta)$  order we can derive the modified Teukolsky equation
  
\begin{widetext}
\begin{align}
\label{modifTeukapp}
 \mathcal{O}^{(0,0)} \Psi_4^{(1,1)}+\mathcal{O}^{(1,0)} \Psi_4^{(0,1)}+ \mathcal{O}^{(0,1)} \Psi_4^{(1,0)}+\mathcal{K}^{(1,1)}(\Psi_3^{(0,1)},\Psi_3^{(1,0)})=\mathcal{T}^{(1,1)}.
\end{align}
The operator $\mathcal{O}^{(1,0)} $ can then be split into two parts
  \begin{align}
  \label{O10operdef}
    \mathcal{O}^{(1,0)}=\mathcal{O}^{(1,0)}_\mathrm{D\mbox{-}vac}+\mathcal{O}^{(1,0)}_{\mathrm{corr}},\hspace{10pt} \mathcal{O}^{(1,0)}_\mathrm{D\mbox{-}vac}= \mathcal{E}_4^{(1,0)} F_4^{(0,0)}+\mathcal{E}_4^{(0,0)} F_4^{(1,0)}-\mathcal{E}_3^{(1,0)} F_3^{(0,0)}-\mathcal{E}_3^{(0,0)} F_3^{(1,0)} -3\Psi_2^{(1,0)},\\
    \mathcal{O}^{(1,0)}_{\mathrm{corr}}=\left(\Psi_2^{(0,0)}\right)^{-1}\left(G_3^{(0,0)}(\Psi_3^{(1,0)})F_3^{(0,0)}-G_4^{(0,0)}(\Psi_3^{(1,0)})F_4^{(0,0)}+S_5^{(1,0)}F_3^{(0,0)}-S_6^{(1,0)}F_4^{(0,0)}\right),
   \end{align}
where $\mathcal{O}^{(1,0)}_{\mathrm{corr}}$ vanishes if the perturbed spacetime remains in the family of Petrov type D vacuum solutions.

The curvature part independent of $\Psi_4$ can also be separated into two parts
\begin{align}
\mathcal{K}^{(1,1)}(\Psi_3^{(0,1)},\Psi_3^{(1,0)})=\mathcal{K}^{(1,1)}_1(\Psi_3^{(0,1)},\Psi_3^{(1,0)})+\mathcal{K}^{(1,1)}_2(\Psi_3^{(0,1)},\Psi_3^{(1,0)}).
 \end{align}
   The first term contains only first-order differential operators acting on $\Psi_3$
\begin{align}\begin{split}
\mathcal{K}^{(1,1)}_1(\Psi_3^{(0,1)},\Psi_3^{(1,0)})=\left(\Psi_2^{(0,0)}\right)^{-1}\big(-G^{(0,0)}_3(\Psi_3^{(1,0)})J^{(0,0)}_3(\Psi_3^{(0,1)})+G^{(0,0)}_4(\Psi_3^{(1,0)})J^{(0,0)}_4(\Psi_3^{(0,1)})-G^{(0,0)}_3(\Psi_3^{(1,0)})S^{(0,1)}_3\\
+G^{(0,0)}_4(\Psi_3^{(1,0)})S^{(0,1)}_4-J^{(0,0)}_3(\Psi_3^{(1,0)})S^{(0,1)}_5+J^{(0,0)}_4(\Psi_3^{(1,0)})S^{(0,1)}_6+(1,0)\leftrightarrow (0,1)\big).
\end{split}\end{align}
The other part containing second-order operators has the form
\begin{align}
\mathcal{K}^{(1,1)}_2(\Psi_3^{(0,1)},\Psi_3^{(1,0)})=\mathcal{K}^{(1,0)}_2\Psi_3^{(0,1)}+\mathcal{K}^{(0,1)}_2\Psi_3^{(1,0)},\hspace{10pt} \mathcal{K}^{(1,0)}_2=\mathcal{E}_3^{(1,0)}J_3^{(0,0)}+\mathcal{E}_3^{(0,0)}J_3^{(1,0)}-\mathcal{E}_4^{(1,0)}J_4^{(0,0)}-\mathcal{E}_4^{(0,0)}J_4^{(1,0)}.
 \end{align}
Alternatively, we can use the commutation relations of the NP derivatives and Ricci identities to eliminate the second derivatives arriving at
\begin{align}\begin{split}
\label{K2tildeapp}
     & \mathcal{K}^{(1,1)}_2(\Psi_3^{(0,1)},\Psi_3^{(1,0)})= \mathcal{\tilde{K}}^{(1,0)}_2\Psi_3^{(0,1)}+\mathcal{\tilde{K}}^{(0,1)}_2\Psi_3^{(1,0)} +20 \Psi_3^{(1,0)}\Psi_3^{(0,1)} \,,
    \\
     &\mathcal{\tilde{K}}^{(1,0)}_2 = \lambda^{(1,0)}\delta^{(0,0)}-\nu^{(1,0)}D^{(0,0)}+2\big(-2D^{(0,0)}\nu^{(1,0)}+2\delta^{(0,0)}\lambda^{(1,0)}-2\bar{\alpha}^{(0,0)}\lambda^{(1,0)}+7\beta^{(0,0)}\lambda^{(1,0)} 
     \\
     & \phantom{\mathcal{\tilde{K}}^{(1,0)}_2 = } -7\epsilon^{(0,0)}\nu^{(1,0)} -2 \bar{\epsilon}^{(0,0)}\nu^{(1,0)}+2\bar{\pi}^{(0,0)}\lambda^{(1,0)}-3\rho^{(0,0)}\nu^{(1,0)}+2\bar{\rho}^{(0,0)}\nu^{(1,0)}+3\tau^{(0,0)}\lambda^{(1,0)}\big) .
 \end{split}\end{align}
From this form it is clear that $\mathcal{\tilde{K}}^{(1,0)}_2=0$ if we stay in a Petrov type D spacetime.

On the right-hand side of eq. \eqref{modifTeukapp} we have the source term $\mathcal{T}^{(1,1)}$  which can be constructed  in terms of the Ricci scalars $\Phi_{AB}$ or in the case of GR in terms of tetrad components of the stress-energy tensor
 \begin{align}
\mathcal{T}^{(1,1)}=\mathfrak{T}^{(1,1)}( T^{(0,1)}_{ab}, T^{(1,0)}_{ab}) +\mathfrak{T}^{(0,0)ab} T^{(1,1)}_{ab}.
   \end{align}
The operator $\mathfrak{T}^{(0,0)ab} $ is the same (eq. \eqref{Sourceoper00app}) as in the standard Teukolsky equation while the other part can be decomposed as 
\begin{align}
\mathfrak{T}^{(1,1)}( T^{(0,1)}_{ab}, T^{(1,0)}_{ab}) =\mathfrak{T}^{(1,0)ab}T^{(0,1)}_{ab}+\mathfrak{T}^{(0,1)ab}T^{(1,0)}_{ab}+\mathfrak{T}^{(1,1)}_q( T^{(0,1)}_{ab}, T^{(1,0)}_{ab}) .
 \end{align}
The linear operator $\mathfrak{T}^{(1,0)}$  and the quadratic part in the stress-energy tensor have the explicit form  
\begin{align}
\mathfrak{T}^{(1,0)ab}&=\mathcal{E}_4^{(0,0)}\mathcal{S}_4^{(1,0)ab}+\mathcal{E}_4^{(1,0)}\mathcal{S}_4^{(0,0)ab}-\mathcal{E}_3^{(0,0)}\mathcal{S}_3^{(1,0)ab}-\mathcal{E}_3^{(1,0)}\mathcal{S}_3^{(0,0)ab} \,,\\
\mathfrak{T}^{(1,1)}_q( T^{(0,1)}_{ab}, T^{(1,0)}_{ab}) &=\left(\Psi_2^{(0,0)}\right)^{-1}\left(\mathcal{S}_3^{(0,0)ab}(T^{(0,1)}_{ab})\mathcal{S}_5^{(0,0)ab}(T^{(1,0)}_{ab})-\mathcal{S}_4^{(0,0)ab}(T^{(0,1)}_{ab})\mathcal{S}_6^{(0,0)ab}(T^{(1,0)}_{ab})\right)+(1,0)\leftrightarrow (0,1).
\end{align}
\end{widetext}

%%%%%%%%%%%%%%%%%%%%%%%%%%%%%%%%%%%%%%%%%%%%%%%%%%%%%%%%%%%%%%%%%%%%%%%%%%%%%%%%%%%%%%%%%%%%%%%%%%%%%%%%%%%%%%%%%%
\section{The background spacetime}\label{app:background}
%%%%%%%%%%%%%%%%%%%%%%%%%%%%%%%%%%%%%%%%%%%%%%%%%%%%%%%%%%%%%%%%%%%%%%%%%%%%%%%%%%%%%%%%%%%%%%%%%%%%%%%%%%%%%%%%%

%%%%%%%%%%%%%%%%%%%%%%%%%%%%%%%%%%%%%%%%%%%%%%%%%%%%%%%%%%%%%%%%%%%%%%%%%%%%%%%%%%%%%%%%%%%%%%%%%%%%%%%%%%%%%%%%%
\subsection{Justification of the pole-dipole approximation}
\label{app:pole-dipole-justification}
%%%%%%%%%%%%%%%%%%%%%%%%%%%%%%%%%%%%%%%%%%%%%%%%%%%%%%%%%%%%%%%%%%%%%%%%%%%%%%%%%%%%%%%%%%%%%%%%%%%%%%%%%%%%%%%%%

\begin{figure}
    \begin{center}
        \includegraphics[width=0.45\textwidth]{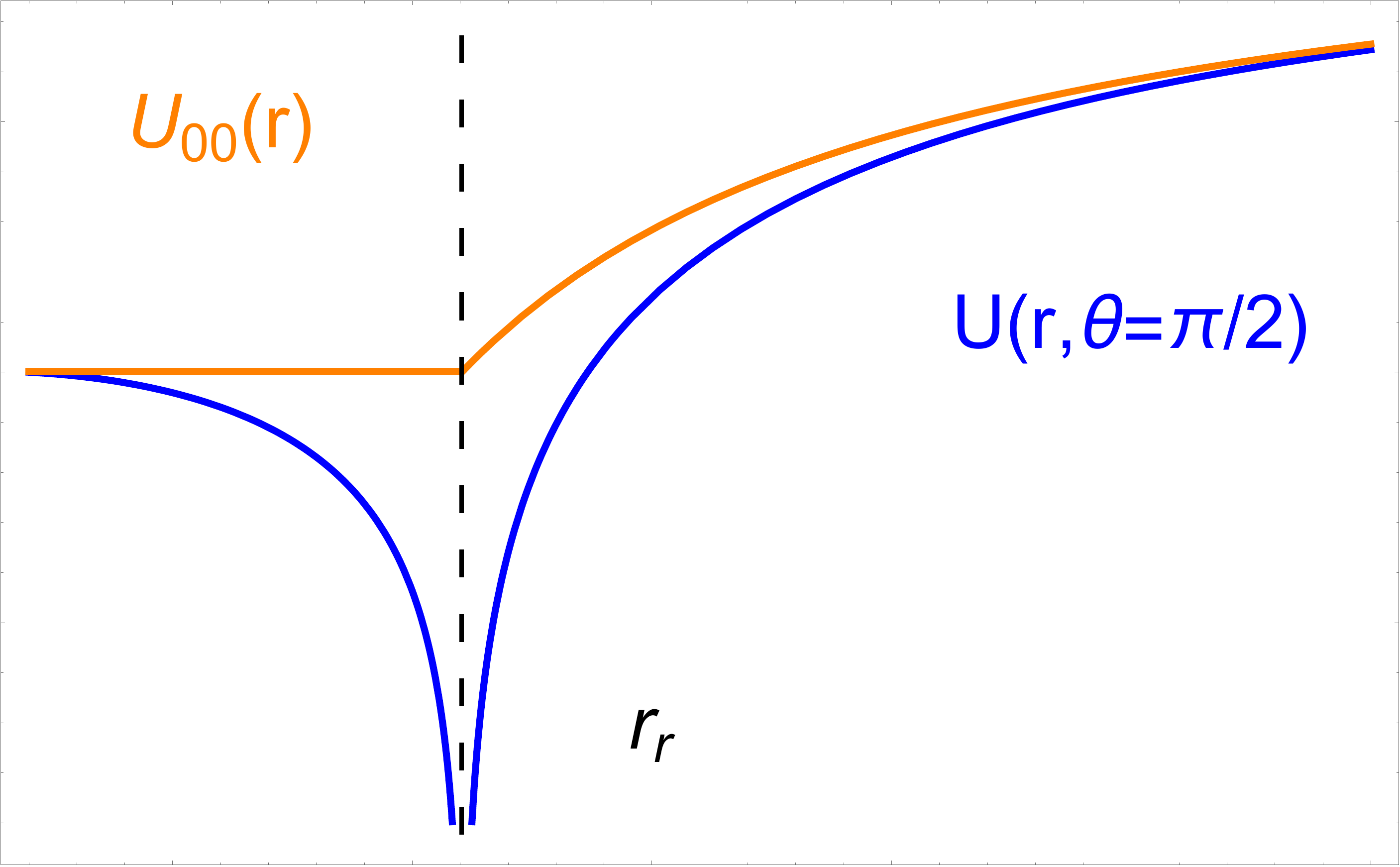}
    \caption{Gravitational potential $U(r,\theta)$    of a ring located at $r=r_{\rm r}$ compared with its monopole truncation  $U_{00 }(r)$.  }     
    \label{fig:monopole}
    \end{center}
\end{figure}

The pole-dipole approximation employed in this work can be justified by examining the exact solution for a rotating ring in general relativity derived by \citet{Will} and \citet{Cizek}. The complete metric for a ring of matter takes the form given by
\begin{align} \begin{split}    
\label{CTB metric}
\mathrm{d}s^2= 
& -e^{2U}\mathrm{d}t^2+ B^2 e^{-2U}R^2\sin^2\theta(\mathrm{d}\phi+ \omega\mathrm{d}t)^2
\\
& +e^{2k-2U}(\mathrm{d}R^2+R^2\mathrm{d}\theta^2 )\,,
\end{split} \end{align} 
where the metric functions $B$, $U$, $k$, and $\omega$ depend on the coordinates $R$ and $\theta$ only. The Schwarzschild metric functions read 
\begin{align} 
\begin{split}    
\label{Schwmetricfunc}
 & B=1-\frac{M^2}{4R^2}, \,
    U_{\rm S} = \ln \left( \frac{2 R - M}{2 R + M} \right), \,
    \\ &\omega_{\rm S} = 0\,,\,
    k_{\rm S} = \ln B.
\end{split}
\end{align}
Inserting these metric functions into \eqref{CTB metric} yields the Schwarzschild metric in isotropic coordinates. We can now linearize the field equations (given in Ref. \cite{Will} or \cite{Cizek}) for the metric functions around the solution \eqref{Schwmetricfunc}.
In total, the perturbed solution reads
 \begin{align}
 \begin{split}
 \label{totalsuperposition}
   \mathrm{d} s^2_{S+r} =& -\left(1 - \frac{2M}{r}\right)(1 + 2 U_{\rm r})\mathrm{d}t^2\\ 
   &+  \frac{1}{1 - \frac{2M}{r}}\left(1 + 2 k_{\rm r} - 2 U_{\rm r}\right) \mathrm{d} r^2 \\ 
   &  + (1 - 2 U_{\rm r})r^2 \left[(1 + 2k_{\rm r}) \mathrm{d}\theta^2 + \sin^2\theta\mathrm{d}\phi^2\right]\\ 
    &-   2\omega_{\rm r} r^2\sin^2\theta\mathrm{d}t\mathrm{d}\phi.
\end{split}
 \end{align}
Here the metric functions $U_{\rm r}$, $\omega_{\rm r}$, and $k_{\rm r}$ are obtained by solving the linearized field equations with a ring source located at $(r = r_{\rm r}, \theta = \pi/2)$. Multipole series for the metric functions can be found in Ref. \cite{Will}, while an exact resummed expression can be found in Ref. \cite{Cizek}.

The exact solutions involve elliptic integrals and exhibit singular behavior at the ring location. More importantly, when expanded in multipoles, the gravitational field shows a clear hierarchy: at distances $r$ from the ring, the $l$-th multipole contribution scales as $(r_{\rm r}/r)^{l+1}$ for $r > r_{\rm r}$ and $(r/r_{\rm r})^l$ for $r < r_{\rm r}$.

For an inspiraling body orbiting at $r_{\rm p} < r_{\rm r}$, the ratio of quadrupole to monopole contributions is approximately
\begin{align}
\frac{|h^{(1,0)}_{\rm quadrupole}|}{|h^{(1,0)}_{\rm monopole}|} \sim \left(\frac{r_{\rm p}}{r_{\rm r}}\right)^2.
\end{align}
Since we consider inspirals where $r_{\rm p} \ll r_{\rm r}$, this ratio is small, justifying the truncation at dipole order. We illustrate the effect of the truncation on the example of the $U$ metric function in Figure \ref{fig:monopole}. By truncating the expansion of the exact ring-hole field at pole-dipole order, one obtains the same ``thin-shell'' metric as derived in the main text through the Israel junction formalism. 

%%%%%%%%%%%%%%%%%%%%%%%%%%%%%%%%%%%%%%%%%%%%%%%%%%%%%%%%%%%%%%%%%%%%%%%%%%%%%%%%%%%%%%%%%%%%%%%%%%%%%%%%%%%%%%%%%
\subsection{Discussion of the thinness condition}
\label{app:thinness-discussion}
%%%%%%%%%%%%%%%%%%%%%%%%%%%%%%%%%%%%%%%%%%%%%%%%%%%%%%%%%%%%%%%%%%%%%%%%%%%%%%%%%%%%%%%%%%%%%%%%%%%%%%%%%%%%%%%%%

We have demonstrated in the previous section that the thin-shell metric is justified at least in a single case --- the infinitely thin ring. Can it be justified also in other cases? In principle, the thin-shell approximation restricts the radial extent of the matter but allows for arbitrary extent in the polar direction. However, considering a physical model for the matter as a gas with approximately isotropic pressure quickly leads to the conclusion that the vertical extent of the cross-section of an axisymmetric and stationary ring must be of the same order of magnitude as the radial extent or smaller. Still, if the configuration requires a ``thin'' ring, which realistic astrophysical configurations will still be satisfactorily approximated within our scheme?

Let us perform a rough scale analysis of a ring with isotropic pressure. The characteristic pressure $P_{\rm r}$ of the ring of physical cross-section diameter $d_{\rm r}$ leads to gradients $ \nabla P \sim P_{\rm r}/d_{\rm r}$. Assuming a rigidly rotating ring with $\Omega_{\rm r} \sim \sqrt{M/r_{\rm r}^3}$, the Euler equations for the fluid dictate that the center of the ring at the pressure maximum moves on a geodesic, while the surface of the ring must fulfill
\begin{align}
    \frac{M}{r_{\rm r}^3} d_{\rm r} \sim \Omega_{\rm r}^2 d_{\rm r} \sim  \frac{P_{\rm r}}{\rho_{\rm r} d_{\rm r}}\,,
\end{align}
where $\rho_{\rm r}$ is the characteristic density in the ring. As a result, we obtain
\begin{align}
    d_{\rm r} \sim \sqrt{\frac{P_{\rm r}}{\rho_{\rm r}} \frac{r_{\rm r}^3}{M}} \,.
\end{align}
The finite-diameter corrections will be controlled by the curvature lengths of the gravitational field in the vicinity of the ring. There are two such lengths, the Schwarzschild curvature length, which is simply $\sqrt{r_{\rm r}^3/M}$, and the curvature length of the GW sourced by EMRI enclosed by the matter ring. The curvature length of the GW is simply its wavelength, which is set by the period of the orbital motion of the EMRI particle, or $\lambda_{\rm GW} \sim T_{\rm orb} \sim \sqrt{r_{\rm p}^3/M}$. For the accuracy of the thin-shell approximation, one must simultaneously require
\begin{align}
    & \frac{d_{\rm r}}{\sqrt{r_{\rm r}^3/M}} \sim \sqrt{\frac{P_{\rm r}}{\rho_{\rm r}}} \ll 1 \,,
    \\
    & \frac{d_{\rm r}}{\sqrt{r_{\rm p}^3/M}} \sim \sqrt{\frac{P_{\rm r}}{\rho_{\rm r}}} \left( \frac{r_{\rm r}}{r_{\rm p}}\right)^{3/2} \ll 1 \,.
\end{align}
The first condition is straightforward; it requires that the fluid should have non-relativistic pressures, or that the sound speeds within the ring are much smaller than the speed of light. Such an approximation is natural and applies to many astrophysical contexts. The second condition is more constraining. It can be recast as
\begin{align}
    \frac{P_r}{\rho_r} \ll \left(\frac{r_{\rm p}}{r_{\rm r}}\right)^3 \,.
\end{align}
Since $r_{\rm p}\ll r_{\rm r}$, this implies that the matter enclosing the BH must be from a cold, essentially dust-like fluid with very low pressures so that it is truly in a collapsed, ring-like state. This is why we refrain from calling our model configuration a ``halo'' or similar, since such extended configurations are out of the scope of our simple model. We hope to generalize our approach to such cases in future work.

%%%%%%%%%%%%%%%%%%%%%%%%%%%%%%%%%%%%%%%%%%%%%%%%%%%%%%%%%%%%%%%%%%%%%%%%%%%%%%%%%%%%%%%%%%%%%%%%%%%%%%%%%%%%%%%%%
 \subsection{Israel junction conditions for the (1,0) metric} \label{app:israel}
 
In this part of the Appendix, we derive the $(1,0)$ metric using the Israel junction conditions \cite{Israel:1966rt}, \cite{Poisson_2004}.
 Recall that we match two metrics describing two regions of our spacetime separated by the matter shell located at $r=r_{\rm r}$. An observer inside the sphere $r=r_{\rm r}$ feels the mass of the black hole $M$ and its spin $a$
\begin{align}
\label{appinnermetric}
\begin{split}
         \mathrm{d} s^2_{r<r_{\rm r}}=
            &  -\left(1 - \frac{2M}{r}\right)\mathrm{d}t^2 + \frac{1}{1 - \frac{2M}{r}}\mathrm{d} r^2
            \\
            &+r^2\mathrm{d}\Omega^2 -\frac{4aM}{r}\sin^2\theta\mathrm{d}t\mathrm{d}\phi \,,
\end{split}
\end{align}
while an observer at infinity sees a combined mass $M+m_{\rm r}$ and  spin parameter $a+a_{\rm r} $    due to the perturbing ring 
\begin{align}
\label{appoutermetric}
\begin{split}
    \mathrm{d} s^2_{r>r_{\rm r}}= 
    & -\left(1 - \frac{2(M+m_{\rm r})}{r}\right)\mathrm{d}t^2 
    \\
    & + \frac{1}{1 - \frac{2M}{r}}\left(1+\frac{2m_{\rm r}}{r-2M}\right) \mathrm{d} r^2
     +r^2\mathrm{d}\Omega^2
    \\
    &-\frac{4\left((a+a_{\rm r})M \right)}{r}\sin^2\theta\mathrm{d}t\mathrm{d}\phi .
\end{split}
\end{align} 
    At this point, let us recall that we treat $a$, $m_{\rm r}$ and $a_{\rm r} $  as $\mathcal{O}(\zeta)$.
 Now, similarly to the metric above, a tensor quantity $\mathcal{A}$ can take values inside the hypersurface $r=r_{\rm r}$ which we shall denote as $\mathcal{A}^-=\mathcal{A}(r<r_{\rm r})$ and outside the shell where we have $\mathcal{A}^+=\mathcal{A}(r>r_{\rm r})$.  An example is the unit normal $n^{({\rm s})}{({\rm r})}_{\mu}$  to the surfaces $r=$constant. 
 
 We can define $[\mathcal{A}]=\mathcal{A}^+\vert_{r=r_{\rm r}}-\mathcal{A}^-\vert_{r=r_{\rm r}}$ as a jump of the tensor $\mathcal{A}$ on the hypersurface $r=r_{\rm r}$ \cite{Poisson_2004}. The first of the two Israel junction conditions then demands that $[g_{\mu\nu}]=0$. Alternatively, one can define the metric on the $r=constant$ surface $h^{({\rm s})}_{\mu\nu}=g_{\mu\nu}\vert_{r=\mathrm{const.}}=g_{\mu\nu}-n^{({\rm s})}_{\mu}n^{({\rm s})}_{\nu}$ and formulate the first condition as follows 
 
    \begin{align}
     \label{appIsrael1}
[h^r_{\mu\nu}]=0, \hspace{15pt} [x^{\mu}]=0.
    \end{align} 
Thus not only the six independent components of $h^{({\rm s})}_{\mu\nu}$ but also all our coordinates $x^{\mu}$ are continuous through the hypersurface $r=r_{\rm r}$. To satisfy this condition, we need to match the components $g_{tt}$ and $g_{t\phi}$ and coordinates $t$ and $\phi$. We shall use the "outer" coordinates $x^{+\mu}:=x^{\mu}$ for our global coordinate chart, while $x^{-\mu}:=x^{\mathrm{ in}\mu}$ is our definition of the "inner" coordinates. The resulting coordinate transformation is then  
   \begin{align}
      \begin{split}
     \label{appcoordmatch}
 t^\mathrm{in}=\left(1-z_{\rm r}\right)t\,, \hspace{15pt}  \phi^\mathrm{ in}=\phi-\Omega_{\rm r}  t\,,\\
 z_{\rm r}=\frac{m_{\rm r}}{r_{\rm r}-2M}\,, \hspace{15pt} \Omega_{\rm r} =\frac{2M a_{\rm r} }{r_{\rm r}^3}\,, 
   \end{split}
    \end{align} 
 with $z_{\rm r}$ being the redshift of the inner region with respect to infinity caused by the shell   while the   angular velocity $\Omega_{\rm r} $ represents a frame dragging with respect to infinity. 
 The nonzero components of the $(1,0)$ metric then read
   \begin{align}
     \label{app10metric}
h_{tt}^{(1,0)} &=\frac{2 m_{\rm r} (r-2 M-(r-r_{\rm r}) \Theta(r-r_{\rm r}))}{r(r_{\rm r}-2 M)} \,,\\
\nonumber h_{rr}^{(1,0)} &=\frac{2 m_{\rm r} r \Theta (r-r_{\rm r})}{(r-2 M)^2} \,,\\
 \nonumber h_{t\phi}^{(1,0)} &= \frac{2Ma_{\rm r}\left((r^3 - r_{\rm r}^3)\Theta(r - r_{\rm r}) - r^3\right)\sin^2{\theta}}{rr_{\rm r}^3}\,.
    \end{align}  
The jump in the metric is realized through the Heaviside step function $\Theta$ and since in $h_{tt}^{(1,0)}$ and $h_{t\phi}^{(1,0)}$ the function $\Theta$ is multiplied by $r-r_{\rm r}$  one can see that these functions are continuous which is not true for $ h_{rr}^{(1,0)}$. This is, however, in full agreement with the first junction condition   \eqref{appIsrael1}  which we have satisfied. In principle, one could also match the $r$ coordinate to get rid of the delta function in the Christoffel symbols component $\Gamma_{rrr}\sim \partial_r  g_{rr} $. However, this would make all the expressions more complicated. What is actually important is to avoid expressions of the form $\Theta(r-r_{\rm r})\delta(r-r_{\rm r})$ in our quantities, since such a product is not well defined in terms of distributions.
Fortunately, since we are interested in linear perturbation no such product appears in our calculation as it would be of the order $\mathcal{O}(\zeta^2)$ or $(2,0)$, which makes the matching \eqref{appcoordmatch} sufficient for our purposes.
In the following, we shall set $a=0$ as it does not play any role in the non-trivial results of this paper.

Now we can derive the effective  stress-energy tensor $T^{(1,0)}_{\mu\nu\,\, {\rm shell}}$,    describing a thin matter shell   located on the sphere $r=r_{\rm r}$, using the second Israel junction condition. The extrinsic curvature of the hypersurface $r=r_{\rm r}$ is given by $K^{({\rm s})}_{\mu\nu}=h\indices{^{({\rm r})}_\mu^\rho}\nabla_{\rho}n^{({\rm s})}_{\nu}$. The surface stress-energy tensor then reads 
     \begin{align}
     \label{appIsrael2}
S^{(1,0)({\rm s})}_{\mu\nu}=-\frac{1}{8\uppi}\left([K^{({\rm s})}_{\mu\nu}]-[K^{({\rm s})}]h^{({\rm s})}_{\mu\nu}\right),
    \end{align} 
where $K^{({\rm s})}=K\indices{^{(s)\mu}_\mu}$. Obviously, we expand the terms on the right-hand side of this formula to the order $\mathcal{O}(\zeta)$.

 The complete 4D stress-energy tensor is proportional \cite{Poisson_2004} to $\delta(\mathfrak{r})$ where $\mathfrak{r}$ is the proper distance from the shell in direction $n^{({\rm s})}_{\nu}$ or explicitly $\mathfrak{r}(r)=\int_{r_{\rm r}}^r\sqrt{g_{rr}(r')}\mathrm{d}r'$ and so we have
   \begin{align}
     \label{app10stressenergyttotal}
     \begin{split}
    T^{(1,0){\rm shell}}_{\mu\nu}
    &=S^{(1,0)({\rm s})}_{\mu\nu}\delta(\mathfrak{r})
    \\
    &=S^{(1,0)({\rm s})}_{\mu\nu}\left(g^{(0,0)}_{rr}\right)^{-\frac{1}{2}}\delta(r-r_{\rm r}).
    \end{split}
\end{align} 
We can now write down its components 
\begin{subequations}
\label{app10stressenergycomponents}
\begin{align}
    T^{(1,0){\rm shell}}_{tt}&=\frac{m_{\rm r} (r_{\rm r}-2 M)}{4 \uppi  r_{\rm r}^3}\delta(r-r_{\rm r}) \,,
    \\ 
    T^{(1,0){\rm shell}}_{t\phi}&=-\frac{3 a_{\rm r}  M   (r_{\rm r}-2 M)\sin ^2\theta}{8 \uppi  r_{\rm r}^3}\delta(r-r_{\rm r}) \,,
    \\
    T^{(1,0)\rm shell}_{\theta\theta}&=\frac{m_{\rm r} M }{8 \uppi(   r_{\rm r}-2M)}\delta(r-r_{\rm r}) \,,
    \\ 
    T^{(1,0)\rm shell}_{\phi\phi}&=\sin ^2(\theta) T^{(1,0){\rm shell}}_{\theta\theta} \,.
\end{align}
\end{subequations}
 From these components one can compute the total energy $E_{\rm s}$ and angular momentum $J_{\rm s}$, obtaining expected results 
\begin{align}
\label{appenergyangmomentum}
    & E_{\rm r}=-\displaystyle\int_{t=\mathrm{const.}} T^{(1,0){\rm shell}}_{\mu\nu}\xi^{(t)\mu} n^{(t)\nu} \sqrt{-g}  \mathrm{d}^3 x=m_{\rm r}, 
    \\  
    & J_{\rm r}=\displaystyle\int_{t=\mathrm{const.}}  T^{(1,0){\rm shell}}_{\mu\nu}\xi^{(\phi)\mu} n^{(t)\nu}\sqrt{-g}  \mathrm{d}^3 x=M a_{\rm r} ,
\end{align}
 where $n^{(t)}=\mathrm{d}t$ is the normal  to the hypersurface $t=\mathrm{const.}$ while $\xi^{(t)}=\partial_t$ and $\xi^{(\phi)}=\partial_\phi$ are the two independent Killing vectors. One can also verify that this stress-energy tensor satisfies $\nabla_\mu T^{(1,0)(s)\mu\nu}=0$.

%%%%%%%%%%%%%%%%%%%%%%%%%%%%%%%%%%%%%%%%%%%%%%%%%%%%%%%%%%%%%%%%%%%%%%%%%%%%%%%%%%%%%%%%%%%%%%%%%%%%%%%%%%%%%%%%%
\subsection{NP quantities on the background} 
\label{app:background NP quantities}

Here we give a complete list of NP quantities introduced in Section \ref{backgroundNP} of the main text.
 The Newman-Penrose tetrad of the background metric \eqref{background} can be expanded as
      \begin{align}
       \label{appBGtetrad}
e^{\mu(BG)}_{a}=(e^{\mu}_{a})^{(0,0)}+\zeta (e^{\mu}_{a})^{(1,0)}.
    \end{align}  
At the $(0,0)$ order we have the Kinnersley tetrad of the Schwarzschild spacetime ($a=0$)
    \begin{align}
       \label{app00tetrad}
l^{\mu(0,0)}&=\left(\frac{r^2}{r(r-2M)},1,0,0\right) ,\\
n^{\mu(0,0)}&=\left(\frac{1}{2},-\frac{r(r-2M)}{2 r^2},0,0\right) ,\\
m^{\mu(0,0)}&=\left(0,0,\frac{1}{\sqrt{2} r},\frac{\mathrm{i} }{\sqrt{2} r \sin (\theta)}\right).
    \end{align}  
    
 From this tetrad one can calculate the $(0,0)$ spin coefficients, the nonzero of which are
 
 \begin{align}
       \label{spincoef00}
  \rho^{(0,0)}&=-\frac{1}{r} \,, \enspace & \mu^{(0,0)}& =  -\frac{r-2 M}{2 r^2}\,,\\
  \gamma^{(0,0)}&=  \frac{M}{2 r^2}\,, \enspace&   \beta^{(0,0)}& = \frac{\cot(\theta)}{2 \sqrt{2} r}\,,\\
    \alpha^{(0,0)}&=- \beta^{(0,0)}\,,
    \end{align}  
  and the only nonzero Weyl scalar  
    \begin{align}
       \label{Weyl00}
\Psi_2^{(0,0)}=-\frac{M}{r^3}.
    \end{align}  
\begin{widetext}
The  $(1,0)$ tetrad expressed using the Heaviside step function reads 

    \begin{align}
    \label{app10tetrad}
       \begin{split}
        l^{\mu(1,0)} &= \left(\frac{m_{\rm r} r \left[r-2 M-(2 M+r-2 r_{\rm r})\Theta (r-r_{\rm r})\right] }{(r-2 M)^2 (r_{\rm r}-2 M)},0,0,\frac{a_{\rm r}  \left[2 M r^2-\left(2 M r^2-r_{\rm r}^3\right)\Theta (r-r_{\rm r}) \right]}{r r_{\rm r}^3 (r-2 M)}\right),
        \\
        n^{\mu(1,0)} &= \left(\frac{m_{\rm r}(1-\Theta (r-r_{\rm r}))}{2(r_{\rm r}-2M) },\frac{m_{\rm r} \Theta (r-r_{\rm r})}{r},0,\frac{a_{\rm r}  \left(2 Mr^2+ \left(r_{\rm r}^3-2 M r^2\right)\Theta (r-r_{\rm r})\right)}{2 r^2 r_{\rm r}^3}\right) ,
        \\
        m^{\mu(1,0)} &= \left(\frac{\mathrm{i} a_{\rm r}  \sin(\theta) \Theta (r-r_{\rm r})}{\sqrt{2} r},0,-\frac{\mathrm{i} a_{\rm r}  \cos (\theta) \Theta (r-r_{\rm r})}{\sqrt{2} r^2},\frac{a_{\rm r}  \cot (\theta) \Theta (r-r_{\rm r})}{\sqrt{2} r^2}\right).
           \end{split}
    \end{align}
We can now write down the complete list of $(1,0)$ spin coefficients
\begin{align}
\label{appspincoef10}
    \kappa^{(1,0)}&= -\frac{\mathrm{i} a_{\rm r}  \sin (\theta) \delta (r-r_{\rm r})}{\sqrt{2} r_{\rm r}}, 
    & 
    \tau^{(1,0)}&= -\frac{\mathrm{i} a_{\rm r}  \sin (\theta) \Theta (r-r_{\rm r})}{\sqrt{2} r^2}+\frac{\mathrm{i} a_{\rm r}  \sin (\theta ) (r_{\rm r}-2 M) \delta (r-r_{\rm r})}{2 \sqrt{2} r_{\rm r}^2},
    \\
    \rho^{(1,0)}&=  -\frac{\mathrm{i} a_{\rm r}  \cos (\theta) \Theta (r-r_{\rm r})}{r^2},
    &\pi^{(1,0)}&=  \frac{\mathrm{i} a_{\rm r}  \sin (\theta ) \Theta (r-r_{\rm r})}{\sqrt{2} r^2}-\frac{i a_{\rm r}  \sin (\theta) (r_{\rm r}-2 M) \delta (r-r_{\rm r})}{2 \sqrt{2} r_{\rm r}^2},
    \\
    \nu^{(1,0)}&=  \frac{\mathrm{i} a_{\rm r}  \sin (\theta ) (r_{\rm r}-2 M)^2 \delta (r-r_{\rm r})}{4 \sqrt{2} r_{\rm r}^3},
    &\mu^{(1,0)}&=  \frac{ (2 m_{\rm r} r-\mathrm{i} a_{\rm r}   (r-2 M)\cos (\theta ))\Theta (r-r_{\rm r})}{2 r^3},   
    \\
    \epsilon^{(1,0)}&=  \frac{ (m_{\rm r} -r_{\rm r}-\mathrm{i} a_{\rm r}(r_{\rm r}-2 M) \cos (\theta)  )\delta (r-r_{\rm r})}{2 r_{\rm r}( r_{\rm r}-2 M )}, 
    &\sigma^{(1,0)}&= 0, 
    \\
    \gamma^{(1,0)}&=   \frac{ (m_{\rm r} r-\mathrm{i} a_{\rm r} (r-2 M) \cos (\theta) )\Theta (r-r_{\rm r})}{2 r^3}
    &\lambda^{(1,0)}&= 0,\\
    &-\frac{(m_{\rm r} r_{\rm r}-\mathrm{i} a_{\rm r} (r_{\rm r}-2 M)\cos (\theta) )\delta (r-r_{\rm r})}{4 r_{\rm r}^2},     
    \\
    \beta^{(1,0)}&= -\frac{\mathrm{i} a_{\rm r}  \cos (\theta ) \cot (\theta) \Theta (r-r_{\rm r})}{2 \sqrt{2} r^2},
    &\alpha^{(1,0)}&= -\frac{\mathrm{i} a_{\rm r}  (3 \cos (2 \theta )-1) \Theta (r-r_{\rm r})}{4 \sqrt{2} r^2 \sin (\theta)},
\end{align}
\end{widetext}

and also the three nonzero Weyl scalars
\begin{align}
       \label{app:Weyl10}
\Psi_1^{(1,0)}&=-\frac{3 \mathrm{i} a_{\rm r}  M \sin (\theta) \delta (r-r_{\rm r})}{2 \sqrt{2} r_{\rm r}^3} \,,\\
\Psi_2^{(1,0)}&=-\frac{ (m_{\rm r} r+3 \mathrm{i} a_{\rm r}  M \cos (\theta))\Theta (r-r_{\rm r})}{r^4} \,,\\
\nonumber&+\frac{m_{\rm r} (2 r_{\rm r}-3 M) \delta (r-r_{\rm r})}{6 r_{\rm r}^2 (r_{\rm r}-2 M)} \,,\\
\Psi_3^{(1,0)}&=\frac{3 \mathrm{i} a_{\rm r}  M(r_{\rm r}-2 M) \sin (\theta)  \delta (r-r_{\rm r})}{4 \sqrt{2} r_{\rm r}^4}.\\
    \end{align}

The Ricci tensor corresponding to the matter shell is described by the following scalars
\begin{align}
    \label{Ricci10}
    & \Phi_{00}^{(1,0)}=\frac{m_{\rm r} \delta (r-r_{\rm r})}{r_{\rm r} (r_{\rm r}-2 M)} \,,\\
    & \Phi_{11}^{(1,0)}=\frac{m_{\rm r} (r_{\rm r}-M) \delta (r-r_{\rm r})}{4 r_{\rm r}^2 (r_{\rm r}-2 M)} \,,\\
    & \Phi_{22}^{(1,0)}=\frac{m_{\rm r} (r_{\rm r}-2 M) \delta (r-r_{\rm r})}{4 r_{\rm r}^3} \,,\\
    & \Phi_{01}^{(1,0)}=-\frac{3 \mathrm{i} a_{\rm r}  M \sin (\theta) \delta (r-r_{\rm r})}{2 \sqrt{2} r_{\rm r}^3} \,,\\
    & \Phi_{02}^{(1,0)}=0 \,,\\
    & \Phi_{12}^{(1,0)}=-\frac{3 \mathrm{i} a_{\rm r}  M \sin (\theta) (r_{\rm r}-2M) \delta (r-r_{\rm r})}{4 \sqrt{2} r_{\rm r}^4} \,,\\
    & \Lambda^{(1,0)}=\frac{m_{\rm r} (r_{\rm r}-3 M) \delta (r-r_{\rm r})}{12 r_{\rm r}^2 (r_{\rm r}-2 M)} \, . 
\end{align}
Using simultaneously type I and II tetrad rotations \eqref{app:tetradrotation} one can in principle find a tetrad for which all Weyl scalars of the background but $\Psi_2$  vanish. The two rotation parameters leading to the tetrad are of the order $ (1,0)$ 
\begin{align}
 \upsilon^{(1,0)}& = -\frac{\mathrm{i} a_{\rm r}  \sin (\theta) (r_{\rm r}-2 M) \delta (r-r_{\rm r})}{4 \sqrt{2} r_{\rm r}} \,,\\
 \varpi^{(1,0)}&=-\frac{\mathrm{i}  a_{\rm r}  \sin (\theta) \delta (r-r_{\rm r})}{2 \sqrt{2}}.
     \end{align} 
In our calculation, however, we use the tetrad \eqref{app10tetrad}.

%%%%%%%%%%%%%%%%%%%%%%%%%%%%%%%%%%%%%%%%%%%%%%%%%%%%%%%%%%%%%%%%%%%%%%%%%%%%%%%%%%%%%%%%%%%%%%%%%%%%%%%%%%%%%%%%%
\subsection{Multipole expansion of the ring } 
 \label{app:ring multipoles}

In this section we would like to provide a clarification of what we mean by ring multipoles. In Newtonian gravity, the Laplace operator is separable in spherical harmonics. It is thus natural to expand the gravitational potential and density in them. If we were doing a Newtonian calculation, we would just have

    \begin{align}
    \rho_{ \rm ring}(r,\theta)=\displaystyle\sum_{l=0}^{\infty}\rho_{(\mathrm{ring})l0}(r) Y_{l0}(\theta)\,,
      \end{align} 
where we have $m=0$ since the ring is axially symmetric. In general relativity, however, we have 10 components of the stress-energy tensor instead of a single scalar $\rho_{ \rm ring}$. Since we work in the NP/GHP formalism, we can use the fact that each tetrad component $T^{(1,0)ab}_{ \rm ring}$ has its own spin weight and expand each component in the spin-weighted spherical harmonics ${}_{s}Y_{l0}(\theta)$

    \begin{align}
    \label{app:multipolesofthering}
     T^{(1,0)ab}_{ \rm ring}(r,\theta)=\displaystyle\sum_{l=\vert s\vert}^{\infty}\mathcal{R}T^{(1,0)ab}_{ \mathrm{(ring)} l0}(r) {}_{s}Y_{l0}(\theta)\,,
      \end{align} 
where the index $l$  can be summed to an arbitrarily high order.    In analogy with the Poisson equation, the linearized Einstein equations can be written in the GHP form (see, for example, \cite{Pound_2021} where the metric is in the Lorenz gauge; a completely general form is presented in Ref. \cite{Green:2019nam}). On the Schwarzschild background, these equations respect spherical symmetry and are thus separable in ${}_{s}Y_{lm}$, leaving us with a system of ordinary differential equations. Then, in principle, one can solve these equations for each $l$ to find the metric components in the form
      
  \begin{align}
    h^{(1,0)ab}= \displaystyle \sum_{l=\vert s\vert}^{\infty}  \mathcal{R}h^{(1,0)ab}_{l0}(r) {}_{s}Y_{l0}(\theta) .  
      \end{align}     
   For the purpose of this work we considered only a monopole and a dipole term for the ring, which means that we have only $l=0,1$. With this restriction, some components such as $T^{(1,0)mm}_{ \rm ring}$ do not contribute to \eqref{app:multipolesofthering} as they have spin weight $\vert s\vert=2$.    The coordinate form is related to the tetrad form as 
      
   \begin{align}
    T^{(1,0)\mu\nu}_{ \rm ring}=T^{(1,0)ab}_{ \rm ring}(e^{\mu}_{a})^{(0,0)}(e^{\nu}_{b})^{(0,0)}\,,
      \end{align} 
which will allow us to relate the stress-energy tensor of the shell \eqref{app10stressenergycomponents} with the stress-energy tensor of the ring  

   \begin{align}
    T^{(1,0)\mu\nu}_{ \rm shell}= \displaystyle \sum_{l=0,1}\mathcal{R}T^{(1,0)ab}_{ (\mathrm{ring})l0} {}_{s}Y_{l0}  (e^{\mu}_{a})^{(0,0)}(e^{\nu}_{b})^{(0,0)}.
      \end{align}    
  The final part is then to identify the relation between the spin parameter $a_{\rm r} $ and the angular momentum of the ring $J_{\rm r}$ using the formula \eqref{appenergyangmomentum}, which leads to the result
  
  \begin{align}
     \label{app:rinangularmom}
J_{\rm r}=M a_{\rm s}=\frac{m_{\rm r}\sqrt{M}(r_{\rm r})^{\frac{3}{2}}}{r_{\rm r}-2M}.
    \end{align} 

%%%%%%%%%%%%%%%%%%%%%%%%%%%%%%%%%%%%%%%%%%%%%%%%%%%%%%%%%%%%%%%%%%%%%%%%%%%%%%%%%%%%%%%%%%%%%%%%%%%%%%%%%%%%%%%%%
\subsection{The transition from \texorpdfstring{${}_sS_{jm}$}{sSjm} to \texorpdfstring{${}_sY_{jm}$}{sYjm}}
\label{app:new angularn expansion}

In this Appendix we shall write down the coefficients used in the expansion of the angular part of $\Psi_4^{(+)}$ (eq.  \eqref{angularexpansion}) into the spin-weighted spherical harmonics.
Starting with the spheroidal harmonics  ${}_s S_{lm\omega}(\theta)$, these can be expanded in powers of spin parameter $a_{\rm r}$  as \cite{ShahWhiting}

  \begin{align}
{}_s S_{lm\omega}(\theta)={}_s Y_{lm}(\theta)+a_{\rm r}\sum_{j=l-1}^{j=l+1}b_{jm\omega}{}_s Y_{jm}(\theta)+\mathcal{O}(a_{\rm r}^2)\,,
   \end{align} 
where the coefficients can be expressed using a single function  ${}_s \beta_{lm\omega}$
  
\begin{align}  
\label{app:bcoefficients }
{}_s b_{l-1m\omega}= {}_s\beta_{lm\omega},\enspace{}_s b_{lm\omega}= 0,\enspace {}_s b_{l+1m\omega}=-{}_s\beta_{l+1m\omega}\,,
  \end{align}
which reads  
  \begin{align}  
{}_s \beta_{lm\omega}= \omega\frac{s \sqrt{l^2-m^2} \sqrt{l^2-s^2}}{l^2 \sqrt{2 l-1} \sqrt{2 l+1}}.
  \end{align} 
  To formulate the matching conditions \eqref{massmatching} and \eqref{spinmatching}, we also use the following identity \cite{seibert2018spin}
  
    \begin{align}
    \label{app:cosY }
\cos (\theta ) {}_sY_{lm}(\theta)=\sum_{j=l-1}^{l+1}{}_sc_{jm}  {}_sY_{jm}(\theta)\,,
  \end{align} 
where for the $c_{jm}$ we have
\begin{align} 
 {}_s c_{l-1m}&={}_s\gamma_{lm},\enspace {}_s c_{l+1m}={}_s\gamma_{l+1m}\\
{}_s c_{lm}&=-\frac{sm}{l(l+1)}.
   \end{align} 
Similarly to the function  ${}_s \beta_{lm\omega}$ \eqref{app:bcoefficients }, the coefficients are given  by $  {}_s\gamma_{lm}$

  \begin{align} 
  {}_s\gamma_{lm}=\frac{1}{l}\sqrt{\frac{(l-m) (l+m)}{(2 l-1) (2 l+1)}} \sqrt{(l-s) (l+s)}.
     \end{align} 
  The identity    \eqref{app:cosY } can also be in principle obtained using the identity     \eqref{Yproduct} since  ${}_0Y_{1,0}(\theta )\sim \cos\theta $.  Finally, the coefficients  $  {}_s d_{lm\omega}(r)$ in the expansion 
   \begin{align}
      \begin{split}
 \label{app:angularexpansion}
\xi^{-4} {}_sS_{lm\omega}(\theta)=&\frac{1}{r^4}\left({}_sY_{lm}(\theta)+a_{\rm r}\sum_{j=l-1}^{l+1}d_{jm\omega}(r) {}_sY_{jm}(\theta)\right)\\+&\mathcal{O}(a_{\rm r}^2)
\end{split}
  \end{align}  
read
      \begin{align}
  {}_s d_{lm\omega}(r)= {}_s b_{lm\omega}+\frac{4 i  {}_sc_{lm}}{r}.
      \end{align}
 
\nocite{*}

\bibliography{apssamp}% Produces the bibliography via BibTeX.

\end{document}